\documentclass[11pt]{article} 

\pdfoutput=1
\usepackage{amsmath}

\usepackage[no-natbib-sort]{jheppub}

\usepackage{verbatim}

\usepackage{amsmath, amssymb}
\usepackage{environ}
\usepackage{mathrsfs}
\usepackage{array,arydshln}

\usepackage{graphicx,epsfig}
\usepackage{epic}
\usepackage{color}
\usepackage{youngtab}
\usepackage{float}

\usepackage{slashed}

\usepackage[nodayofweek]{date time}

\newcommand{\be}{\begin{equation}}
\newcommand{\ee}{\end{equation}}

\newcommand{\sfg}{\sf g}

\newcommand{\ads}{\AdS$_5\times S^5$\ }

\newcommand{\la}{\longrightarrow}

\newcommand{\DD}{{\mathcal D}}

       \newcommand{\beq}{\begin{equation}}
    \newcommand{\eeq}{\end{equation}}
    \newcommand\bea{\begin{eqnarray}}
    \newcommand\eea{\end{eqnarray}}

\def \N {{\cal  N}}
\def \del{ \partial}
\def \la {\label}
\newcommand{\rf}[1]{(\ref{#1})}
\def\ov{\over}
\def\no{\nonumber} \def \aa {{\rm a}}

\def \ci {\cite}
\def \p {\phi}
\def \m {\mu}\def \n {\nu} 

\def \ed {\small
\bibliography{JT-biblio}
 \bibliographystyle{JHEP-v2.9}
\end{document}}

\def \edu {%%%
\end{document}}

\def \foot {\footnote}
\def \b {\beta}

\def \tr {{\rm tr}}
\def \D {\Delta} 
\def \vp {\varphi} 
 \def \ha {{{1 \ov 2}}}

\def \cc {{\rm c}}
\def \aa {{\rm a} }
 
\def \OO  {{\mc O}}   \def \tr  {{\rm tr }}

\newcommand{\aat}[1]{{#1}}

\newcommand{\bpm}{\begin{pmatrix}}
\newcommand{\epm}{\end{pmatrix}}

\newcommand{\rt}{\right}
\newcommand{\PBK}[1]{\ensuremath{\begin{pmatrix}#1\end{pmatrix}}}

\newcommand{\beqn}{\begin{eqnarray}}
\newcommand{\eeqn}{\end{eqnarray}}

\usepackage{slashed}

\def \iffa  {\iffalse}

\begin{document}

\title{On co-dimension 2  defect  anomalies  in $\N=4$ SYM\\   and (2,0) theory  via    brane  probes in AdS/CFT}
%%%%%%%%%%%%%%%%%%%%%%%%%%%%%%%%%%%%%%%%%%%%

\ \ \author{Hongliang Jiang  and   Arkady A. Tseytlin\footnote{Also at the   Institute for Theoretical and Mathematical Physics (ITMP) of Moscow U. and Lebedev Institute. E-mail: tseytlin@imperial.ac.uk}} 

%\iffa 

\affiliation{Blackett Laboratory, Imperial College, London SW7 2AZ, U.K.}
                      
% \emailAdd{h.jiang@imperial.ac.uk}
 %\emailAdd{tseytlin@imperial.ac.uk} %\fi

\abstract{\small  We  consider a  { $\ha$-BPS}  %supersymmetric 
solution for a D3 brane probe  in AdS$_5 \times S^5$ that 
 has world-volume geometry of AdS$_3 \times S^1$. It  intersects the boundary over a  surface
that represents  a   dimension 2 defect  in the boundary $\N=4$  SYM theory.  The  effective  action 
of the probe brane is proportional to the logarithmically divergent volume of AdS$_3$  and  may  thus be interpreted as 
computing   conformal anomaly of  supersymmetric  $S^2$ defect. 
The   classical action scales as $N$.  
We compute the 1-loop  correction  to it due to quantum 
fluctuations of the D3 brane world-volume   fields 
 and compare the result to an earlier  
  suggested expression for the defect anomaly. %prediction. 
%%% v4:
 {With the boundary condition for the zero mode of the world-volume vector required by supersymmetry 
 %and by fixed defect parameters,
 the sum over the modes  gives the 1-loop coefficient $-3$ instead of the expected $-6$.
 We suggest that the remaining part is due to the classical response of the background to the probe.}
 We also perform a  similar  analysis of a   { $\ha$-BPS} %supersymmetric  
  M5 brane probe  solution 
   in AdS$_7 \times S^4$
 with the  world-volume geometry of AdS$_5 \times S^1$ that represents   a dimension 4  defect in the boundary (2,0)   6d theory.  
 Here the classical M5 brane action  computes the  leading  order  $N^2$ term in $a$-anomaly  of the 
%%% v4:
 supersymmetric  $S^4$ defect. 
 {The fluctuations form a tower of chiral multiplets and a flavour current multiplet of the 4d $\N=2$ superconformal symmetry.
 We perform a detailed computation of the
   1-loop   correction to the M5 brane  effective action  and 
   discuss the dependence of their contribution to the $S^4$ defect $a$-anomaly coefficient on the summation prescription.
     %prescription that is supported by a comparison with the orbifold theories. 
 We also point out that the defect anomaly  may  contain a term of order $N$.
 }
   
 }

% \allowdisplaybreaks

% ===============================================================
%_____ Main text  _________________________________________________________

\date{\currenttime}
% \begin{flushleft} \boxed{\small{\tt \today \ \ - \ \  \currenttime }}\end{flushleft}

%%% v4:
%\preprint{\nw{v4 (11/11/26, 10:53): corrections to v3 are marked in boldface}}

 \maketitle

%\flushbottom
%%%%%%%%%%%%%%%%%%%%%%%%%%%%%%%%%%%%
\def \De {\Delta} 
\def \ads {\AdS$_{5}$\ }
\def \te {\textstyle} \def \iffa {\iffalse} 

\def \ha {{\te {1 \ov 2}}}
\def \AdS  {{\rm AdS}}

 \def  \ba { \begin{align} }
 \def  \ea { \end{align} }

\def \gg   {{\rm g}}
\def \cc    {{\rm c}} 
\def \aa  {{\rm a}}

\def \ep {\epsilon}
 \def \k {\kappa} \def \r {\rho} 
\def \rt {{\rm T}}
\def \RR {{\rm R}}
\def \OO {{\cal O}} 
\def \edd {\end{document}} 
\def \td {\tilde} 
\def \rX {{\rm X}}
\def \tO {{\td \OO}}\def \rZ {{\rm Z}}
\def \eps {\epsilon}
\def \vol {{\rm vol}}
 \def \Deltat  {{\bar \OO}}
 
\def \de {\delta} 
\def \KK {{\rm K}}

\def \rmR {{\cal R}}  \def \rmH {{\cal H}}
\def \adsz  {\AdS$_{4}\times S^{7}/\mathbb{Z}_{k}$}
\def \adsc  {\AdS$_{4}\times {\rm CP}^3$}
\def \adssf  {\AdS$_{7}\times S^{4}$}
\def \adssb  {\AdS$_{7,\b}\times S^{4}$}
\def \RR {\mathbb R} \def \ZZ {\mathbb Z}
\def \four  {\tfrac{1}{4}}
\def \AdS {{ \rm AdS}}
  \def \tS  {\tilde S}
\def \wZ  {{\wh Z}} 
\def \V  {{\rm V}}  \def \G {\Gamma} \def \Ze {{\cal Z}} 
\def \ha {{1\ov 2}}  \def \nb {n_{_\b}}
 \def \rZ  {{\rm Z}}
\def \Ze {{\cal Z}}
\def \A  {{\cal A}} 
\def \y  {{\rm y}}  \def \rb  {{\rm b}}
\newcommand{\cX}{\mathcal X}
\newcommand{\cM}{\mathcal M}
\newcommand{\cD}{\mathcal D}
\newcommand{\cF}{\mathcal F}
\newcommand{\cA}{\mathcal A}
\newcommand{\cG}{\mathcal G}
\newcommand{\cT}{\mathcal T}
\newcommand{\cJ}{\mathcal J}
\newcommand{\cO}{\mathcal O}
\newcommand{\cV}{\mathcal V}
\newcommand{\cR}{\mathcal R}

\newcommand{\cS}{\mathcal S}
\newcommand{\cW}{\mathcal W}
\newcommand{\cQ}{\mathcal Q}
\newcommand{\cE}{\mathcal E}
\newcommand{\cH}{\mathcal H}
\newcommand{\cK}{\mathcal K}
\newcommand{\cL}{\mathcal L}
\newcommand{\cI}{\mathcal I}
\newcommand{\cN}{\mathcal N}

\def \LR {{\Lambda}}
\def\ba#1\ea{\begin{align}#1\end{align}}	
\def \I  {I} \def \bm {  } \def \tAdS {\rm AdS}\def \zZ  {Z} \def \ie  {i.e.}
\def \D  {\Delta} 
 \def \g {\gamma} \def \l {\lambda}
 \def \adsss {\AdS$_7 \times S^4$  }
 \def \vp {\varphi} \def \const {{\rm const}}
  \def \adss {\AdS$_5\times S^5$ }\def \a  {\alpha} \def \b {\beta} \def \cF {{\cal F}}
 \def \vol {{\rm vol}}
\def \rg {{g}}\def \pa {\del}
    \def \sfg {\rg}
  \def \p {\del} \def \vtheta {\vartheta}
 \def \un {\underline}
 \def \DD  {{\rm D}}
 \def \MM {{\rm M}}
\def \ha {\tfrac{1}{2}}
\def \vtheta {\vartheta}\def \s {\sigma}
\def \un {\underline} 
\def \oo {10}
\def \na { \nabla} \def \hD {\hat \Delta}  \def \hDelta{\hD}
\def \Z  {{\cal Z}}
 \def \cR {{\cal R}}
 \def \GS {\hat{\G}} 
\def \cbM {{{\bar{ \cal M}}}}

 %%%%%%%%%%%%%%%%%%%%%%%%%%%%%%%%%%%
 \def \np {\newpage}
 
 \section{Introduction}
 
 %V3

 Study of defects    plays an important role in the investigation of properties   of QFT's  in various dimensions. 
  In particular, it  reveals more information  about  aspects of CFT's.
%v3:  In general,   conformal defects are characterized by a  set  anomaly coefficients  and OPE data 
 In general,   conformal defects are characterized by a  set {of} anomaly coefficients  and OPE data 
 %which are much richer due to its 
 determined by interplay of a defect with an ambient CFT   (see, e.g., \ci{Andrei:2018die,Jensen:2018rxu,Chalabi:2020iie,Wang:2021mdq,
 Drukker:2020swu,Chalabi:2021jud,Drukker:2023jxp,Capuozzo:2023fll}  
  and refs. there).

 In  the context of  the AdS/CFT duality 
 the  properties of defects   may  be described in terms of brane probes intersecting the boundary of AdS 
 on which the dual CFT  lives.
 In particular, in  \ci{Drukker:2020swu}  the subleading   contribution  to $S^2$  (co-dimension 4) defect anomaly in 6d (2,0)  theory 
 was   computed by quantising M2  brane probe in AdS$_7 \times S^4$   background. The induced   geometry 
 was AdS$_3$ and the  classical and  1-loop correction were   proportional to the  log divergent 
  volume of AdS$_3$ with $S^2$ as its boundary
  %. It was    thus  log   IR divergent  
   and  thus determined  the  leading and subleading  contributions  to the   defect  $a$-anomaly.
 
 Our aim  below will be to  perform  similar  
 computations in  the case of co-dimension 2 spherical defects in  4d $\N=4$ SYM 
 and 6d (2,0) theories  using the dual  brane probe setup. 
 In the SYM   case  we will   consider  a $\ha$-BPS supersymmetric 
 D3 brane probe in AdS$_5 \times S^5$  
 wrapped on AdS$_3$   and  $S^1$  in AdS$_5$ and  also  on $S^1$ in $S^5$  with  the resulting 
  induced geometry  being AdS$_3 \times S^1$  \ci{Constable:2002xt,Drukker:2008wr,DHoker:2008wvd}.\foot{\la{cod2}In general, the   AdS  factor of the bulk geometry may be parametrized  as 
$
ds_{\AdS_{p+2}}^2=z^{-2}({dz^2+dx^2+x^2ds^2_{ S^ {p-1}}+dy^2}) =
 du^2+\cosh^2 u \; ds^2_{\AdS_p}+\sinh^2 u\; d\psi^2
 $ 
where
 $
 ds^2_{\AdS_p}=d\rho^2+\sinh^2\rho\; ds^2_{ S^{p-1}}$.  The  two metrics are related by 
$
z=r/(\cosh u \cosh\rho-\sinh u\cos\psi), \quad 
x=z \cosh u \sinh\rho, \quad y=z \sinh u \sin \psi  
$.
When $u$ or $\rho$ goes to infinity, we  have $z\to 0$, i.e.  reach the $\mathbb R^{p+1}$ boundary   containing %giving rise to  
a   co-dimension 2 defect  $S^{p-1}$ with radius $r$. It may be described in terms of a probe  brane  in 
 AdS$_{p+2}\times S^q$.  For  closely related discussions see   \ci{Drukker:2008wr,Drukker:2008jm,Gutperle:2020gez,Gupta:2021hko,Drukker:2023bip}. 
% Another   example  of  1-loop   consideration   of D3-brane probe  in   such  background 
 % was discussed in 
 %\ci{Drukker:2005kx,Faraggi:2011bb,Buchbinder:2014nia}.  
}
 Its effective   action   is proportional to vol(\AdS$_3$) = $-2\pi \log (r \LR )$  and thus  should  capture the 
  anomaly  coefficient of  an $S^2$ defect in $\N=4$ SYM theory.
  
  In the    (2,0) theory  case we will   consider   a $\ha$-supersymmetric 
 M5  brane probe in AdS$_7 \times S^4$  
%v3:  which is wrapped on $S^1 \subset$ AdS$_5$  and $S^1\subset S^4$     with 
 which is wrapped on $S^1 \subset$ {\AdS$_7$}  and $S^1\subset S^4$     with   %%% v4: was AdS_5
 the  induced geometry  being AdS$_5 \times S^1$ \ci{DHoker:2008wvd} (cf. also \ci{Lunin:2007ab}). 
 The M5 brane  effective   action   is then  proportional to vol(\AdS$_5$) = $\pi^2 \log (r \LR )$  and  should  capture  the 
 $ a$-anomaly of  an $S^4$ defect in $A_{N-1}$  (2,0)  theory  (cf. \ci{Chalabi:2021jud}).
 %  Both configurations  are $\ha$   BPS.
 
  Let us   note  that a  similar   M2  brane probe solution   in AdS$_4 \times S^7/\mathbb Z_k$  determines 
 % the leading  correction to 
  the vortex defect expectation   value  in the ABJM theory  
  %is given  by a  1-loop  computation  for
   \ci{Drukker:2023bip} 
  (where   the induced geometry  is AdS$_2 \times S^1$   which has finite volume). %    so has different interpretation. 
  %v4
  {For even $k$, an effective field theory calculation of the expectation value $g$ of the vortex loop
that corresponds to a single M2 brane gives
$\log g=F_N(k)-F_{N-1}(k)$, where $F_N(k)$ is the free energy of the ABJM theory on $S^3$ \ci{Cuomo:2026kuo}.
Its large $N$ expansion has no term of order $N^0$, in agreement with the vanishing
of the 1-loop M2 brane correction found in \ci{Drukker:2023bip}.
This is the 3d analog of the relation \rf{7008} discussed below.}

     Other similar 1-loop  computations for  M-branes  in AdS  backgrounds  were recently discussed in \ci{Giombi:2023vzu,Beccaria:2023ujc,Beccaria:2023sph,Beccaria:2023cuo}.
     %aat8
     One of our motivations  here  is to provide more examples when 
     semiclassical  quantization of  supersymmetric branes in curved spaces leads to  consistent results.

 In all of these  cases the  spectrum of fluctuations   on a $p$-brane  brane embedded into AdS$_{p+2} \times S^{q}$ ($q=8-p$ or $9-p$)   so that  the   world-volume metric  is  AdS$_{p}  \times S^1$  %with equal radii
 will  contain  2   scalar modes  corresponding to fluctuations in the    transverse directions of 
 AdS$_{p+2} $  and also  $q-1$ scalar fluctuations    from  $S^{q}$. All scalars propagating  on AdS$_{p}  \times S^1$ 
 will  be  conformally coupled.\foot{\la{scal} Explicitly, their kinetic operator  will  be 
 $-\nabla^2  + { d-2 \ov 4(d-1)}R $    with  $d=p+1$ 
   and  $R$ of unit-radius AdS$_{p}  \times S^1$, i.e.  $R=R(\AdS_p)=- p(p-1)$.
   Expanding in Fourier modes in $S^1$  coordinate   gives a tower of scalars on AdS$_p$  with 
   operators $-\nabla^2_{\AdS_p}   + m^2 $, where $m^2= n^2  + m_0^2= n^2  - {1\ov 4} (p-1)^2$. 
   Assuming the Dirichlet  boundary conditions, 
   the  dimension of the corresponding dual operators at the boundary of AdS$_{p}$  is then 
   $\D[\D - (p-1)] = m^2 = n^2 - {1\ov 4} (p-1)^2$ or $\D- \ha (p-1)= |n|$.} 
  In addition, the  first two scalars will be  mixed, or equivalently, 
  coupled  to an effective   abelian   constant gauge    potential   in $S^1$ 
  direction. This will result in  a shift of   their    $S^1$ mode number:   $n\to n \pm \ha (p-1)  $. 
  
  The fermions will   be  massless  in AdS$_{p}  \times S^1$ but also  coupled  to  the same  constant 
  gauge potential (with  half the charge) 
  and thus  having    $S^1$ mode  number shifted   as   $n\to n  \pm {1\ov 4}  (p-1)  $.\foot{Similar spectrum
   for an M2  probe in AdS$_4 \times S^7/\mathbb Z_k$    was found in  \ci{Sakaguchi:2010dg} 
   and used in \ci{Giombi:2023vzu,Drukker:2023bip}
   (direct analogy with the present case 
 is for $k=2$ when   the radii of AdS$_2$   and $S^1$  are equal).}
%%% v4:
 {The fermions will be  periodic on $S^1$ for $p=3$ and antiperiodic for $p=5$,  %(in the frame used below), 
 so that their mode number $n$ is integer in the first case
 and half-integer in the second.}
%v3:  In addition,  there will   fluctuations of the   world-volume   gauge fields   propagating  in AdS$_{p} \times S^1$ geometry:
 In addition,  there will {be}  fluctuations of the   world-volume   gauge fields   propagating  in AdS$_{p} \times S^1$ geometry:
  vector  in the D3 brane case and  self-dual tensor in the M5  brane case.

 Having found the  fluctuation spectra   we   compute the 1-loop   contribution to the corresponding 
 effective action  using the standard 
 expressions for the determinants  of the  scalars, fermions and  world-volume vector and  antisymmetric tensor  fields 
  propagating on  AdS$_{p}  \times S^1$  where $p$ is odd in the present case  (see \ci{Camporesi:1994ga,%Camporesi:1992tm,
  Camporesi:1994-hig,Giombi:2008vd,Giombi:2013yva,Giombi:2013fka,Giombi:2014iua,Beccaria:2014xda}).  The  coefficient of the IR 
   log divergent vol(\AdS$_p$)   factor   determines the 
    1-loop   contribution to the defect conformal anomaly. It   can be  represented 
  as  an infinite sum over the $S^1$   mode number $n$. This sum happens to be finite in D3 case and 
   quadratically 
%v3:    divergent in the M5  case.  In the latter case we  use 
%v3:    the standard Riemann $\zeta$-function regularization  to  define  it  (like in %providing its   natural  regularization 
%v3:     similar  examples  of  M-brane   computations with  AdS$_p$ with  even $p$  discussed in  
%v3:    \ci{Giombi:2023vzu,Beccaria:2023ujc,Beccaria:2023cuo}).
%%% v4:
   divergent in the M5  case.
   Similar  examples  of  M-brane   computations with  AdS$_p$ with  even $p$ were  discussed in   \ci{Giombi:2023vzu,Beccaria:2023ujc,Beccaria:2023cuo}.
   \iffa
   \foot{Let us note that 
    %The resulting  value depends on how the modes are summed  together  (see Appendix \ref{7001}).
   %In the M5 case its finite part depends on the prescription.
   one may  combine the modes into supermultiplets and compare with the $\N=2$ orbifolds AdS$_5\times S^5/{\mathbb Z}_k$,
whose twisted sector contains the same multiplets and whose  free energy reproduces conformal anomalies 
of the  dual gauge theory. 
   With the prescription for the untwisted sector fixed,
    this rules out the Riemann $\zeta$-function value for the twisted tower.
   We discuss two prescriptions that reproduce the orbifold anomalies but differ in the finite part of the twisted sector (see Appendix \ref{7003}).}
   \fi

 %  We combine the modes into supermultiplets and use the sharp cutoff on the label of a multiplet.
%   This prescription is supported by a comparison with the exactly known conformal anomalies of the ${\mathbb Z}_k$ orbifold theories
%   (see Appendix \ref{7003}).}
 %  We then  comment on the dual gauge theory interpretation of the resulting defect anomaly  expressions.  
   
   %However, the resulting numbers do not seem   to   be reasonable somehow ...   so far. 
   
   \

 This paper is organized as follows. 
 In section \ref{s2D3} we consider a D3 brane probe solution  in \adss  with AdS$_3\times S^1$ world volume metric 
 that should be representing  the supersymmetric $S^2$ defect in  the  boundary $\N=4$  SYM. We  first compute its  classical action 
 that gives the leading  order $N$ term in the   defect anomaly. We  then  find   the  quadratic fluctuations of 
 the   probe  brane fields  near this  brane   configuration and find  the  value of the 1-loop  correction to  its effective  action 
%v3:  that  contributes  a  finite  constant term to the defect anomaly. The  value of this  constant  that 
%v3:  we find does not appear to match 
%v3:  the  expression  suggested  earlier  % found  on the gauge theory side  
%v3:  in   \ci{Jensen:2018rxu,Chalabi:2020iie}. 
%%% v4:
 that  contributes  a  finite  constant term to the defect anomaly. 
 %v4
 {With the supersymmetric boundary condition for the zero mode of the vector field
 this constant is $b^{(1)}=-3$. % when the sum over the modes is taken at fixed $n$.
 This is one half of the value $-6$ that follows from the expression found in \ci{Jensen:2018rxu,Chalabi:2020iie} (see also \ci{Choi:2024ktc}).
 We suggest that the other half comes from the classical 
 response of the background to the probe, which is of the same order, i.e. from the shift $N\to N-\ha$
 in the classical term.}
 
 %v4
{ In section \ref{s3M5} we perform a similar  computation  in the case of a  supersymmetric 
 M5 brane probe in AdS$_7 \times S^4$
 that should be related to an $S^4$   defect in 6d (2,0)  theory. Here the world-volume  metric is AdS$_5 \times S^1$
%v3:  and the classical  value of the action   scales as $N^2$.  We  compute the 1-loop correction  to the effective action  and 
%v3:   regularizing it  using  the $\zeta$-function  get  a prediction
%v3:    for the  subleading  constant  term in the $S^4$ defect anomaly.  
%%% v4:
 and the classical  value of the action   scales as $N^2$.
 We compute the 1-loop correction to the effective action and find that the contribution of the M5 brane fluctuations
to the constant term in the $S^4$ defect anomaly is $a^{(1)}=-\frac{5}{48}$ if the sum over the multiplets is defined with the cutoff on their label,
and $a^{(1)}=-\frac{1}{12}$ if it is defined as in the superconformal index approach of \ci{ArabiArdehali:2014cey}.
 The classical response of the background to the probe is expected to produce also a term of order $N$.}
 % and further constant terms, so that this  may  not yet a prediction  for the full constant term.
 
   %A10 %v2
   
    It is not  clear at the moment how to compare the result we  found  in the probe brane   setup 
   with the   holographic computation of  the $S^4$ defect anomaly in  \ci{Capuozzo:2023fll}
   which   considered only  the supergravity  bubbling geometry  approach  where the numbers of all  M5 
   branes are of the same 
   order and  thus  the leading  anomaly term  superficially scales as $N^3$.
%v3:    A special  choice   of the parameters of the solution which corresponds to a  particular probe limit with leading scaling   being $N^2$   remains to be understood.}
%%% v4:
   {A specialization of this solution with one of the ranks kept fixed has the leading scaling $N^2$, as shown in Appendix \ref{7004}.
   Its precise identification with our probe defect and the matching of the normalizations and of the subleading terms remain to be understood.}
    % which studied the defect from the bubbling geometry from the supergravity point view.}

%   The latter    is  not  currently  available from   alternative    considerations based on supersymmetry or holography, cf. \ci{Chalabi:2020iie,Gutperle:2022pgw,Gutperle:2023yrd,Capuozzo:2023fll}.  \jhl{Is it right to say this? They just give a different class, probably.. Maybe footnote: \footnote{Note the holographic computation of defect anomaly in  \ci{Capuozzo:2023fll} is based on the bubbling geometry which  is different from the probe brane here. The resulting defects a priori may be different.}}

 Section 4 contains concluding remarks on the boundary conditions and on the summation prescription.
 Appendix \ref{scalarflu}  provides  a general derivation  of the  scalar   quadratic  fluctuation  Lagrangian for similar BPS brane    configurations. The explicit form of the 
 spin connection  and the  fermion covariant derivative  are  given in Appendix \ref{spincon}.
 In Appendix \ref{D3massu}   we show that  like the scalar fluctuation  action,    
  the fermion mass   matrix  does not have a non-trivial  dependence on the  value of the ``radial"
   position $u_0$ of the  probe brane in AdS. 
  The supersymmetry of the probe M5 brane solution studied  in section \ref{s3M5}
   is demonstrated in Appendix \ref{kisp}. 
%%% v4:
   {Appendix \ref{7000} contains comments on the boundary conditions, on the supermultiplet structure of the fluctuations, on the prescription for the sum over the $S^1$ modes
   and on the terms of order $N$ in the M5 brane case.}

 \np

 \section{$S^2$ defect  anomaly in $\N=4$ SYM from    D3  probe  in  AdS$_5 \times S^5$  }\la{s2D3}

 Let us first  consider  the  conformal anomaly associated with  a surface defect in $\N=4$ SYM from 
  the dual \adss perspective (see \ci{Drukker:2008jm,Gutperle:2020gez}). 
 We shall parametrize  the \adss   background as  (cf. footnote \ref{cod2})
 \ba
&ds_{10}^2=L^2 \big( du^2+\cosh^2 u \; ds_{\AdS_3}^2 +\sinh^2 u \; d\psi^2 \big) 
+ L^2 \big(d\theta^2+\sin^2\theta\, d\phi^2+\cos^2\theta\, ds^2_{S^3}\big) \ , \la{1}\\
%\ee
%\be
&\la{2} F_5=dC_4  \ , \qquad \qquad 
C_4=L^4(\cosh^4 u -1) \vol_{\AdS_3}\wedge d \psi+\cdots\ , \qquad L^4 = 4\pi g_s N \a'^2 \ , 
\ea
 where in \rf{2}  dots  stand for  ``magnetic"  terms   that make  $F_5=dC_4$   (anti-)self-dual.
 The  bosonic part of the action for  a D3 brane probe  in this background is
\be\la{33}
S=-T_{3}\Big[\int d^4 \xi \;  \sqrt{-\det(G_{\a\b}+2\pi \a' \cF_{\a\b})} -\int C_4\Big]\ , \ \ \ \ \ \ \ \ 
T_{3}=    {1 \ov (2 \pi)^3 g_s  \a'^2 } = {N\ov 2 \pi^2 L^4} \ ,
\ee
where $G_{\a\b} = G_{mn} (X) \del_\a X^m \del_\b X^n$.
Here we are    considering the case of  Minkowski signature  $(-+...+$) 
but  will rotate to the 
 Euclidean one    when discussing the value 
 of the classical and 1-loop free  energy as we will be interested in the case  when the boundary of AdS$_3$ representing the defect 
  is $S^2$.\foot{Note that the formal structure of the spectrum  of  fluctuations of the brane  near the  classical   configuration 
   will   not depend on  the signature choice.}

 We shall consider the  solution for which  the 
 probe D3 brane is  wrapped  on  AdS$_3$     and also on the circle   parametrized  by the $2\pi$ periodic angles  
 $\psi$ and $\phi$  so that 
  % in the following configuration
\be\la{3} 
u=u_0~, \qquad\qquad  \theta={\pi\ov 2}~, \qquad\qquad  \phi=\psi \ . 
\ee
The world-volume gauge field $\cF_{\a\b}$ will have  vanishing background. 
We will identify the  AdS$_3$ coordinates  with  the first 3   world-volume coordinates $\xi^\a$ ($\a=\hat 0,\hat 1,\hat 2$)    
and $\phi=\psi$   with $\xi^{\hat 3}$. 
\foot{\la{not}We shall use the following notation:
$m,n,... $  will stand for  target space indices ($0, 1, ...9$);
 $\alpha, \beta, ...$  will be  the  brane world-volume indices; $i,j,k$   will denote the AdS$_{5}$  indices. 
 Explicit values of the  world-volume indices will be indicated  with hats, e.g.,  $\hat { 1}$, etc. 
  Indices along the  tangent  space directions  will be  underlined, e.g.,  $\underline 9$ or $\underline{\hat 1}$, etc. }

Such   configuration  preserves   half of supersymmetry  \ci{Skenderis:2002vf,Drukker:2008jm,Gupta:2021hko}
(this   may be shown  along the same lines as for the  similar M5 brane case  in Appendix \ref{kisp}).
The induced metric on the brane is then that of  AdS$_3 \times S^1$  with equal radii
\be\la{4} 
ds^2_{D3}=L^2 \cosh^2 u_0 (ds^2_{\AdS_3}+d\psi^2)=  L^2 \cosh^2 u_0\  {\sf g}_{\a\b}(\xi)   d \xi^\a d \xi^\b  \ , 
\ee
where $ {\sf g}_{\a\b}$  is the metric of the unit-radius AdS$_3 \times S^1$.

The defect is represented   by  the boundary of AdS$_3$  that we shall assume to be $S^2$ (see footnote \ref{cod2}). 
The leading large $N$ contribution to the  corresponding free energy  is given by  the   classical value of the Euclidean action 
$S_E$ of the D3  brane probe  
 that is found to be\foot{\la{min}Note 
 %aat8
 that   for the   choice  of $C_4$  in \rf{33} 
 (different from  one in  \ci{Drukker:2008wr}  but  the same as  in \ci{Gutperle:2020gez}) 
 we  have  this potential regular at $u=0$:   for small $u$   the metric  in \rf{1} contains
 $du^2 + u^2 d \psi^2$   so that $C_4 \sim  u^2  d \psi \wedge \vol_{\AdS_3}$ is 
analytic when expressed in terms of Cartesian coordinates  (cf. also footnote \ref{mini}).
 % implying that the potential  is regular  and vanishes  at $u=0$
   As a result,   the classical  value of the action is non-zero. }
\be\label{55}
F^{(0)}={S_E}= T_3 \, L^4\,  2 \pi\,   \text{vol}({\AdS_3} )=-\frac{N}{2\pi^2 }  (2\pi)^2 \log(r \LR ) =-2N\log(r \LR )\ .
\ee
Here we used that  the regularized volume  of   unit-radius AdS$_3$ whose boundary is $S^2$ 
\be\la{5}
\text{vol}({\AdS_3} )=-2\pi\log(r \LR ) \ , 
\ee
where $\LR \to \infty $  is an  IR cutoff  (or UV   cutoff in  the dual gauge theory side)  and $r$ is the radius of $S^2$. 
Note that the  dependence on the position $u_0$ cancelled  out in \rf{55}. The dependence on this  modulus
parameter  will be absent  also in the quantum corrections discussed below.  %One may thus fix  it  to be $u_0=0$.

 On the gauge theory side, the  corresponding $S^2$ defect  free energy $F_{\rm def}\equiv F$ 
  defined  in terms of the partition function as  $Z_{\rm def}= Z \, e^{-F}$ 
   may be  represented as 
  % parametrized by central charge $b$ defined as follows: 
\be
F%_{\rm def}
=  -\frac{1}{3}b \log(r \LR ) \ , \la{6}
\ee
where $b$   is the corresponding defect conformal anomaly coefficient.
%aat3
\iffa 
The discussion in \ci{Jensen:2018rxu,Chalabi:2020iie} implies that 
%  based on  information  about related  supergravity solutions,  
  the case of the  $S^2$ defect 
 corresponds to the  Levi-group $S(U(1) \times U(N-1))$  and so the 
  prediction for the value of the defect  anomaly 
  $b$  in the $SU(N)$   SYM case is 
  \fi 
  %%%%%%%%%%%%%%%
%A10 %v2 
   {The  general expression for the coefficient $b$   was found 
in   \ci{Jensen:2018rxu,Chalabi:2020iie}  %, the anomaly coefficient $b$ has been computed and
and  was shown to  depend only on  the   Levi group of the defect.
 For a single probe D3 brane, the resulting defect  on the boundary corresponds to the     Levi group 
 $S(U(1) \times U(N-1))$, as  discussed in  \cite{Drukker:2008wr}.  In this case  
 we get\foot{%v4
 { The value of $b$ in \rf{7} is the same for the $SU(N)$ and for the $U(N)$  boundary theory.
For the gauge group $G$ and the Levi group $L$ of the defect one has $b=3(\dim G-\dim L)$
\ci{Jensen:2018rxu,Chalabi:2020iie}. For $G=U(N)$ the Levi group is $L=U(1)\times U(N-1)$,
and the overall $U(1)$ factor cancels in the difference,
$b=3\big[N^2-1-(N-1)^2\big]=6N-6$.
Equivalently, $b$ is the central charge of the 2d sigma model with the target space $T^*(G/L)$
(cf. \ci{Chalabi:2020iie}), and $G/L$ is the same in the two cases.
This is different from the case of the surface defect in the (2,0) theory described by a probe M2 brane.
There the defect is charged under the overall $U(1)$, and the values of $b$ in the $SU(N)$ and the $U(N)$ theories
differ by a term of order $1/N$ \ci{Beccaria:2025ahf}.
In the present case the only component of the defect along the overall $U(1)$ is the trace part of its parameters.
It corresponds to a defect in the free abelian theory, for which $b=0$.}}
 % for the anomaly coefficient arising from the D3 brane
  }
\be\la{7} 
b=3\big[ N^2-1-(N-1)^2\big]=6N - 6\ . 
\ee
The defect   free energy  in \rf{6} is expected to be 
  matched to  the  corresponding  value of the  
   effective  action of the D3 brane probe in \adss   background.  Indeed, the leading   order $N$ term in \rf{6},\rf{7}
   is in agreement with the one following from the classical value of the D3 brane action in \rf{55}, i.e. 
   \be \la{888}   b^{(0)} = 6 N \ . \ee
%v3:    The subleading $ b^{(1)} = -6$ term  in \rf{7}  should come from the quantum 1-loop   contribution of the D3 brane    
%v3:     fluctuations   around the above  background.  
%%% v4:
   {The subleading $-6$ term  in \rf{7}  should come from the  order $N^{0} $ corrections.
   These include the quantum 1-loop   contribution of the D3 brane fluctuations   around the above  background and also the classical response of the background to the probe,
   which is of the same order (see the discussion below \rf{41}).
%v3:      According to \rf{7}  all higher (2-loop, etc.)   D3 brane  corrections should be absent   which should be a
%v3:     consequence of the supersymmetry  of this  problem. 
%%% v4:
     Eq. \rf{7} implies that all   higher order $N^{-k}$   corrections should cancel.} % should cancel in the complete defect anomaly.
  %   Whether this follows from the separate vanishing of the higher world-volume loops or from cancellations with the background response
 %    and mixed contributions requires a further analysis.}

    Our aim will be  to compute the   1-loop  correction to \rf{55}. 
    We will need to add together   the fluctuations  of  the ``transverse" scalars, fermions and world-volume gauge vector.
    The structure of the 1-loop  computation is   similar to the one described  in 
    \ci{Faraggi:2011bb,Buchbinder:2014nia} in the case of the  solution of \ci{Drukker:2005kx} 
    where  the induced metric on D3 brane in \adss was    AdS$_2 \times S^2$.

\subsection{Scalar fluctuations}\la{D3scalaraflu}

%The spectrum of  scalar fluctuations  in this case was  previously found in \ci{Constable:2002xt}. 
Choosing the static  gauge  
\be
 \AdS_3=\{\xi^{\hat 0},  \xi^{\hat 1}, \xi^{\hat 2} \} \,,  \qquad \qquad \psi=\xi^{\hat 3}\in [0, 2\pi] \ , \la{8} \ee
 and considering fluctuations of the  two  ``transverse''  AdS$_5$ coordinates
$u =u_0+ \delta u, \  \phi=\xi^{\hat 3}+ \delta \phi$  one finds 
  for the   quadratic  fluctuation part of the D3 brane action in \rf{33} (see Appendix~\ref{scalarflu} for   details 
  of the   derivation of the scalar fluctuation action)
  %explicit computations.
\ba
 \int  (\sqrt{-G}-C_4) & \to c \int  d^4 \xi \sqrt { - \sf g} 
 \Big({\sf g}^{\alpha\beta}\pa_\alpha \delta u \pa_\beta \delta u+{\sf g}^{\alpha\b}\tanh^2 u_0\pa_\alpha \delta \phi \pa_\beta \delta\phi
 +4\tanh u_0 \delta u \pa_{\hat 3} \delta\phi \Big) \no \\
 &= c  \int  d^4 \xi \sqrt { - \sf g} \Big[ { \sf g}^{\a\b}\pa_\a \bar\chi \pa_\b   \chi   +i(\bar \chi \pa_{\hat 3}  \chi-\chi\pa_{\hat 3}  \bar\chi)
  \Big] \ ,   \la{9}\\
%v3:   & \chi\equiv \delta u+i\tanh u_0\, \delta\psi\ , \qquad \qquad  c\equiv   \frac12 \cosh^2 u_0\, L^4 \ .\la{99}
  & \chi\equiv \delta u -  i\tanh u_0\, {\delta\phi}\ , \qquad \qquad  c\equiv   \frac12 \cosh^2 u_0\, L^4 \ .\la{99}  %%% v4: \delta\psi -> \delta\phi
\ea
%{}
Note that  in the second line   $u_0$   enters  only via  the overall factor  $c$
that can be rescaled away.  
Since 
$ { \sf g}^{\a\b}\pa_\a \bar\chi \pa_\b   \chi   +i(\bar \chi \pa_{\hat 3}  \chi-\chi\pa_{\hat 3}  \bar\chi)
=    { \sf g}^{ij}\pa_i \bar\chi  \pa_j  \chi +
  (\pa_{\hat 3}  \chi -i  \chi)(\pa_{\hat 3}  \bar\chi +i \bar\chi)-\bar\chi\chi
$
(where $i,j$ label AdS$_3$ directions) 
we get a   conformally coupled  complex scalar  in AdS$_3 \times S^1$\foot{To recall, 
  the conformally coupled  4d   scalar operator 
 $-\nabla^2 + {1\ov 6} R$  for $R(\AdS_3 \times S^1) = -6$    becomes    $-\nabla^2 -1$  (see also  footnote \ref{scal}).}
   also  coupled 
 to an effective  constant $U(1)$   gauge field  with $A_{\hat 3} =1$ with charge $q=1$. 

Expanding in  Fourier modes  in $\xi^{\hat 3}$  we get 2  towers of   scalar modes  in AdS$_3$   with masses\foot{Note that  
our D3 brane embedding and fluctuation spectrum is different from the one in \ci{Constable:2002xt}   where 
the  brane  was not wrapping $\psi$  of AdS$_5$ or effectively $u_0$ was set to zero from the start.}
\be \la{10}
m^2= n^2  \pm   2n = (n  \pm  1)^2 -1   \ , \qquad \qquad  n=0, \pm 1, \pm 2, .... \ . 
 \ee
 The corresponding values of the AdS$_3$   boundary  field dimensions $\Delta (\Delta -2) = m^2 $ 
   are thus  given by    
   \be \la{111}
    \Delta -1 = |n \pm 1|\ . \ee
    %\jhl{should we have $n\pm 1$ for $\chi, \bar \chi$ respectively. the same for appendix and M5.. just to be consistent with multiplet strucutre and 1-loop det}
    Here we assume the Dirichlet boundary conditions  for the scalars (as appropriate for a  defect interpretation) 
    so that $\Delta-1 \geq 0$.

 To find the  quadratic action of the   4 transverse  fluctuations in $S^5$  
 let us   set $\theta={\pi\ov 2}+v $  and $ds^2_{S^3}=d\varphi _1^2+\cos^2\varphi_1  (d\varphi_2^2+\sin^2\varphi_2\, d\varphi_3^2)$ 
 and introduce 4 cartesian coordinates ${ \sf X}^a=\{x,y,z,w\}$   such that 
 \be\la{13}
v= \sqrt{x^2+y^2+z^2+w^2}~,\quad
\tan \varphi_1= \frac{x}{\sqrt{y^2+z^2+w^2}}~, \quad
\tan \varphi_2= \frac{\sqrt{w^2+z^2}}{ {y }}~, \quad
\tan \varphi_3= \frac{w}{ {z}}~. 
\ee
 Then from \rf{33} we get   ($C_4$ does not contribute at this order, cf. \rf{99}) 
 \be \la{14}
 \int  \sqrt{-G}\  \to  \ \ 
c \int d^4\xi \sqrt { -\sf g}\, \sum_{a=1}^4\big(  
 {\sf g}^{\alpha\beta}\pa_\alpha{ \sf X}^a \pa_\beta{ \sf X}^a  -  { \sf X}^a   { \sf X}^a  \big) ~.
\ee 
 Thus we get  4 conformally coupled scalars  in AdS$_3 \times S^1$. 
 Expanding in $S^1$ modes we get 4 towers of scalar operators with masses 
 \be\la{144}
   m^2=n^2 -1 \ , \qquad \qquad  n=0, \pm 1, \pm 2, ...\ .  \ee
   Again  assuming the Dirichlet boundary conditions  the corresponding 2d scaling  dimensions  are 
 \iffa
 or $(\D-1)^2 = n^2$.
 Here for $n=0$   we get 4 modes   with $\D=1$.   For  higher modes 
 \fi 
  \be 
 \la{15} 
  \D-1 = |n|      %; \qquad \qquad    n=0:  \ \ \ \   \D=  1 , \ \   \D\stackrel{?}{=}0
 \ .  \ee
% Indeed, for $n=\pm 1$  we get   again massless   scalars  so total of   4+4 modes with $\D=2$. 
% Thus \rf{15}    applies to all values of $n$. 

 \subsection{Fermionic  fluctuations}
 
 The quadratic fermionic  part of the D3 brane   action in a  target space   with
  a non-trivial  $F_5$  background    and no  world-volume gauge field 
   may be  written as 
 (see, e.g., \ci{Martucci:2005rb})\foot{Here   we ignore the overall constant factor of $\frac{1}{2}T_3$. 
  The complete  form of the action for a  D3  brane in \adss    background was given in 
 \ci{Metsaev:1998hf}.}
 \be\la{16}
S_f=\int d^4 \xi \; \sqrt{- g }\,g^{\alpha \beta}\,  \bar \Theta (1-\tilde\Gamma_{D3}) \, 
\tilde \Gamma_\a \widehat D_\b \Theta \ , 
\ee
where $g_{\a\b}$ is the   induced metric in the static gauge \eqref{4} (i.e. AdS$_3\times S^1$) % one   in the present case)  
 and 
\ba\la{17}
&\widehat D_\alpha = \pa_\alpha X^m D_m~ , \qquad
D_m = \nabla_m +\frac{1}{16} \slashed F_5 \Gamma_m \otimes (i \sigma_2) \ , \qquad \slashed F_5=\frac{1}{5!}F_{mnklp}\Gamma^{mnklp}\ , \\
&
  \nabla_m = \pa_m +\frac14 \Omega_m^{\underline {n}\underline {k}} \G_{\underline n\underline k} \ , \ \ \ \ \ \ 
  \Gamma_\alpha=\p_\alpha X^m  \G_m \ , \qquad 
   \G_m = E^{\underline m }_m\Gamma_{\underline m }\ , \qquad 
   E^{\underline m }_m E^{\underline n }_n \eta_{\underline m \underline n}=G_{mn}\ , \no \\
   &\qquad \ \ \  \{ \Gamma_{\underline m }, \Gamma_{\underline n}\}=2\eta_{\underline m \underline n} \ , \qquad 
   \{\Gamma_\alpha, \Gamma_\beta\}=2g_{\alpha \beta} \ . \la{18}
\ea
$\Omega_m^{\underline {n}\underline {k}}$ is the  spin   connection in the  10d target space.
For index  notation see footnote \ref{not}.  

 The fermion field  is 
{\small $\Theta=\PBK{\theta_1 \\ \theta_2}$ } where $\theta_{I} $ are 10d  positive chirality MW spinors satisfying
 $ \theta^*= i C\Gamma^{\underline 0}\theta$  and (we use  $\epsilon^{\underline{012\cdots 9}}=\aat{1}$)
 \be \la{222}
 \Gamma_{10} \theta_{I} =\theta_{I}\ , \qquad \ \ \ \  
 \Gamma_{10}=\aat{-} {1\ov 10!} \epsilon^{\underline{m_0}....\un{m_9}} \Gamma_{{\underline m_0}}...\Gamma_{{\underline m_9}} 
 = %\jhl{\pm}
 \Gamma^{\underline 0}\Gamma^{\underline 1}\cdots \Gamma^{\underline 9}\ . \ee
  Also, we use  
$\bar\Theta=i \Theta^\dagger \tilde\Gamma^{\underline 0}=\Theta^T C$.
 In a proper basis, both $\theta_I$ have 32 real components.  
 $\tilde \Gamma_\a$   stands for  $\Gamma_\a\otimes 1_2$
 where $1_2= \delta_{IJ}$.  In \rf{17} we suppressed $1_2$ factors    and   $\sigma_2$  acts on the index $I=1,2$.
  % with size $32\times 2$. 
  Finally, $\tilde\Gamma_{D3}$ in \rf{16} is defined as (here $\epsilon^{\underline {  \hat 0}\underline {  \hat 1}\underline {  \hat 2}\underline{  \hat 3}}= 1$)
\be\la{19}
 \tilde\Gamma_{D3}=\Gamma_{D3}\otimes(-i\sigma_2) \ , \qquad \qquad 
\Gamma_{D3}=\frac{\epsilon^{\alpha_1\cdots \alpha_4}}{4! \sqrt{-g}}\Gamma_{\alpha_1\cdots \alpha_4}=\Gamma_{\underline {  \hat 0}\underline {  \hat 1}\underline {  \hat 2}\underline{  \hat 3}} \ , \qquad \ \    (\tilde\Gamma_{D3})^2 =1 \ . 
%\jhl{\epsilon^{\underline {  \hat 0}\underline {  \hat 1}\underline {  \hat 2}\underline{  \hat 3}}= 1}
\ee
Following \ci{Martucci:2005rb}   we will fix the $\kappa$-symmetry  gauge 
by imposing 
\be \tilde \Gamma_{10}\Theta =\Theta\ , \ \ \ \ \ \ \ \ \  \tilde \Gamma_{10}=\Gamma_{10}\otimes \sigma_3 \ , \qquad \ \ \ 
\Gamma_{10}= %\jhl{\pm}
 \Gamma^{\underline 0 \underline 1...\underline 9}\ .  \la{20}
 \ee
 This is equivalent to    $\theta_2=0$  so   from now on we set    $\theta_1\equiv \vtheta$. 
 Then we get for \rf{16}
\be \la{21}
 S_{f}=\int d^4 \xi \; \sqrt{-  g }\; 
  \bar \vtheta\,   \Gamma^\alpha\Big( \nabla_\alpha    
+\frac{1}{16}  \Gamma_{D3}  \slashed F_5 \Gamma_\alpha  \Big)  \vtheta\ , 
\ee
where we used that  $\{\Gamma_{D3}, \Gamma_\alpha\}=0$.
 
 We shall set $L=1$   and label the 10d coordinates as 
 \be
 X^{0,1,2}=AdS_3~,\quad X^3=u~,  \qquad X^4=\psi~,  \quad 
X^5=\theta, \quad X^6=\phi~, \quad X^{7,8,9}=S^3\ , \la{22}
\ee
where in the static gauge 
$ X^{0,1,2}=\xi^{\hat 0, \hat 1, \hat 2}, \quad X^3=u=u_0,
 \quad X^4 =\psi =\xi^ {\hat 3}, \ 
 \quad X^5=\theta={\pi \ov 2}, \ 
 X^6 =\phi =\xi^ {\hat 3},  \  X^{7,8,9}=$const. 
  The corresponding  spin connection  components are given  in Appendix~\ref{spincon}. In the  present case of \adss
  we  find that     
 the  induced   covariant derivative in \rf{17}  is given by (see \eqref{b15})
\be\la{23} 
\Gamma^\alpha\p_\alpha X^m \nabla_m   =  \slashed \nabla   +2 \tanh u_0  \Gamma_{  \underline 3}  
 +  \frac{1}{2\cosh  u_0 } \Gamma_{\underline 6\underline 4 \underline 3}  \ , 
 \ee 
 where $  \slashed \nabla $    denotes  the Dirac operator on AdS$_3 \times S^1$. 
As a result,  we get from \rf{21}  the  action  for $\vtheta$   of the standard  Dirac  form 
\ba\la{244}
& S_{f} =  \int d^4 \xi \; \sqrt{- g }\,  \bar\vtheta\big(\slashed\nabla+\MM\big) \vtheta\ , \ \ \ \\
&
\MM= 2 \tanh u_0\,   \Gamma_{  \underline 3}  
 +  \frac{1}{2\cosh  u_0 } \Gamma_{\underline 6\underline 4 \underline 3}
  -\frac{1}{16}  \Gamma_{D3}  \Gamma^\alpha   \slashed F_5 \Gamma_\alpha \ . \la{245}
\ea
 To compute the contribution of the $F_5$ term  in \rf{17}   to $\MM$   we note that  the  self-dual 
$F_5$ corresponding to $C_4$ in \rf{2}   is given by 
\be\la{233}
F_5=-4L^4(1+*)\vol_{\AdS_5}=- 4 L^{-1}  \big(E^{\underline{  0}}\wedge E^{\underline{  1}} \wedge E^{\underline{  2}}\wedge E^{\underline{  3}}\wedge E^{\underline{  4}}
+ E^{\underline{  5}}\wedge E^{\underline{  6}} \wedge E^{\underline{  7}}\wedge E^{\underline{  8}}
\wedge E^{\underline{  9}}\big)\ . 
\ee
Setting $L=1$  we get 
$
 \slashed F_5
 = - {4} \big(\Gamma^{\underline 0 \cdots \underline 4}+\Gamma^{\underline 5 \cdots \underline 9}\big),
 $
  and thus 
 \be \la{25}
\MM= 2 \tanh u_0\,   \Gamma_{  \underline 3}  
 +  \frac{1}{2\cosh  u_0 } \Gamma_{\underline 6\underline 4 \underline 3}
+\frac{1 }{4  }  \Gamma_{D3} \Gamma^\alpha   \big(\Gamma^{\underline 0 \cdots \underline 4}+\Gamma^{\underline 5 \cdots \underline 9}\big) \Gamma_\alpha \ ,  
\ee 
where $ \Gamma_{D3} $  is given in 
%according to 
 \rf{19}. 
We have checked   explicitly that   the  result  for  the fermionic spectrum does not depend on $u_0$:
 the dependence on $u_0$ can be  absorbed into a  rotation of $\vtheta$  by a $u_0$ dependent phase in the $(\underline 4 \underline 6)$ plane.\foot{This   rotation    reflects the fact that   on  the classical solution  both $X^4=\psi$ and $X^6=\phi$ 
  are equal to $\xi^{\hat 3}$   and thus  have   the same   projections  on the world volume,  leading to an  effective
  mixing of the $\G$-matrices in the $(\underline 4 \underline 6)$ directions.}
  Thus we may  thus  take the limit $u_0\to 0 $   and keep only the leading terms  (as is easy to see, they are non-singular). 
  
%   without loss of generality.  \jhl{I think it is misleading to consider the case of $u_0=0$ exactly. Instead we may say: For simplicity, we can first consider the limiting case $u_0\to 0$ and keep the leading terms. The spectrum is independent of $u_0$ as we will show later. }
  
Using   that $ 
\Gamma^i   (\Gamma^{\underline 0 \cdots \underline 4}+\Gamma^{\underline 5 \cdots \underline 9}) \Gamma_i=3   (\Gamma^{\underline 0 \cdots \underline 4}-\Gamma^{\underline 5 \cdots \underline 9})$    
 for  $\alpha= i=\hat 0, \hat 1, \hat 2$  and 
$
\Gamma^6   (\Gamma^{\underline 0 \cdots \underline 4}+\Gamma^{\underline 5 \cdots \underline 9}) \Gamma_6=-   (\Gamma^{\underline 0 \cdots \underline 4}-\Gamma^{\underline 5 \cdots \underline 9})$   for  $\alpha =\hat 3
$, together with $  \Gamma_{D3}
 = \Gamma_{\underline {  \hat 0}\underline {  \hat 1}\underline {  \hat 2}\underline{  \hat 3}}
 =  \Gamma_{\underline { 0}\underline { 1}\underline {  2}\underline{ 6}} $, we get
\be\la{26}
\MM= \frac{1}{2  } \Gamma_{\underline 6\underline 4 \underline 3}+\frac{1 }{4}   (3-1)\Gamma_{D3} \big(\Gamma^{\underline 0 \cdots \underline 4}-\Gamma^{\underline 5 \cdots \underline 9}\big)
=\frac{1}{2  } \Gamma_{\underline 6\underline 4 \underline 3}+\frac{1 }{2} \Gamma_{\underline { 0}\underline { 1}\underline {  2}\underline{ 6}}   \big(\Gamma^{\underline 0 \cdots \underline 4}-\Gamma^{\underline 5 \cdots \underline 9}\big)
=\frac12  \Gamma_{\underline 0\underline 1\underline 2\underline 5\underline 7\underline 8\underline 9}  
 \ . 
\ee
Due to the chirality constraint  in \rf{222}, i.e. $\Gamma^{\underline 0 \underline 1\cdots\underline 9}\vtheta =\vtheta $,     we  conclude 
 that  acting on $\vtheta$  the mass operator takes the  following simple form 
\be \la{27} 
\MM     = \frac12  \Gamma_{\underline 0\underline 1\underline 2\underline 5\underline 7\underline 8\underline 9}\Gamma^{\underline 0 \underline 1\cdots\underline 9}
  = -\frac{1}{2  } \Gamma_{\underline 6\underline 4 \underline 3}
    = \frac{1}{2  } \Gamma_{\underline 3\underline 4 \underline 6} \ .
  % \ \qquad \qquad \Gamma_{\underline 6 \underline 4 \underline 3}^2=-1 \ . 
\ee 
Thus  the contribution   of the $F_5$ term in \rf{245}   is twice opposite    that    of the normal component of the spin connection 
 in \rf{23}, i.e.   it effectively reverses the sign of the former.\foot{The same  mass operator is found also for generic $u_0$, 
 see  Appendix~\ref{D3massu}.}
 
%  \jhl{For generic $u_0$, the same mass matrix can be obtained. See appendix~\ref{D3massu} for details. }

 Expanding $\vtheta$ in modes  in $\xi^{\hat 3}$  the Dirac  operator in \rf{244} on AdS$_3\times S^1$ 
 reduces to that  on AdS$_3$ (we use that $\G^{\hat {  \underline 3}} = \G^{\underline 6}$) 
 \be \la{28} 
i ( \slashed\nabla+\MM )  = i\slashed\nabla_{\AdS_3}   +i \G^{\hat 3}  \del_{\hat 3}  +i \MM\ \ \to \ \ 
 i \slashed\nabla_{\AdS_3}   -   \hat M \ , \qquad 
  \ \ \    \hat M =  n \G_{\underline 6}  +   \frac{1}{2  } i \Gamma_{\underline 6\underline 4 \underline 3} \ . 
  \ee 
   Equivalently, we  may write the operator in  \rf{28} as  
 $ i\slashed\nabla_{\AdS_3}   +i \G^{\hat 3}  (\del_{\hat 3}  - \ha \Gamma_{\underline 4 \underline 3})$, i.e. we get 
 a set of 4  massless  fermions in AdS$_3 \times S^1$  coupled to a constant  $U(1)$ gauge  potential  in $\hat 3$
  direction.\foot{Since $\Gamma_{\underline 4 \underline 3}$   commutes with $\G^{\hat {  \underline 3}} = \G^{\underline 6}$
 it can be diagonalized with $\pm i$ as   eigenvalues, i.e.  the $U(1)$ gauge  field is $A_{\hat 3}=1$ with the fermion charges  
  being $q=\pm \ha $.} 
 
 Since $\G_{\underline 6} ^2 =1,\ \  (i\Gamma_{\underline 6 \underline 4 \underline 3})^2=1$
 and $[\G_{\underline 6}, \Gamma_{\underline 6\underline 4 \underline 3}]=0$ 
 we conclude that $\hat M$   has eigenvalues  $m_f=\pm n \pm \ha$  ($n=0, \pm 1, ...$).
 Thus we find 4 towers of 3d   fermions   with  such   masses. The  corresponding dimensions  of the boundary operators are then  (assuming again the standard, i.e. the Dirichlet, boundary conditions) 
 \be\la{29} 
 \Delta-1=|m_f | =|n\pm\tfrac12|  \  , \qquad \qquad n=0, \pm 1, \pm 2, ...  \ . 
 \ee
% For $n <0$ one  is to reverse the sign   in front of  $n\pm\frac12$ getting equivalent set of $\Delta$'s.   \jhl{more precise, sings? e..g  $\Delta-1=|n\pm\tfrac12| $}

 \subsection{Vector field   contribution} %  and multiplet structure}
 
 As  the  world-volume  vector  gauge field in \rf{33}  has no background   value  its contribution to 1-loop partition  function  is 
 the same as of a Maxwell  field  propagating on AdS$_3 \times S^1$   background  with the  standard action 
 $\int d^4 \xi \sqrt{-g} \, \cF^{\a\b} \cF_{\a\b}$.  The  partition function  of a Maxwell vector  on 
  a general  curved 4d   background   may be  written as 
 \be \la{30}
 Z_1= { \det ( - \nabla^2) \ \big[\det ( - g_{\a\b}\nabla^2   +  R_{\a\b} ) \big]^{-1/2}  }\ . 
 %\jhl{\nabla^2_{\a\b}  =\nabla_\a\nabla_\b?}
 \ee
 In the  unit-radius  AdS$_3 \times S^1$    case we have $R_{\a\b} = (- 2 g_{ij}, 0)$   where $g_{ij}$ is the AdS$_3$ metric.
 Then $\det ( - g_{\a\b}\nabla^2   +  R_{\a\b} )= \det ( - \nabla^2) \, \det ( - \nabla^2   -2 )_{ij}$. Splitting $A_\m$ into
 the   longitudinal and transverse parts  we get $ \det ( - \nabla^2   -2 )_{\m\n} =  \det ( - \nabla^2) \det ( -\nabla^2  -2 )_{ij,\perp}$   where $( -\nabla^2  -2 )_{ij,\perp}$ is defined on a transverse $A_\m$  depending on AdS$_3\times S^1$  coordinates. 
 Thus 
 \be \la{31}
 Z_1= \big[\det ( - \nabla^2  -2  )_{ij,\perp} \big]^{-1/2}  \ , \qquad \qquad 
 \nabla^2 =\nabla^2_{_{\AdS_3}} + \del_{\hat 3}^2\ . 
 \ee
 Expanding the transverse  vector $A_\m$   in modes in $\xi^{\hat 3}$  we thus get a tower of transverse 
   3d vectors   with masses
 \be 
 \la{322}
 m^2_1 = n^2-2\ . \ee
  The   corresponding   boundary  dimension  is found from 
  %\foot{Similar  relations  are  found  in  a general case of a 
  %higher spin field, i.e.  $\Delta(\Delta-2) = m^2 _s + s$, etc. (see, e.g., \ci{Giombi:2013yva,Giombi:2013fka} and refs. there).} 
 $\Delta(\Delta-2) = m^2_1 + 1 = n^2 -1 $  and   thus  with the Dirichlet boundary condition choice 
 \be  \la{32} 
  \Delta -1   = |n|  \  ,  \qquad \qquad n=0, \pm 1, \pm 2 , ...\ .\ee 
 % Note in particular  that for $n^2=1$   the relation $\Delta(\Delta-2) = n^2 -1 $ implies that we   should choose  the ``physical" solution 
 % $\Delta=2$, in agreement with \rf{32}. 
  One  can check directly that for $n=0$  the same result   is found by first dimensionally reducing the 
 Maxwell action to AdS$_3$  (i.e. getting a 3d Maxwell field  plus a massless scalar) 
 and  then quantizing in the $\nabla_i A^i=0$ gauge.
%%% v4:
 {Eq. \rf{32} thus describes the modes with $n\not=0$ and provides a formal continuation to $n=0$.
 The boundary condition for the $n=0$ mode that is appropriate to fixed defect parameters 
 requires a separate discussion, see below.}

%v3:  The resulting  fluctuation spectrum   in  \rf{111},\rf{15},\rf{29},\rf{32}) 
 The resulting  fluctuation spectrum   in  \rf{111},\rf{15},\rf{29},\rf{32}  %%% v4: unmatched bracket removed
  is  that of the  supersymmetric $\N=4$  vector multiplet defined on   AdS$_3 \times S^1$. 
  It  can  be  indeed 
 organized into $\N=2$  supermultiplets on AdS$_3$ as  described in \ci{Aharony:2015hix}. 
 Recall   that in flat  4 dimensions    the $\N=4$  vector multiplet 
  is a superposition of  one $\N=2$   vector multiplet  (vector,  2 real  scalars,  2 Weyl fermions)
 % \footnote{\jhl{Note that the vector and Weyl fermion both contains 2 real degrees of freedom.}}) 
 and one $\N=2$  hypermultiplet (4 real  scalars and 2  Weyl fermions). 
 In the present  AdS$_3\times S^1$ case we also  get a   collection of $\N=2$    vector 
 multiplet   and a  hypermultiplet with masses of fields  as given  above.  In terms of AdS$_3$ 
 towers of fields we  have:\foot{In the notation of \ci{Aharony:2015hix}  adapted to the 2d boundary theory  
 these are    (2,2)   vector  multiplet and (2,2)  hypermultiplet.
% \jhl{Note that  in the special case of $n=0$, the vector here actually corresponds to a 3d Maxwell field and a massless scalar field with $\Delta=2$. }
 }
 % supersymmetry of 2d boundary theory 
% we get  (assuming $n\geq 0$):

%\noindent
  (i)   vector   multiplet 
  containing   1 vector  with  $\Delta-1=|n|$, 
  2 scalars  with  $\Delta-1=|n|$, 
   2  fermions  with $\Delta-1=|n\pm   {1\ov 2}| $;   %and  2 fermions $\Delta-1=|n-{1\ov 2}|$; 
   
  % \noindent 
   (ii) 
hypermultiplet   containing 
2 scalars  with $\Delta-1=|n|$, 2 scalars  with 
 $\Delta-1=|n \pm 1|$,   2  fermions $\Delta-1=|n\pm {1\ov 2}|$.  

%%% v4: new paragraphs
{The world-volume translation $\del_{\hat 3}$ corresponds to a central extension of the superconformal algebra $psu(1,1|2)\oplus psu(1,1|2)$ preserved by the defect
(cf. \ci{Aharony:2015hix}).
For $n\not=0$ the above fields form short multiplets with
\ba
\Delta-1:\quad &|n|-1\ \ (1\ {\rm scalar}), \qquad |n|-\ha\ \ (2\ {\rm fermions}), \no \\
&|n|\ \ (4\ {\rm scalars\ and\ the\ vector}), \no \\
&|n|+\ha\ \ (2\ {\rm fermions}),\qquad |n|+1\ \ (1\ {\rm scalar}) \ . \la{7005}
\ea
For $n=0$ the central charge vanishes and the modes should form the short multiplet 
\ba
&4\ {\rm scalars\ with}\ \Delta=1\ , \qquad \qquad 4\ {\rm fermions\ with}\ \Delta=\tfrac{3}{2}\ , \no \\
&4\ {\rm singlet\ scalars\ with}\ \Delta=2 \ . \la{7006}
\ea
The four operators with $\Delta=2$ are exactly marginal. They correspond to the four parameters $(\alpha,\beta,\gamma,\eta)$ of the surface defect of \ci{Gukov:2006jk} (see also \ci{Choi:2024ktc}).
In the probe description $\beta+i\gamma$ is the position of the brane, i.e. the zero mode of $\delta u$ and $\delta\phi$.
The parameter $\alpha$ is the holonomy of the world-volume gauge field, i.e. the zero mode of $A_{\hat 3}$,
and $\eta$ is the holonomy of the dual gauge field (cf. \ci{Drukker:2008wr}).
The latter is the boundary value of the scalar $\sigma$ that is dual to the $n=0$ mode of the 3d vector, $d\sigma=*dA$.

Let $z$ and $x_i$ with $i=1,2$ be the Poincar\'e coordinates of AdS$_3$, with the metric $z^{-2}(dz^2+dx_idx_i)$.
A massless scalar behaves as $\sigma=\sigma_0(x)+{\cal O}(z^2\log z)$ near the boundary, and in the gauge $A_z=0$ one finds
\be\la{7007}
A_i=\epsilon_{ij}\del_j\sigma_0\, \log z+a_i(x)+\dots \ .
\ee
For the defect with fixed $\eta$ the scalar $\sigma$ should satisfy the Dirichlet condition, with $\Delta=2$.
This fixes $\sigma_0$, i.e. the constant $\eta$ and the coefficient of the logarithmic branch, and leaves $a_i$ free.
It is the Neumann condition for $A_i$ with the boundary value integrated over (cf. \ci{Chen:2023lzq}).
The Dirichlet condition for $A_i$ would instead fix $a_i$ and correspond to the Neumann condition for $\sigma$, with $\Delta=0$, which is not consistent with \rf{7006}.
Thus for $n=0$ the vector is equivalent to the two massless scalars $A_{\hat 3}$ and $\sigma$ with $\Delta=2$.}
% (see also  Appendix \ref{7001}).}

\subsection{1-loop free energy}
 
 The expressions  for  1-loop determinants 
  of  the   relevant  fields in AdS$_3$   can be    found, e.g., in \ci{Camporesi:1994ga,%David:2009xg,
 Giombi:2008vd,Giombi:2013fka}. 
%v3:   We shall always assume Dirichlet boundary conditions so  that $\Delta-1 \geq 0$.\foot{In general, if,. e.g., for a scalar 
  We shall always assume the Dirichlet boundary conditions so  that $\Delta-1 \geq 0$ 
  %v4
  {(for the $n=0$ mode of the vector this is the Dirichlet condition for the dual scalar $\sigma$ in \rf{7007})}.\foot{In general, if, e.g., for a scalar 
   in AdS$_{ p+1}$
 we have $(\Delta - {p\ov 2})^2 = m^2$,  then for  the standard 
 Dirichlet case $\Delta=\Delta_+$ where   $\Delta_+ - {p\ov 2}= |m| \geq 0$.}
  %%%%%%%%%%%%%%%%%%%%%%%
%%%%%%%%%%%%%%%%
In particular,  for  a real scalar  
we  get\foot{\la{dua}Let us  recall that   for an operator defined on 
  symmetric traceless transverse  spin $s$  
field in AdS$_3$  \ci{Giombi:2013fka}
\ba
F_s=  & \frac12 \log \det  (-\nabla^2_{_{\AdS_3}} +m^2)_{{\rm T},\perp} 
=   -g_s  \frac{\vol(\AdS_3)}{\vol(S^2)\, [2\Gamma({3\ov 2})]^2}  \lim_{z\to 0}\frac{\pa}{\pa z}
\int_0^\infty d\lambda \frac{\lambda^2+s^2}{[\lambda^2+(\Delta-1)^2 ]^z}\no 
\\ &=
 - \frac{1}{12 \pi } g_s\,  (\Delta-1) \Big[(\Delta-1)^2-3 s^2\Big]  \, \vol(\AdS_3) \ , \qquad \qquad 
 (\Delta-1)^2 = m^2 + s + 1 \ . \no 
\ea
where $g_0=1$ and $g_s=2$ for $s>0$. To get  the free energy for a massless gauge field in AdS$_3$ one is to add the contribution of the ghost  operator. As discussed above, 
here  for $n=0$  the vector contribution $F_1$ 
 is different from the one for a  3d   $s=1$ gauge  field    
 as it also contains an extra massless scalar  part  that cancels the ghost  determinant  contribution.
 %aat8
%v3:  Note also   that if one dualises the  massless  3d vector in AdS$_3$ 
%v3:   to a  massless  scalar   the latter   will  be subject to the  Neumann 
%v3:  boundary condition so will have the opposite  sign of the free energy contribution   compared to 
%v3:  the standard Dirichlet massless scalar. Then  the total  contribution   of a vector in AdS$_3\times S^1$ 
%v3:   dimensionally reduced to AdS$_3$  (i.e. a  combination of a 3d vector and a massless scalar) will be zero, in 
%v3:   agreement with the vanishing of the   vector contribution in \rf{37} or \rf{39}  for $n=0$.}
%%% v4:
}
\be\la{333}
F_0 =\frac12  \log \det (-\nabla^2 _{_{\AdS_3}}+m ^2)=
 -\frac{ 1}{12 \pi }(\Delta-1)^3\,  \vol(\AdS_3)~, \qquad (\Delta-1)^2 =m^2+1\ . 
\ee
%Note that   such   determinants depend only on $m^2$  or $(\Delta -1)^2$ 
% odd powers of $\Delta -1$  should appear  in general with modulus.
% This corresponds to using  $\Delta_+$   dimension corresponding to Dirichlet  boundary conditions 
% as appropriate for a defect  anomaly computation.
For the  vector  contribution we find (see \rf{31})  %, we have $-\nabla^2_{ab}+R_{ab}=-\nabla^2_{ab}-2g_{ab}$, and 
\be\la{34}
F_1=\frac12 \log \det  (-\nabla^2_{_{\AdS_3}} +m ^2)_{ij,_\perp}
= -\frac{2 }{12 \pi } (\Delta-1) \big[(\Delta-1)^2-3  \big]\, \vol(\AdS_3) \ , \ \ \ \ \ \ \ 
(\Delta-1)^2 = m^2+2. 
\ee
 For the 2-component spin $\ha$ fermion  the kinetic operator $i\slashed \nabla_{_{\AdS_3}}  +m_f  $ has 
its square  given by 
%v3:   $-\nabla^2_{_{\AdS_3}}+\frac14 R+m_f^2= -\nabla^2_{_{\AdS_3}}+m_f^2 - \ha $    and thus 
  $-\nabla^2_{_{\AdS_3}}+\frac14 R+m_f^2= -\nabla^2_{_{\AdS_3}}+m_f^2 - {\tfrac{3}{2}} $    and thus   %%% v4: -1/2 -> -3/2
\be\la{35}
F_{1/2}=\frac12 \log \det (-\nabla^2_{_{\AdS_3}}+\tfrac14 R+m_f^2) =-\frac{2}{12\pi }(\Delta-1) \big[(\Delta-1)^2-\tfrac34 \big]\,  \vol(\AdS_3)\ , \ \ \  (\Delta-1)^2=m_f^2.   
\ee
Here $\vol(\AdS_3)$  is given   by \rf{5}, i.e.  each contribution scales as $\log(r\LR )$. 

Let us   first consider  the 1-loop   result found for the  collection of fields 
of the standard $\N=4$   vector multiplet on AdS$_3 \times S^1$, i.e.  
%v3: for   6 conformally coupled   scalars,  a  gauge vector   and 4 Weyl fermions.\foot{Here  do not introduce a  coupling to an 
for   6 conformally coupled   scalars,  a  gauge vector   and 4 Weyl fermions.\foot{Here {we} do not introduce a  coupling to an 
extra $U(1)$  gauge field   so the global supersymmetry on AdS$_3 \times S^1$ is not preserved.}
%  \jhl{Note that the SUSY may be broken in this case.} 
  Upon expansion in $S^1$ modes 
that gives a set of AdS$_3$ fields  for each value of $n=0, \pm 1, ...$: 
 6 scalars  with $m^2=n^2 +{1\ov 6} R = n^2 - 1$,   a  vector with $m^2=n^2-2$  and 4 fermions with $m^2=n^2$. 
Then  using \rf{333},\rf{34},\rf{35}  the total  1-loop   free energy  is found to  vanish 
\ba
& F^{(1)} = 6F_0 + F_1 - 4 F_{1/2}= -  \frac{1}{ 12\pi}P\, \vol(\AdS_3) \ , \ \ \ \ \ \ \ \ \ \, 
P=   \sum_{n=-\infty}^{ \infty}   P_n ~, \la{36}  \\
&P_n =  6  |n|^3   + 2 |n| ( n^2-3 )  -   8 |n| ( n^2 - \tfrac{3}{4} )      = 0 \ . \la{37}
  \ea     
%%% v4:
{Here the $n=0$ mode of the vector is included with the formal zero value of the vector term $2 |n| ( n^2-3 )$.
This corresponds to the Dirichlet condition imposed on the massless 3d vector in AdS$_3$.
Then the dual massless scalar is subject to the Neumann boundary condition and has the opposite sign of the free energy contribution
compared to the standard Dirichlet massless scalar. 

Thus 
 the total contribution of a vector in AdS$_3\times S^1$ dimensionally reduced to AdS$_3$
(i.e. of a combination of a 3d vector and a massless scalar) is zero.
For the defect with fixed parameters the dual scalar satisfies instead the Dirichlet condition (see \rf{7007}).
Then the $n=0$ mode of the vector is equivalent to two massless scalars with $\Delta=2$,
and one would get $P_n=2\delta_{n,0}$ in \rf{37}. This is taken into account in \rf{39} below.}

The cancellation of the  $n^3$  terms is  due to the  balance of the  numbers of bosonic and fermionic  degrees of freedom. 
The   cancellation of the linear in $n$ terms (which  would   produce a quadratic divergence  in the sum in \rf{36}) 
   is    related to   the general fact that   for  the  $\N=4$ vector multiplet defined on a   curved 4-space 
    the coefficient of  quadratic UV  divergence  (determined for $\log \det (-\nabla^2 + X)$  by the  Seeley coefficient 
    $b_2= \tr (  {1\ov 6} R - X) $)   can be shown to vanish:   the 
    conformally coupled scalars    have $b_2=0$ while 
    the vector  contribution  cancels against the fermionic one.\foot{Note also that  in the special case of AdS$_3 \times S^1$  the $b_4$  Seeley coefficient  also vanishes (in agreement with no log UV divergence in \rf{37}):
    this space is conformally flat, i.e. Weyl tensor is zero and also the 4d Euler density $R^* R^*$ vanishes for any $M^3 \times S^1$  space.}

    This   observation should also apply  to the case of the fluctuation spectrum we have found above:
    it  corresponds  to the fields  of the $\N=4$  vector multiplet on AdS$_3 \times S^1$ 
     coupled   also in a specific way  to a constant 
    $U(1)$ gauge potential   in 
    $\hat 3$ direction (which in the present  case  originates from  
     a non-trivial embedding of the D3 brane into  the target space  \adss  background). Its presence  
      shifts  the values of $n$ for 2 scalars  and the fermion modes. It cannot alter the cancellation of 
       UV divergences  but  may  contribute a non-trivial constant  term to the analog of the sum in \rf{36}. 
       
Combining the contributions of the   AdS$_3$  modes in  \rf{111},\rf{15},\rf{29},\rf{32}, i.e.  2 conformally coupled 
 scalars with shifts $\pm 1$, 4 scalars with shift 0,  a vector and 2 sets of fermions with shifts $\pm \ha$   we get the following 
counterpart of \rf{36},\rf{37} 
\ba  &  F^{(1)}= {1\ov 6} P\,  \log (r\LR ) \ ,  \ \ \qquad  \ \ \ \ \ \  P=   \sum_{n=-\infty}^{ \infty}   P_n~ , \la{38}  \\
&\qquad P_n =|n+1|^3 + |n-1|^3  +   4  |n|^3   + 2 |n| ( n^2-3 ) \no\\
& \qquad \ \ \ \ \ \ \ 
%v3:  -   4 |n+\ha | \big[ ( n+\ha)^2 - \tfrac{3}{4} \big]  - 4 |n-\ha | \big[ ( n-\ha)^2 - \tfrac{3}{4} \big]    \ . \la{39}
 -   4 |n+\ha | \big[ ( n+\ha)^2 - \tfrac{3}{4} \big]  - 4 |n-\ha | \big[ ( n-\ha)^2 - \tfrac{3}{4} \big]  {+\, 2\, \delta_{n,0}}  \ . \la{39}   %%% v4: the last term added
  \ea     
For $n\not=0$  we learn that  $P_n= 0$, i.e.   all order  $n^3$  and $n$ terms cancel 
which should be a  consequence of underlying supersymmetry.
%%% v4:

The  non-trivial  contribution  thus comes  just from  the
 $n=0$ level states:    from 2  ``shifted" scalars,  the  fermions
 {and the $n=0$ mode of the vector. As discussed above, the latter is equivalent to the two massless scalars $A_{\hat 3}$ and $\sigma$ with $\Delta=2$.
 They contribute 2 instead of the formal zero value of the term $2 |n| ( n^2-3 )$, which is accounted for by the last term in \rf{39}. Thus}
\be
P=P_0 = 2 {\, +\, 2}  - 4(- \tfrac{1}{4} - \tfrac{1}{4} )   = {6} \ , \qquad \qquad \ \ \ \  b^{(1)}=- \ha P   = {-3} \ .  \la{40} \ee
Adding  $ F^{(1)}$ in \rf{38},\rf{40}  to the classical contribution in \rf{55}
we  finish with
 \be \la{41}
F= F^{(0)}+ F^{(1)} = -2\big(N-  {\tfrac{1}{2}}\big) \log (r \LR )
\ . \ee
%v4
{This is to be compared with the
 prediction  in  \ci{Jensen:2018rxu,Chalabi:2020iie}, i.e.
  $F=  -2(N- 1) \log (r \LR )       $  as  given in \rf{6},\rf{7}.
 Thus the 1-loop term \rf{40}  found from the sum at fixed $n$
 is one half of the expected  value $-6$.

 Let us note that \rf{40} can be written as $b^{(1)}=-12\, \aa_{_{U(1)}}$, where $\aa_{_{U(1)}}=\tfrac{1}{4}$ is the conformal anomaly coefficient of the
 $\N=4$ Maxwell multiplet (see Appendix \ref{7001}). For the $\N=4$ SYM theory with the gauge group $G$ the conformal anomaly coefficient is $\aa(G)=\tfrac{1}{4}\dim G$.
 Then the general expression $b=3(\dim G-\dim L)$ of \ci{Jensen:2018rxu,Chalabi:2020iie} for the defect with the Levi group $L$, which reduces to \rf{7}
 in the present case, can be written as  (cf. \ci{Cuomo:2026kuo})
 \be\la{7008}
 b=12\,\big[\aa(SU(N))-\aa(SU(N-1))\big]-12\, \aa_{_{U(1)}}=(6N-3)-3 \ .
 \ee
  This suggests that the remaining $-3$ is not a fluctuation effect and that the classical term \rf{888} should be computed with
 \be\la{7009}
 N\ \to\ N-\ha \ .
 \ee
 Indeed, the bulk action scales as $N^2$ and the probe action as $N$, so that the classical response of the background to the probe
 contributes at the same order $N^0$ as the 1-loop term.
 The probe carries one unit of the 5-form flux and separates the region $u>u_0$ with $N$ units of the flux from the region $u<u_0$ with $N-1$ units.
 A charged shell feels the mean of the fields on its two sides, which would give \rf{7009}.}\foot{%v4
 {To derive
 \rf{7009} would require the solution of the coupled problem for the metric and the 5-form at fixed asymptotic flux $N$,
 for example from the expansion of the bubbling solution of \ci{Gomis:2007fi} to the second order near the probe limit.
 One should also take into account that the open-string loop and the closed-string exchange describe the same order.}}

%v4
 {An effective field theory derivation of the relation \rf{7008} was recently given in \ci{Cuomo:2026kuo}.
 At large generic values of the scalar parameters, the Gukov--Witten defect is described by the Coulomb-branch theory.
 Its free Abelian sectors have vanishing normalized defect anomaly, while the supersymmetric Wess--Zumino action
 gives $b=12\Delta\aa$, where $\Delta\aa$ is the difference between the Euler anomalies of the original theory
 and the complete massless theory on the Coulomb branch.
 For the Levi group $S(U(1)\times U(N-1))$ this reproduces \rf{7008}.

 In \ci{Cuomo:2026kuo} the dependence of the defect partition function on the radius is expressed
 as the difference of the conformal anomalies of the massless theory on the Coulomb branch and of the original theory on $S^4$.
 In this form the second term in \rf{7008} is the contribution of the free Maxwell multiplet,
 i.e. of the D3 brane fluctuations, and the first term is the difference between the theories with $N$ and $N-1$ colours,
 in agreement with \rf{7009}.
 The same result can also be written with the whole of $b$ coming from the Wess--Zumino action
 and no contribution of the Maxwell multiplet.
 Thus the total is fixed, while its separation into the probe fluctuation
  contribution and the response of the background
 depends on the definition of the 1-loop term.  }
% For example, the sum in \rf{38},\rf{39} taken at fixed $n$ differs by a finite term
% from the result of the 4d proper-time regularization (see Appendix \ref{7001}).}

 \iffa
 This suggests that the remaining $-3$ is not a fluctuation effect and that the classical term \rf{888} should be computed with
 \be\la{7009}
 N\ \to\ N-\ha \ .
 \ee
 Indeed, the bulk action scales as $N^2$ and the probe action as $N$, so that the classical response of the background to the probe
 contributes at the same order $N^0$ as the 1-loop term.
 The probe carries one unit of the 5-form flux and separates the region $u>u_0$ with $N$ units of the flux from the region $u<u_0$ with $N-1$ units.
 A charged shell feels the mean of the fields on its two sides, which would give \rf{7009}.}\foot{{To derive 
 \rf{7009} would require the solution of the coupled problem for the metric and the 5-form at fixed asymptotic flux $N$,
 for example from the expansion of the bubbling solution of \ci{Gomis:2007fi} to the second order near the probe limit.
 One should also take into account that the open-string loop and the closed-string exchange describe the same order.
 Then the separation of the correction into the two parts depends on the definition of the 1-loop determinant.
 %In particular, the sum in \rf{38},\rf{39} taken at fixed $n$ differs by a finite term from the result of the 4d proper-time regularization (see Appendix \ref{7001}).}
 
 \fi

%%%%%%%%%%%%%%%%%%%%%%%%%%%%%%%%%%%%%%%%%%%%%%%%%%%%%%%%
  \section{$S^4$ defect anomaly  in (2,0) theory from   M5 probe in  AdS$_7 \times S^4$  }\label{s3M5}
 
Let us  now consider a similar   computation  in the case of $\ha$  BPS configuration of 
 M5 brane  in  AdS$_7 \times S^4$   with induced metric AdS$_5 \times S^1$. % (see \ci{Gutperle:2020gez}).
% Here subleading result is not known so we will make a prediction. 
 %discuss UV div for (2,0) on AdS$_5 \times S^1$: what we get.  no log.   but what about $b_4?  b_0, b_2  =0$
% \subsection{Classical solution}
 Following 
\ci{Gutperle:2020gez}  we  parametrize  \adsss  as  
($u\in (0, \infty); \  \psi,\phi \in [0, 2\pi]$)\foot{Note that the  $2\pi$ periodicity of $\psi$ guarantees that there is no singularity for  $u\to  0$. %We formally  start   with  Minkowski  signature of AdS$_7$
%and will continue to Euclidean signature when computing free energy.  
Here   $\vol(S^4) =\int  \vol_{S^4} = {8\ov 3} \pi^2$. % \jhl{
}  
% we have   M5  probe in  AdS$_7 \times S^4$ \foot{Choice of $2\pi$ 
%periodicity  of $\psi$  is  required to have regularity near $u=0$. 
%Here in contrast to \rf{1}   we have $\vol(S^4) =\int  \vol_{S^4} = {8\ov 3} \pi^2$. }  
\be\la{60}
ds_{11}^2=L_A^2 \Big(du^2+\cosh^2 u\; ds^2_{\AdS_5}+\sinh^2 u \;d\psi^2\Big) 
+ L^2  \Big( d\theta^2+ \sin^2 \theta\,  d\phi^2  +\cos^2\theta\,  ds^2_{S^2}   \Big) \ ,\ee
\be\la{62}
F_4=dC_3 = {3} L^3  \, \vol_{S^4}= {3}  L^3  \sin\theta \cos^2\theta\,  d\theta\wedge d\phi \wedge \vol_{S^2}\ ,\ee %\qquad 
%C_3 =   - L^3  \cos^3\theta\,  d\phi \wedge \vol_{S^2}\ , \qquad \ee
 \be   L_A =2 L \ , 
\qquad \ \ \  \ \ L^3_A=8\pi N \ell_P^3\ . \la{622}
\ee
% $F_7 =dC_6 =*_{11}F_4$ with 
We will   assume  the Minkowski    signature of this 11d background  but at the end will rotate to the Euclidean one 
  as will be interested in the case   when  the boundary of AdS$_5$ is $S^4$.

The bosonic part of the   action of  an  M5 brane in a  11d supergravity   background 
  may be written as  \cite{Bandos:1997ui,Aganagic:1997zq,Howe:1997fb}  %(see also \cite{Howe:1996yn,Bandos:1997ui,Claus:1998fh})
\ba
\la{3.6}
S =& - T_{5}\Big\{ \, \int d^{6}\xi\,\Big[\sqrt{-\det(G_{\a\b}+i \widetilde{\rm H}_{\a\b})} - \tfrac{1}{4}\,\sqrt{-G}\, \widetilde{\rm H}^{\star\, \a\b}\widetilde{\rm H}_{\a\b}\Big]
%a28
% \aat{+} 
 -  \int\big(C_{6}+\tfrac{1}{2}{\rm H}\wedge C_{3}\big)\Big\}\ , \\
& H_{\m\n\l} = 3\partial_{[\m}A_{\n\lambda]}~, \qquad {\rm H}_{\m\n\l} = H_{\m\n\l}-C_{ \m\n\l}, \qquad \ {\rm H}^{\star\, \m\n\l} = \frac{1}{6\,\sqrt{-G}}\,\eps^{\m\n\l \a\b\g }{\rm H}_{\a\b\g}~,   \\
& \widetilde {\rm H}_{\m\n} = {\rm H}^{\star}_{\m\n\l}\,U^{\l}~, \qquad \ \ \ \widetilde {\rm H}_{\m\n}^{\star} = {\rm H}_{\m\n\l}U^{\l}, 
\qquad \ \ \  U_{\l}(\xi) \equiv   \frac{\partial_{\l}a(\xi) }{\sqrt{(\partial_{\m} a)^{2}}} \ ,\no\\
& \ \ \ \qquad
   T_5= {1\ov (2 \pi)^5 \ell_P^6}= {2N^2\ov \pi^3 L^6_A} \ . \la{355}
\ea
Here 
 $G_{\a\b} = \del_\a X^m \del_\b X^n G_{mn} (X(\xi))$   ($X^m$ are 11d coordinates),   $C_{\m\n\l} = C_{mnk} \del_\m X^m \del_\n X^n\del_\l X^k     $   and 
 $H_{\a\b\g}$ (which is self-dual on shell) is the  field strength of the world-volume  antisymmetric gauge field $A_{\a\b}(\xi)$.
 % on the world-volume
% and $\theta$ denotes  the   11d  fermions ($S_{\rm F} (\theta)$  is given in  \rf{3.41}  below). 
 The auxiliary scalar $a(\xi)$    may  be fixed   by a gauge choice    $a(\xi)=\xi^{\hat 5}$ \cite{Pasti:1997gx}  and will play no role below. 
 The 6-form potential 
 $C_{6}$  is defined by\footnote{$F^\star _{4}$ is the 11d dual of $F_4$.
 %  defined 
% originally {in Minkowski space} where $F_{4}$ is real.  In the present case 
%  Here  $\star F_{4}$ is purely spatial and  thus remains real  after  rotation to Euclidean space. 
 Note also that $d(dC_{6})=0$ on the equations of motion  for  $C_{3}$  (assuming   there is no 11d  gravitino  
 background). }% see, e.g.,  \cite{Candiello:1993di}).}
\be
\la{3.5}
dC_{6} =  F^\star _{4}-\frac{1}{2} C_{3}\wedge F_{4} \ , 
\ee
%and fermionic terms were discussed in \cite{Howe:1996yn,Bandos:1997ui,Aganagic:1997zq}.
Then  from \rf{62} we get\foot{\la{mini} In general, the  %About  possible ambiguity of   shifting $C_6$ by a constant: 
 WZ term  in \rf{3.6}    should be defined  in terms  of an  integral  of $F_7   + \tfrac{1}{2}{\rm H}_3 \wedge  F_4$ over 
 7-space with 6d boundary  (cf. also  \ci{Drukker:2005kx}).
   Then it is invariant under    ``large''  gauge transformations that change 
$C_6$  and may in principle  change its integral. 
The result does not depend on a choice of 7-space as long as the 
charge  quantization condition is satisfied. 
%This implies   that a  shift of  by a constant  should not matter. 
Note  that  the shift  by $-1$ in \rf{349} is  required for the  potential  $C_6$ not to be singular near the origin $u\to 0$ (cf.  footnote \ref{min}).}
%By analogy with suggestion of Drukker  (in email) to follow  \ci{Drukker:2005kx}    we may try 
\iffa 
Thus let us   write $\int C_6 = \int  F_7 = 6 \int \sinh u \cosh^5 u  du \wedge \vol_{\AdS_5}\wedge d\psi $
and integrate  over 7-volume with boundary AdS$_5 \times S^1$.  This  may be interpreted as  6 times the 
 volume of AdS$_7$ that upon regularization as in \rf{3111} gives $- {1\ov 3} \pi^2 \log \Lambda r $. 
 This is same  that comes from -1  term  in $C_6$ in \rf{349}:      we get 
 $-  2 \pi \times \pi^2 (\Lambda r)$. 
 Again,  this requires accepting $\int F_7 $ as a fundamental definition of the WZ term -- 
 it is clear that shift of $C_6$ by a constant does  change the result regardless the above argument unless we change the definition of WZ term. }
 \fi 
% using \rf{3.5}   we get\foot{  {\bf   why?   should be extra  i 
% here unless signature is Minkowski -- see  footnote \ref{31777}}}
\be\la{349}
C_6 =% \aat{-}
L_A^6 (\cosh^6 u -1) \vol_{\AdS_5}\wedge d\psi \ .
\ee
%{what is the reason of -1 above? This choice ensures that the potential vanishes at $u=0$.}
The  BPS solution    for the M5  brane wrapped  on AdS$_5\subset $ AdS$_7$  and also on  two circles 
$\psi$ and $\phi$     has vanishing  3-form ${\rm H}_{\m\n\l}$
  and  is a direct analog of the D3 brane solution in \rf{3}
\be u=u_0~, \qquad \ \   \theta =\frac{\pi}{2} \ , \qquad \ \   \phi= 2 \psi \ . \la{51}  \ee
Here  the  factor of 2 in the relation between $\phi$ and $\psi$   is related to  the 
 factor of  2 ratio of the AdS$_7$ and $S^4$ radii  in \rf{622}. 
  $u_0$ is  an arbitrary modulus. As both $\psi$ and $\phi$ are 2$\pi$ periodic   the   brane wraps  twice  around the  $\phi$ circle of $S^4$. 
  This M5 embedding  preserves  half of supersymmetry 
%v3:   of the AdS$_4 \times S^7$ background (see Appendix \ref{kisp}). 
  of the {\AdS$_7 \times S^4$} background (see Appendix \ref{kisp}).  %%% v4: was AdS_4 x S^7

%Explicitly,   the M5 brane solution is  % ($\xi^\a$ are world-volume  coordinates; $i=2, ..., 6$)
%\be 
%\psi= \xi^1 \ , \qquad  X^i(\AdS_5) = \xi^i \ , \ \ \ \ \ \  u=u_0 \ , \qquad 
%   \theta= {\pi\ov 2} \ , \ \ \ \   \vp=0\ , \qquad  Y^r(S^2) = \const \ .  \ee
The induced metric  %$ ds_{M5}^2=  g_{ab} d \xi^a d \xi^b$ 
on M5 brane is then   %(\jhl{what is the period of the circle $\psi, \phi$?}): 
\ba
ds_{M5}^2=  &L^2_A  \big( \cosh^2 u_0\; ds^2_{\AdS_5}+\sinh^2 u_0 \, d\psi^2 \big)+  4L^2  d\psi ^2 =
L^2_A \cosh^2 u_0\big(   ds^2_{\AdS_5 }  + d \psi^2 \big)\no\\= &   g_{\a\b}  d\xi^\a d \xi^\b =  L^2_A \cosh^2 u_0 \,  {\sf g}_{\a\b} d \xi^\a d \xi^\b   \ ,   \la{64} 
\ea
where   ${\sf g}_{\a\b}$    is the metric of unit-radius  AdS$_5 \times S^1$.
% where  both factors have the same radius

The classical  value of the Euclidean M5 brane action $S_E$   corresponding to 
the Minkowski one $S=- T_5\big(\int   d^6 \xi \sqrt{- \det G}  - \int  C_6\big) $   found from   \rf{3.6} is 
%\foot{\bf 
%Here we  formally assume  Minkowski signature   on the world volume  so that $C_6$ is real.
%It is better to switch to Euclidean signature.} 
\be\la{644}
F^{(0)} = S_E%=T_5\Big(\int   d^6 \xi \sqrt{\det G}  -  \int  C_6\Big)
=2\pi  L_A^6 \, T_5 \, \text{vol}(\AdS_5)
%=128\pi L^6\times \frac{1}{(2\pi)^5 \ell_p^6}\pi^2 \log (\Lambda r)
%=? 16\pi L^6 \text{Vol}(\AdS_5\times S^1) 
= 4N^2\log (r\LR) \ . 
\ee
 Here we  assumed that AdS$_5$ has  $S^4$ boundary   and  used that  for global  odd-dimensional AdS space one has 
  (cf. \rf{5})
\be\la{3111}
\text{vol}(\AdS_{2n+1})=\frac{2(-1)^n \pi^n}{\Gamma(n+1)}\log (r\LR )
\xrightarrow{n=2}  \text{vol}(\AdS_{5}) = \pi^2 \log (r\LR)~,
\ee
where  $\Lambda$ is  an  IR cutoff and  $r$ is the radius of  $S^4$. 
The   value of the classical action does not depend on $u_0$ and the same will be true also for the   contribution of the quantum fluctuations. 

As usual, we express the free energy on $S^4$ 
 in terms of the $a$-anomaly coefficient 
 as 
 \be \la{fff}
 F=4a\log (r\LR)   \ . \ee
Then the expression in \rf{644}  corresponds to  the leading large $N$ value of the 
$a$-anomaly for the $S^4$ defect  in  (2,0)   theory 
being\foot{Incidentally,   this is 4 times the  (large $N$ part of)   conformal  anomaly of $SU(N)$ $\N=4$ SYM   theory on $S^4$.
There 
 should not be   any direct 
 connection   to  the $\N=4$ SYM anomaly   which has  dual description in terms of  the 10d 
 supergravity (string theory)  on AdS$_5 \times S^5$.}
% We can then compare it to \eqref{aanomaly}, then we get the anomaly coefficient
\be
a^{(0)}= N^2 \ . \la{001}
\ee
%that is obtained from (2,0) theory by naive dimensional reduction on 2-torus. 
The leading $N^2$   scaling is consistent  with the expectation that  it should be effectively determined by gauge theory degrees of freedom (cf. \ci{Capuozzo:2023fll}).\foot{This is  also consistent with the discussion in Appendix B of \ci{Santilli:2023fuh} 
 although there the  defect had shape $S^1\times  S^3$ and  thus the coefficient of   conformal anomaly  was zero.}

%v3: Our aim will  be to compute the  subleading   correction  to  \rf{644} or \rf{001}   coming from the quantum  M5 brane 
%v3: fluctuations  near the above classical solution.  There appears to be no  alternative 
%v3: (2,0)  theory result   for this   subleading coefficient  known at the moment (cf. \ci{Capuozzo:2023fll}) 
%v3:   so  our  1-loop 
%v3:   M5 brane   computation will provide a prediction for it. 
%%% v4:
Our aim will  be to compute the  contribution of the quantum  M5 brane
fluctuations  near the above classical solution to the subleading terms in  \rf{644} or \rf{001}.
%v4
{This contribution depends on the boundary conditions and on the summation prescription discussed below.
% It is not, by itself, the full constant term
%in the defect anomaly of the coupled system of the brane and the background.
}
There appears to be no  alternative
(2,0)  theory result   for {the subleading terms} in $a$   known at the moment (cf. \ci{Capuozzo:2023fll}).

 %There is  a puzzle that this constant comes from constant in $C_6$   which was not 
%included in similar  discussion in the D3 brane context in \ci{Drukker:2008wr}  
%where  result for  2d defect anomaly was zero as a result.
%As expected,  since $u_0$ is a modulus,  the value of the action   does not depend on it.  
%The same will be true also for fluctuation action so we may just set
%\be u_0 =0 \ . \la{606} \ee

The set of fluctuations will  be the same as for a (2,0)  6d  multiplet (5 scalars, 4 fermions  and self-dual 2-form) 
on  AdS$_5 \times S^1$   background with all scalars    conformally   coupled  to the   metric 
and 2 scalars and fermions  coupled also 
  to a constant $U(1)$   potential  in $S^1$   direction 
  reflecting the presence of a non-trivial target space geometry and  the $F_4$ flux.
  As in the D3 brane case, the latter  should be also   responsible for preservation of the global supersymmetry
  of the world-volume  theory  defined on AdS$_5 \times S^1$ 
  at the level of the quadratic fluctuation action.

\subsection{Scalar  fluctuations}
  
  The derivation  of the scalar fluctuation   action is  directly analogous to 
  the one in section  \ref{D3scalaraflu}  in D3 brane case. 
Let us  choose a static gauge  where AdS$_5$  and $\psi$ coordinates are not fluctuating, i.e. 
\be
 AdS_5=\{\xi^{\hat 0}, ..., \xi^{\hat 4}\} ~,
  \, \qquad u =u_0+ \delta u~, \qquad \psi=\xi^{\hat 5} ~, \qquad \phi=2\xi^{\hat 5} + \delta \phi ~, \la{71}
\ee
and also $S^2$   coordinates are fluctuating near trivial vacuum values.  Here $\xi^{\hat 5} $  is $2\pi$ periodic.

Specializing  the discussion in Appendix \ref{scalarflu} to the case of $p$-brane with $p=5$     
 we get the following action for the  fluctuations  of  $u$ and $\phi$  
 (ignoring overall constant factor $\sim \cosh^4 u_0$) 
\ba
& S \to    \int d^6 \xi \sqrt {- \mathsf g} \, {\cal L} (\chi) 
\ , \ \ \qquad   {\cal L} = 
\mathsf  g^{\a\b}\pa_\a \chi \pa_\b \bar \chi  +  2i    ( \bar\chi\pa_{{\hat 5}} \chi -\chi \pa_{{\hat 5}} \bar\chi)  \ , \la{72}
\\ 
& 
\chi=\delta u - \tfrac{i }{ 2} \tanh u_0 \delta\phi \ , \qquad \qquad 
ds^2 = \mathsf g_{\a\b} d \xi^\a d\xi^\b  =  \mathsf g_{ij}  d \xi^i d\xi^j  + (d \xi^{\hat 5})^2 \ , \la{73}
\ea
where $\mathsf g_{ij}$  is the unit-radius  metric  on  AdS$_5$. 
Eq. \rf{72} is  a direct analog of \rf{9}   and thus we conclude that   we get  a complex   scalar   which is conformally coupled\foot{In 6d  a  conformally coupled scalar  has kinetic operator $-\nabla^2 + {1\ov 5} R$   and $R(\AdS_5 \times S^1) = - 20$ (cf. footnote~\ref{scal}).}
 to curvature of  
AdS$_5 \times S^1$  and is also  coupled to a constant  $U(1) $   gauge potential $A_{\hat 5} = 1$  with  charge 2
 \be \la{74}
  {\cal L} = 
\mathsf  g^{ij}\pa_i \chi \pa_j  \bar \chi  + ( \pa_{{\hat 5}} \chi -  2i  \chi) (\pa_{{\hat 5}} \bar \chi +  2i \bar  \chi)  - 4 \bar \chi \chi 
\ . \ee
Expanding  in modes  in $S^1$  coordinate  $\xi^{\hat 5}$ we get 2 towers of real  scalars on AdS$_5$  with masses
\be 
m^2 = (n \pm 2)^2  - 4  \ , \qquad \ \ \   n=0, \pm 1, \pm 2, .... \ , \la{75}  \ee 
and thus  with 4d   boundary dimensions $\Delta ( \Delta -4) = m^2$. Assuming  as in the D3 brane  case (cf. \rf{111}) the Dirichlet boundary conditions, i.e. $\Delta=\Delta_+$   we  get 
\be \la{76} 
\Delta-2= |n \pm  2|  \  .   %\jhl{modulous?}
\ee
%where  may  need  exclude special value of $n$   to satisfy the  condition $\Delta >0$
%(i.e. to exclude logarithmic scalar mode in 4d). 
 The analysis   of the remaining fluctuations in $\theta$ and $S^2$ directions  is again analogous to the D3 brane case in \rf{13},\rf{14}: setting $\theta={\pi\ov 2}+v $,  $ds^2_{S^2}=d\varphi _1^2+\cos^2\varphi_1\ d\varphi_2^2$ and 
introducing 3 Cartesian coordinates ${ \sf X}^a=\{x,y,z\}$  as 
$
v= \sqrt{x^2+y^2+z^2},\ \ 
\tan \varphi_1= \frac{x}{\sqrt{y^2+z^2}}, \ \ 
\tan \varphi_2= \frac{z}{ {y}}\ , 
$
we  find that the quadratic fluctuation  action  for ${ \sf X}^a$ is 
 \be \la{344}
%v3:  \int d^4\xi \sqrt { -\sf g}\, \sum_{a=1}^3\Big(  
 \int d^{{6}}\xi \sqrt { -\sf g}\, \sum_{a=1}^3\Big(    %%% v4: d^4 -> d^6
 {\sf g}^{\alpha\beta}\pa_\alpha{ \sf X}^a \pa_\beta{ \sf X}^a  -  4{ \sf X}^a   { \sf X}^a  \Big) \ , 
\ee 
which describes  3 real  conformally coupled scalars on AdS$_5 \times S^1$.
Expanding in modes in $\xi^{\hat 5}$
 we get 3 towers of scalar operators with masses  and   scaling dimensions given by 
  {(assuming Dirichlet boundary conditions)} 
 \be\la{1445}
   m^2=n^2 -4 \ , \ \ \ \ \ \ \ \   \D-2 = |n| \ , \qquad  n=0, \pm 1, .... \ .   \ee

\subsection{Fermionic  fluctuations}

The   quadratic fermionic part  of the  $\kappa$-symmetric M5  brane  action
in a general  background  which is solution of 11d  supergravity 
may be written as   \ci{Bandos:1997ui,Aganagic:1997zq,Howe:1997fb,Bandos:1997gm,Claus:1998fh}\foot{Note that 
in the particular  cases of the   maximally supersymmetric AdS$_4 \times S^7$
 or   AdS$_7 \times S^4$ backgrounds   the  fermionic part of  the M5 brane action may be written in an explicit form 
including also   higher orders in $\vtheta$ \ci{Claus:1997cq,deWit:1998yu,Claus:1998fh,Sakaguchi:2004bu}.}  
    \ba
 & S_f =  \int d^6\xi\;\sqrt {-G} \, \Big[    G^{\a\b}   \, \del_\a X^m  \,   \bar \vtheta \,  \Gamma_m \widehat D_\b \vtheta \nonumber\\
&\qquad \qquad \qquad\qquad\quad \ \ \   - \tfrac{1}{ 5!}  \eps^{\m\n\k\l\s\r}  \del_\m X^m  \del_\n X^n \del_\k X^k  \del_\l X^l  \del_\s X^r  \,  \bar \vtheta \,  
 \Gamma_{mnklr}  \widehat D_\r \vtheta   \Big]
  \ , \la{3.41} \\
  & G_{\alpha\beta} = \del_\alpha X^m  \del_\beta X^n   G_{mn} (X)~, \quad \
   G_{mn} = E^{\underline a} _m E^{\underline b}_n \eta_{\underline a \underline b} ~, \quad  \ \ 
  \Gamma_m  =   E_m^{\underline a}(X) \Gamma_{\underline a}\ , \ \ \   \G_\a = \del_\a X^m \G_m \,,  \no 
  \\
  &   \widehat D_\alpha = \del_\alpha X^m    D_m , \ \qquad 
   D_m = {\nabla}_m  - \tfrac{1}{288} (\G^{pnkl}_{\  \ \   \ m} + 8  \G^{pnk}\delta^l_m ) F_{pnkl} \ .  \la{3.43}
\ea
Here we ignored dependence on ${\rm H}_{\m\n\l}$ that is not relevant in the present case. 
%\jhl{is there an $i$ factor? If yes, we may absorb it into the definition of $\bar\theta$. As in D3 case, we can suppress the overall %factor.}
As in \rf{16} we omitted  the overall factor  of  brane tension that can be  absorbed into a rescaling of  $\vtheta$. 
%Following \ci{Bandos:1997ui}  
We use Minkowski notation with 
$\vtheta$  being  a 32 component 11d Majorana spinor.\foot{To recall (cf. footnote \ref{not}),   we use Latin letters $m,n,\cdots$ to label the spacetime coordinates, and Greek letters $\alpha, \beta, ...$ to label the world-volume coordinates.  When numbering the  coordinates,   $0, 1, \cdots, 9, \oo$ will be   used for the spacetime, while
 hatted numbers $\hat 0, \hat 1, \cdots $ --  for the  world-volume indices.
 For both spacetime and world-volume indices  we  use underlined letters to denote   indices  along the tangent  directions.}
  $  D_m $  is the   generalized 11d  spinor covariant derivative  \ci{Cremmer:1979up} 
and  ${\nabla}_m=\del_m  + \four \G_{\un{ab}} \Omega^{\un {ab}}_m $. 

Specifying to  the bosonic background \rf{51} we may use 
 the background value $g_{\a\b}$  of the   induced metric  action \rf{3.41}  may be  rewritten as  (cf. \rf{16},\rf{19}) 
\ba
&\qquad \qquad \qquad \qquad S_f=   \int d^6 \xi  
\sqrt{-g}\,  g^{\alpha\beta} \, \bar\vtheta (1-\Gamma_*) \Gamma_\alpha \widehat D_\beta \vtheta \ , \la{765}\\
&  \Gamma_*\equiv   \frac{\epsilon^{\alpha   \beta \gamma   \mu\nu\sigma}}{6!\sqrt{-g}}  \del_\a X^m \del_\b X^n \del_\g X^k  \del_\m X^l \del_\n X^p \del_\s X^q \Gamma_{mnklpq}=
 \frac{\epsilon^{\alpha   \beta \gamma   \mu\nu\sigma}}{6!\sqrt{-g}}%\epsilon^{\alpha   \beta \gamma   \mu\nu\sigma}
   \Gamma_{\alpha   \beta \gamma  \mu\nu\sigma}
 = \Gamma_{ \underline {\hat 0}\un{\hat 1}\un{\hat 2}\un{\hat3} \un{\hat4}  \underline {\hat 5}}\ , \la{766}
\ea
where  $\G_*\equiv \G_{M5}$ is the analog of $\td \G_{D3}$   in \rf{16}  with $\G_*^2=1$.
We  shall label the coordinates as follows:
\ba
 &AdS_5= \{X^{ 0}, ..., X^4\}= \{\xi^{\hat0}, ...,\xi^{\hat4}     \}~,
 \qquad X^{5 }=u=u_0~,
   \qquad X^{ { 6}}=\psi=\xi^{\widehat{ 5 }}~,  \no \\  & \qquad 
X^{7}=\theta={\pi\ov 2}~ , \qquad X^{8}=\phi =2\xi^{\widehat{ 5 }}\ , \qquad  \ \   S^2 = \{ X^{9}, X^{\oo}\}\ . \la{78}
\ea
Then $F_4$ in \rf{62}   may be written as 
\be
F_4=3L^3 \sin\theta \cos^2\theta\, d\theta\wedge d\phi \wedge \vol_{S^2}
={3}{L}^{-1}\ E^{\underline  7}\wedge E^{\underline  8}\wedge E^{\underline 9}
\wedge E^{\underline  \oo}\ . \la{79}
\ee
For  $m<7$, the second term in brackets in $D_m$ in \rf{3.43}  does not contribute  and we get 
\ba\la{80}
&  D_m=\nabla_m-\frac{1}{288}\Gamma^{pnkl}{}_m  F_{pnkl}
%=\nabla_m-\frac{1}{288}4!\Gamma^{789,11}{}_m  F_{789\,11}
=\nabla_m-\frac{1}{ 12}\Gamma^{789\,\oo}{}_m  F_{789\,\oo}~, \qquad m <7 \ , \\
&
 \sum_{\alpha=\hat 0}^ {\hat 4} \Gamma^\alpha D_\alpha =\sum_{m=0}^4 \Gamma^m D_m 
 % \Gamma^m\nabla_m-\frac{1}{ 12} \Gamma^m\Gamma^{789,11}{}_m  F_{789,11}
% =\slashed \nabla-\frac{5}{ 12}\Gamma^{789\,11}F_{789\,11}
    =\sum_{m=0}^4 \Gamma^m \nabla_m  -\frac{5}{ 4L}  \;\Gamma^{\underline 7\underline 8\underline 9\, \underline{\oo}}
    \la{fluxAdS5}
\ ,       \\&%  \qquad 
  D_6% =  \nabla_6-\frac{1}{12}\Gamma^{789\,11}{}_6  F_{789\,11}
  =  \nabla_6 - \frac{1}{4L} \; \Gamma^{\underline 7\underline 8\underline 9\, \underline{\oo}}{}_6   \ .\la{3322}
\ea
%%%%%%%%%%%%%%%%%%%%%%%
\iffa 
\be
  \Gamma^6 D_6
  =  \Gamma^6 \nabla_6-1/4/L\; \Gamma^{\underline 6\underline 7\underline 8\underline 9,11}  \;
\ee
  \fi 
  %%%%%%%%%%%%%%%%%
For  $m=8$   the first term in brackets in $D_m$ in \rf{3.43}  does not contribute  and we get 
\ba
&D_8 %=\nabla_8-\frac{8}{288} \Gamma^{pnk } \delta_8^l  F_{pnkl}
 =\nabla_8-\frac{ 1}{ 36} \Gamma^{pnk } F_{pnk8}   
%=\nabla_m-\frac{ 3!}{ 36} \Gamma^{7 9,11}{}  F_{789,11}
=\nabla_8-\frac{ 1}{  6}\Gamma^{7 9\;10}{}   F_{789\;10}
=\nabla_8-\frac{ 1}{2L}E_8^{\underline 8}\, \Gamma^{\underline 7\underline  9\, \underline{\oo}} \ , \la{3333}
\\
&\la{91}
\p_{\hat 5} X^m D_m =D_6+2D_8= \nabla_6+2\nabla_8-{1\ov 4L}\; \Gamma^{\underline 7\underline 8\underline 9\,\underline{\oo}}{}_ 6-\frac{ 1}{L} \Gamma^{\underline 7\underline 8\underline  9\,\underline{\oo}} {}_8 \ , 
\\
&\Gamma^{\hat 5}\p_{\hat 5} X^m D_m=\Gamma^{\hat 5} (D_6+2 D_8)\ , \qquad \ \  \Gamma^{\hat 5}
  =\frac{\sinh u_0\, \Gamma_{\underline 6}+ \Gamma_{\underline 8}}{ L_{A}   \cosh^2 u_0 } 
 %(D_6+2 D_8)
  \ . \la{3.35}
\ea
 Also,   computing the spin connection   gives  (see \eqref{b15} in Appendix \ref{spincon})
 %Meanwhile, from \eqref{b15}
\be\la{spinconM5}
\Gamma^\alpha\p_\alpha X^m \nabla_m  
=\slashed \nabla 
+
\frac{3}{2 L } \tanh u_0\Gamma_{  \underline 5}
 +  \frac{1}{4L\cosh  u_0 }  \Gamma_{\underline  {8}\underline  {6} \underline 5}  \ , 
 \ee
 where $\slashed \nabla $ corresponds to AdS$_5 \times S^1$.
 
 %One can check that  like in the D3 case  the resulting fermionic  action  has no nontrivial dependence on $u_0$ 
 % so to simplify the presentation  we may set $u_0=0$. \jhl{Again I feel it is better not say in this way, maybe: 
Like in the D3 case  one expects that the resulting fermionic  action   should  have  no nontrivial dependence on $u_0$.
 Indeed, in Appendix~\ref{M5massu} we will show that    the fermion spectrum is the same for all  values of $u_0$. 
 Thus, to simplify the presentation,  here we may just consider the limiting case of $u_0\to 0$  ignoring the subleading terms.
  %}% it can be set  to 0. 
Then in total 
  \be
\Gamma^\alpha\p_ \alpha  X^mD_m  
 =\slashed \nabla 
  +    \frac{1}{ 2 L   }\Gamma_{\underline  {8 }}  \Big(\frac12   \Gamma_{  \underline  {6} \underline 5}  
+\frac32 \Gamma_{\underline  {7 }\underline  {9}\,  \underline{\oo}}  \Big) \ . \la{87}
    \ee
 The fermionic  action \rf{765} then  may be written as\foot{We scale out $L_A = 2L$  so that $g_{\a\b}\to \sf g_{\a\b}$ is the unit-radius AdS$_5 \times S^1$   metric and ignore the overall constant factor.}
   \ba
& \la{3361} S_f =    \int d^6 \xi  
\sqrt{-\sf g }  \,  \bar\vtheta (1-\Gamma_*)  \big(\slashed \nabla+\MM  \big)  \vtheta  \ ,\\  &   
\Gamma_* 
  =  \Gamma_{ \underline {\hat 0}\cdots \underline { \hat 4}   \underline {\hat 5}}
    =   \Gamma_{ \underline {  0}\cdots  \underline {  4} \underline {  8}}\ , \qquad \ \ \
 \MM=\frac12\Gamma_{\underline  {8 }}  \Big(   \Gamma_{  \underline  {6} \underline 5}  
+3 \Gamma_{\underline  {7 }\underline  {9}\,  \underline{\oo}}  \Big)  \ . \la{88}
\ea
%v3:  Note that $\G_* $ in \rf{765} {anti}commutes with $\slashed \nabla$  and $M$.
 Note that $\G_* $ in \rf{765} {anti}commutes with $\slashed \nabla$  and {$\MM$}.
 We shall  fix the  $\kappa$-symmetry  gauge by 
 \be
 (1-\Gamma_*)\vtheta=0 \ . \la{77}
 \ee
 % \jhl{corrected the sign..}
Then  acting on $\vtheta$ we have  $  \Gamma_{ \underline {  0}\cdots  \underline { 4}  \underline { 8}} =\aat{+}1$.
We may also use that in the  conventions assumed in \rf{3.41}   one has  %$\Gamma_{\un{\oo}} =
$  \Gamma_{ \underline {  0}\cdots \un{4} \un{56789}\, \un{\oo} }=\aat{+}1$
(see  \rf{ddd},\rf{d200}).  % or 
%$\Gamma_{ \underline {  0}\cdots  \underline {  9} \, \underline { 11}}=  -1$.
Then $\Gamma_{ \underline {  5}  \underline {  6}  \underline {  7} \underline {  9}     \, \underline {\oo}}=  -1$
or $\Gamma_{  \underline {  7} \underline {  9}     \, \underline {\oo}}= 
\Gamma_{ \underline {  5}  \underline {  6} }$ (see  \rf{555}). 
As a result,   we may simplify $\MM$ in \rf{88} to
\be\la{3.39}
 \MM
 %\Gamma_{\underline  {8 }}  \Big(  \frac12 \Gamma_{  \underline  {6} \underline 5}  
 % +\frac32 \Gamma_{\underline  {7 }\underline  {9} \underline ,11}  \Big) 
 % =  \frac12   \Gamma_{  \underline  {8}  \underline  {6} \underline 5}    -\frac32 \Gamma_{ \underline  {7} \underline  {8 }\underline  {9}\, \underline{\oo}}
= 
  \frac12 (1-3)   \Gamma_{\underline  {8}\underline  {6} \underline 5}   = -   \Gamma_{\underline  {8}\underline  {6} \underline 5}  
\ ,  \ee
which is the same result as found in \rf{666}.

%\jhl{different result in Appendix....}
Like   in the D3 case  (cf. \rf{27})    the final expression for the 
mass operator  is given by a combination of the  contributions of the ``transverse'' 
 part of the spin connection  and of $F_4$  in \rf{3.43}.   After diagonalization of $\MM$  the   resulting action \rf{3361}  or 
 $   \int d^6 \xi  
\sqrt{-\sf g }  \,  \bar\vtheta   \big(\slashed \nabla+\MM  \big)  \vtheta$ thus describes 
2 sets of 6d fermions in AdS$_5 \times S^1$ with masses  $\pm 1$. 
 
 %v4
 {The fermion $\vtheta$   should satisfy the antiperiodic boundary condition in the frame used here.
 This is  as required by the preservation of supersymmetry:
 the Killing spinor \rf{d16} restricted to the brane is antiperiodic in $\psi$
 (see \rf{7010} in Appendix \ref{kisp} and Appendix \ref{7002}).}
 Expanding $\vtheta$ in modes  in $\xi^{\hat 5}$ 
 %aat8
%v3:  (assuming periodic boundary condition as required by preservation of supersymmetry) 
%%% v4:
% \nw{(assuming the antiperiodic boundary condition in the frame used here, 
   the Dirac  operator  on AdS$_5\times S^1$ 
 reduces to that  on AdS$_5$ as (recall  that $\G^{\underline{\hat 5}} = \G^{\underline 8}$ in the case of $u_0\to 0$) 
 \be \la{286} 
i ( \slashed\nabla+\MM )  = i\slashed\nabla_{\AdS_5}   +i \G^{\underline{ \hat 5}}  \del_{\hat 5}  +i \MM\ \ \to \ \ 
 i \slashed\nabla_{\AdS_5}   -   \hat M \ , \qquad 
  \ \ \    \hat M =  n \G_{\underline 8}  -    i \Gamma_{\underline 5\underline 6 \underline 8} \ . 
  \ee 
 %  Equivalently, we  may write the operator in  \rf{28} as  
% $ i\slashed\nabla_{\AdS_3}   +i \G^{\hat 3}  (\del_{\hat 3}  - \ha \Gamma_{\underline 4 \underline 3})$, i.e. we get 
 %a set of 4  massless  fermions in AdS$_3 \times S^1$  coupled to a constant  $U(1)$ gauge  field in $\hat 3$ direction.
 Since $\G_{\underline 8} ^2 =1,\ \  (i\Gamma_{\underline 5 \underline 6 \underline 8})^2=1$
 and $[\G_{\underline 8}, \Gamma_{\underline 5\underline 6 \underline 8}]=0$ 
%v3:  we conclude that $\hat M$   has eigenvalues  $m_f=\pm n \pm 1$.\foot{Since $[\G_{\underline 8}, \Gamma_{\underline 5\underline 6 \underline 8}]=0$  
 we conclude that $\hat M$   has eigenvalues  $m_f=\pm n \pm 1${, where $n$ is half-integer}.\foot{Since $[\G_{\underline 8}, \Gamma_{\underline 5\underline 6 \underline 8}]=0$  
 the  action  corresponding to \rf{286} may be interpreted as that of massless  fermions  in AdS$_5\times S^1$ 
 coupled to  $U(1)$  gauge potential in $\hat 5$ direction   with charge 1.
 This is similar to the  D3 brane case in \rf{28}  where  the charge was $\ha$.
 }
 
 Thus we find 2 towers of 6d    fermions   with  such   masses. The  corresponding scaling 
 dimensions  of the boundary operators are then  (cf. \rf{29})
 \be\la{299} 
%v3:  \Delta-2=|m_f |= |n\pm 1| \ , \qquad \ \ \ \    n=0, \pm 1,\pm 2,  ... \ .
 \Delta-2=|m_f |= |n\pm 1| \ , \qquad \ \ \ \    \ {n=\pm \tfrac{1}{2},\ \pm \tfrac{3}{2},\  ... }\ .   %%% v4: was n=0, \pm 1, \pm 2, ...
 \ee
%%% v4:
 {%In contrast to the D3 brane case the fermions are antiperiodic.
Since in contrast to the D3 brane case the fermions here  are antiperiodic, 
 the dimensions in \rf{299} are half-integer, as required for the superpartners
 of the bosonic fields in \rf{76},\rf{1445} and \rf{199} below, which have integer dimensions.}
 
 %For $n <0$ one  is to reverse the sign   in front of  $n\pm\frac12$ getting equivalent set of $\Delta$'s.  
%where physical states   correspond to $\Delta > \ha$. 

 \subsection{Antisymmetric tensor   field   contribution}
 
Since the  self-dual  $H_{\m\n\l}$ field in the M5 brane action \rf{3.6}  has   no background value 
its contribution to the 1-loop  free energy is the same as   half of that of   rank 2  antisymmetric 
tensor $A_{\m\n}$   propagating on AdS$_5 \times S^1$.

The  partition function for  $A_{\m\n}$  with the standard   action 
$\int d^6 \xi \sqrt{- g} \, H_{\m\n\l} H^{\m\n\l}$   in a general  6d curved background  is given by 
% can   be  found, {\em e.g.},   from the general curved space expression 
 (see, e.g.,   
 %v2
 \ci{Obukhov:1982dt,Fradkin:1982kf,Bastianelli:2000hi}  and refs. there)
\ba \la{550}
Z_2=({\rm det\,}\hD_2)^{-1/2}
\ {\rm det\,}\D_1 \ ({\rm det\,}\hD_0)^{-3/2}\ ,
\ea
Here the Hodge-DeRham operators $\hD_p$ are  %defined as ($\nabla$ is 6d covariant derivative) 
\ba \la{551} 
({\hD}_2)_{\m\n}^{\a\b} = -  \nabla^2 \delta_{\m\n}^{\a\b} 
+2 R_{[\m}^{[\a} \delta_{\n]}^{\b]}
- R_{\m\n}{}^{\a\b}\ , 
  \qquad 
({\hD}_1)_\m^\n = -  \nabla^2  \delta_\m^\n  + R_\m^\n \ , \qquad 
{\hD}_0 = -  \nabla^2\ . 
\ea
%%%%%%%%%%%%%%%%
%%%%%%%%%%%%%%%
Let us specify \rf{550} to AdS$_5 \times S^1$ with equal radii =1. 
For AdS$_5$ we have $R_{ijkl} = - ( g_{ik} g_{jl} - g_{il} g_{jk} )$,  $R_{ij} = - 4 g_{ij}$, $R=R^i_i= -20$. 
Thus  from \rf{551} 
% with now all  covariant derivatives  being along AdS$_5$   and
 splitting 
  $A_{\m\n} =(A_{ij}, A_i=A_{i5}$)  and also  the vector ghost $C_\m= (C_i, C=C_5)$   we get  ($i=0,1, 2, 3, 4$)  
\ba 
&
A^{\m\n} {\hD}_2 A_{\m\n}   = 
 A^{ij} ( -\nabla^2  - 6 ) A_{ij}  +   A^i (- \nabla^2 - 4)  A_i \ ,  \la{580}\\
& C^\m {\hD}_1C_\m  =  C^i ( -\nabla^2  - 4 )  C_i   +   C (- \nabla^2 )  C\ , \qquad \ \ \ 
 \nabla^2  = \nabla^2 _{\AdS_5} + \del_5^2 \ . 
\ea
%This is checked.
We may express  $ \det {\hD}_1 $ as (cf. \rf{30} and discussion below it)\foot{Set   $A_i = A_i^\perp + \del_i \vp$.  We use that  
$\int dA\,   e^{-\int A^2} \to \int   dA_\perp\,  d \vp\,  \sqrt{ \det ( - \nabla^2) } \ e^{-\int A_\perp^2} $ 
and 
$
A^i (-\nabla^2-4) A_i = A^\perp_i  (-\nabla^2-4) A^\perp_i   + \vp (-\nabla^2)^2  \vp $.
Note also that  for $\na_i$ derivatives 
 $- \nabla^i ( -\nabla^2) \nabla_i = \na^4 + \na^i[\na^2, \na_i]=  \na^4 - 4 \na^2 $. 
} 
\be  \det {\hD}_1 = \det ( -\nabla^2  - 4)_\perp \  [\det (- \nabla^2  )]^2 \ . \la{3471}
\ee
Similarly,  applying  the redefinition 
$A_{ij} = A_{ij} ^\perp + \na_i V_j^\perp - \na_j V_i^\perp$  and accounting for the   Jacobian factor  we get 
 \ba 
 Z_2 (\AdS_5\times S^1 ) = 
  \Big[
\frac{(\det \hD_{1}(-4))^2 (\det \hD_{0}(0))^2   }{\det \hD_{2}(-6)\,  \det \hD_{1}(-4)\,  (\det \hD_{0}(0))^3}
\Big]^{1/2}=
%\\ =  &
\la{4001}
 \Big[
\frac{1}
{\det  \hD_{2\perp}(-6)\,   }\Big]^{1/2}\,  .
% \Big[ \frac{\det\hat\Delta_{1\perp}(-4)\ \det\hat\Delta_{0}(0)} { \det\hat\Delta_{1\perp}(-8) \   \det\hat\Delta_{0}(-4)    } \Big]^{1/2} \ . 
\ea
%Check of  dof:  $(\ha 5\times  4 - 4)   = 6 $.
Here $\hD_p(X) \equiv  - \na^2 + X$ are defined on $p$-forms with AdS$_5$ indices   and 
$ \hDelta_{2\perp}(-6) A^\perp_{ij} = ( - \nabla^2_{\AdS_5} - \del_5^2 - 6) A^\perp_{ij} $. 

One can give an alternative derivation of \rf{4001}   as follows. 
Let us  split  $H_{\m\n\l}^2 =H_{ijk}^2 +   3 H_{ij5}^2$ and fix 
the  gauge as  $A_{i5}=0$  so that  $H_{ij5}^2=( \del_5 A_{ij})^2$.  Then 
for $A_{ij} = A_{ij}^\perp   + \del_i C^\perp_j - \del_j C^\perp_i$   we get 
$H_{ijk}^2 = A_{ij}^\perp \hDelta_{2}(-6)  A_{ij}^\perp $. 
From $( \del_5 A_{ij})^2$ we  find  that   determinant of $\del_5^2$ cancels
 against  the  ghost determinant. Including also the contribution of the Jacobian we end up with  \rf{4001}.
% That implies   that we should get $\Delta^\perp_{1}(-4)$   but without $\del_1^2$ in kinetic term.
The  same  expression   \rf{4001} was     given also  in  \ci{David:2020mls,David:2021wrw}.

%%% v4: new paragraph
{The above derivation, and in particular the gauge choice $A_{i5}=0$, applies to the modes with $n\not=0$ in the expansion in $\xi^5$.
For $n=0$ the transverse part of $A_{i5}$ cannot be gauged away, and the zero eigenvalue of $\del_5^2$ cannot be cancelled against the ghost determinant.
For this mode \rf{4001} is only a formal continuation.
The $n=0$ mode of the self-dual tensor is a massless 5d vector and should be treated separately (see the footnote above \rf{3.86}).
With the standard boundary condition for this vector the partition function of the self-dual tensor is
\be\la{7044}
Z_{\rm sd}=Z_{V}\, \Big[ {\det}{}'\, \hD_{2\perp}(-6)\Big]^{-1/4} \ , \qquad \qquad
Z_{V}=\Big[\frac{\det \hD_{0}(0)}{\det \hD_{1\perp}(-4)}\Big]^{1/2} \ ,
\ee
where the prime means that the $n=0$ mode is omitted and the operators in $Z_V$ act on the fields that do not depend on $\xi^5$.}

% On AdS  determinants for p-forms are in \ci{Camporesi:1994-hig}. 

Expanding  $A_{ij}^\perp(\xi^k, \xi^5) $    in  $S^1$  modes   we get a tower of transverse 
  antisymmetric tensor  fields in AdS$_5$  with masses
\be   \la{500}
m^2 =  n^2 -6 \ . \ee
Let us recall  how this  case fits into the general   discussion  of    fields  in AdS$_5$  corresponding to  
 representations of $SO(2,4)$. 
  % of the 5d operators $\OO$   we will be considering  below. 
Let $\phi$  be a    massive ($ \De > 2 + j_1+j_2$  for $j_1 j_2 \not=0$ or $ \De > 1 + j_1+j_2$  for $j_1 j_2=0$) or massless 
($ \De =   2 + j_1+j_2$, \  $j_1 j_2 \not=0$)  field in AdS$_{5}$  corresponding  the $SO(2,4)$ representation 
$(\Delta;\, j_{1}, j_{2})$.
%\footnote{%v2
%Here we   assume  that   $j_{1}j_{2}>0$. For a  general  discussion see Appendix \ref{A:FF} (cf. \rf{a1}).}
%\foot{The associated boundary conformal field   will have  canonical dimension equal to $\De_-= 4- \De$. Thus $\De \ge 4$  cases 
%will correspond to 4d  fields with  higher $2(\De-2) \ge 2( j_1 +j_2) $ derivative kinetic  operators.}
%will correspond to non-local  boundary operators
$h_1 = j_1 + j_2=s$  and $h_2= j_1-j_2$  are  integer for bosons and  half-integer for fermions 
(in the bosonic case, $h_{1}$ and $|h_{2}|$ are the lengths of a two-row Young tableau).
According to \cite{Metsaev:1994ys,Metsaev:2003cu}, the  covariant AdS$_5$ 
equation of motion for a
bosonic transverse   field $\phi$   is (for  $j_{1}\ge j_{2}$)\foot{This equation is also for 
the fermionic fields after squaring the 5d Dirac  operator. For a generic fermion  spinor-tensor field 
 $\Psi $ %= \Psi^\alpha_{m_1, \dots, m_{\rm s}}$,  ${\rm s}\equiv = s -\ha =  j_{1}+j_{2}-\ha $
one has     $(\slashed{\na}+\Delta-2)\,\Psi=0$ \ \ci{Metsaev:1998xg}. 
 After squaring, this turns out to be 
$\big[ -\na^{2}_{\AdS_5} + {1\ov 4} R -2 j_1 +1 +(\Delta-2)^{2}\big] \,\Psi=0$, %  (see  \cite{Metsaev:2013wza}  for details), 
 where $R=R(\AdS_5)=-20$.    This  gives  the 
  same $m^2$ as in (\ref{1.16}). }
%v4
\be
\label{1.16}
 \OO \phi =0 \ ,\qquad \qquad  \OO =-\na^{2}_{\AdS_5} +m^2\,, \quad \qquad    m^2 =   (\Delta-2)^2-4  -  2\,j_{1}  \ .
\ee
%Alternatively, in fermion case another definition of mass is Dirac one:
%$(\slashed{D} + m_D)\, \Psi=0$, so $m_D = \Delta -2$.} %We thank R.R. Metsaev for related explanations.}
The    partition function 
for such  massive  field  %with  the  standard (Dirichlet)   boundary  conditions %corresponding to %$\De=\De_+$ 
is then given by\foot{In the massless   case of $\De= 2 +s$  one  needs to take into  account 
  the contribution  of the corresponding ghosts 
that belong to the representation $(\De+1; j_1-\ha, j_2-\ha) $ (see, e.g., \ci{Giombi:2013yva}).}
%\rf{1.2}  with $\OO$  defined on transverse  fields  in representation $(j_1,j_2)$. We shall denote the massive case 
%quantities with \ $\widehat {}\ $\  in what follows, i.e. 
\be 
  Z(\De; j_1, j_2)  = \big[ \det ( - \na^2_{\AdS_5} + m^2)_\perp \big]^{-1/2} \ . \la{1.19} \ee
%v3:   The antisymmetric tensor case in \rf{4001},\rf{500}  corresponds 
  The antisymmetric tensor case in \rf{4001},\rf{500}  corresponds {(for $n\not=0$)}
  to the  sum of two (self-dual and anti self-dual)  representations (cf. \rf{1.16}) 
  \be 
   (\Delta; 1,0) + (\Delta; 0,1 ) \ , \qquad \qquad   \Delta-2 =   |n| \ . \la{199} \ee
  % with self-duality condition leaving one of the two. 
      The collection   of  2 scalars with masses/dimensions 
       in \rf{75},\rf{76},  3 scalars   in \rf{1445}, the fermions in
   \rf{299}  and the self-dual  rank 2  tensor  \rf{199} 
     form  5d supermultiplets  that  represent the 
      (2,0) multiplet  defined on AdS$_5\times S^1$ 
    in a way consistent   with preservation of supersymmetry,
     i.e. requiring conformal coupling of  all scalars to the curvature 
     and  a particular coupling  of 2 scalars and all fermions to a constant 
    $U(1)$   gauge  potential   %consistent with the discussion in 
   (cf. \ci{Gukov:1998kk,Aharony:2015hix,Aharony:2015zea}).\foot{It
   %aat8
    is interesting to note   that by 
   applying  an  analytic continuation  to $S^5 \times S^1$   with the angle of $S^1$  identified with period $\beta$ 
    one   gets a  similar system of fields  whose   supersymmetric partition function 
      computes the Schur  index \ci{Kim:2012ava}  of the (2,0)  tensor multiplet (see section 5 in \ci{Beccaria:2023cuo}).
      Similar relation is true also in the D3 brane case    in section 2 upon analytic continuation to $S^3 \times S^1$ world volume theory (cf.  \ci{Beccaria:2024vfx}).}

%%% v4:
{Explicitly, the world-volume translation $\del_{\hat 5}$ is the $U(1)_r$ generator of the 4d $\N=2$ superconformal algebra $su(2,2|2)$ preserved by the defect.
The modes form the chiral multiplets ${\cal E}_m$ with $m=2,3,...$, whose primary has $\Delta=r=m$, their conjugates, and one flavour current multiplet
(see Appendix \ref{7002} and Table \ref{7011}).}

\iffa 
{\bf a bit puzzled about the following:  suppose we just consider theory on AdS$_5 \times S^1$   without  expanding 
 on $S^1$. Then it should be susy as a 6d theory,   and then we do need to worry  about constraints on $\Delta$'s  etc....
 Also: alternatively susy should   apply level  by level in $n$}
 \fi

\subsection{1-loop free energy}

%For a generic $SO(2,4)$  representation  
The free   energy  corresponding to \rf{1.19}  can be computed explicitly   like in the AdS$_3$   case in \rf{333}--\rf{35}.
% and is proportional to $\vol(\AdS_5)= \pi^2 \log (r\RL)$.
 In the case of the Euclidean AdS$_{5}$   with boundary $S^4$   it is proportional to $\vol(\AdS_5)= \pi^2 \log (r\LR)$
 and thus is  proportional to the  4d  conformal $a$-anomaly  coefficient 
(see \ci{Camporesi:1994ga,Diaz:2007an,Giombi:2013yva,Giombi:2014iua} and,  in particular,  \ci{Beccaria:2014xda})
\ba
&F(\De; j_1, j_2) = \ha \log \det ( - \na^2_{\AdS_5} + m^2)_\perp 
= - \ha \zeta'_{(\De; j_1, j_2)} (0)  =   4 a(\De; j_1, j_2)\,  \log (r\LR) \la{3532}  \\
&a(\De; j_1, j_2) = 
-  { \frac{1}{96\,\pi} }(-1)^{2(j_{1}+j_{2})}(2j_{1}+1)(2j_{2}+1)\   \lim_{z\to 0}  J'(z) \ , 
 \la{3531} \\
  &  % \lim_{z\to 0} \frac{\partial}{\partial z}
J(z)= \int_{0}^{\infty}d\lambda\,\frac{\left[\lambda ^2+(j_{1}-j_{2})^2\right] \left[\lambda ^2+(j_{1}+j_{2}+1)^{2}\right]}{\big[\lambda^{2}+(\Delta-2)^{2}\big]^{z}}\ , \ \ \ \ \ \   (\Delta-2)^2 = m^2 + 4  + 2 j_1 \ . \no 
 \ea
This gives, assuming Dirichlet boundary conditions so that $\Delta -2 \ge 0$, 
  \ba
\no & a (\Delta;\,j_{1},j_{2}) = \frac{1}{1440} (-1)^{2 (j_1+j_2)} (2 j_1+1)
   (2 j_2+1)\,  (\Delta-2) \\ 
& \quad \quad  \times   \Big[3 (\Delta-2)^4
   -{10} \big(   j_1^2 +  j_2^2 +   j_1 +  j_2   + \ha   \big) (\Delta-2)^2 + 15  (j_1-j_2){}^2 (j_1+j_2+1){}^2\Big].  \label{3.3}
\ea
%This  expression  is  odd   under $\De \to 4- \De$, i.e. 
%anomaly  corresponding to $Z^-$ computed with the alternative boundary condition    has the opposite sign, 
% i.e.    $\widehat \aa=\widehat  \aa^- - \widehat \aa^+ = - 2\widehat  \aa^+$.  
%Here  the modulus $ |\Delta-2|$  is put for generality  but we will always assume the 
%Dirichlet boundary conditions, i.e.  the $\Delta=\Delta_+$ choice  or 
%pick up a  contribution of state   with  $\Delta-2$  that is  positive. 
As a result, we get for a  real AdS$_5$   scalar  contribution
\be 
a(\De; 0,0) 
 =  \frac{1}{ 1440}\, \, (\Delta-2) \,\big[  3\,(\Delta-2)^{4}  - 5(\Delta-2)^2\big]\ .\label{3.85}
\ee
The  contribution of the self-dual antisymmetric tensor  is\foot{%aat8
The contribution of the self-dual  antisymmetric tensor to the 1-loop free energy  is  by definition 
half that of the standard antisymmetric tensor. 
If  one dimensionally reduces the  antisymmetric tensor  action $\int H_{\m\n\l}^2 $ to AdS$_5$ 
(i.e. considers only the $n=0$ mode  of the $S^1$  expansion)  one gets a  collection of 
a rank 2 tensor $\int H_{ijk}^2$  and  massless  vector $\int F_{ij}^2$   ($A_i\equiv A_{i5}$) 
5d actions. Dualizing the former to a vector $A_i'$  (which can be done  by a path integral transformation 
and thus preserves the expression for the partition function modulo zero mode  contribution absent in the present case) 
%%% v4:
 one thus gets a collection of two
5d   vectors. {If their boundary conditions were opposite (cf. footnote \ref{dua}  for a similar remark  in the AdS$_3$   context)
their total contribution  to $a$-anomaly would be zero, as for the formal $n=0$ value of the tensor term in \rf{999}.
For the self-dual tensor the $n=0$ components $A_{ij}$ and $A_{i5}$ are, however, related by the self-duality condition and describe one massless 5d vector.
With the standard boundary condition it is dual to the conserved $U(1)$ current of the flavour current multiplet of the defect and contributes
$a(3;\ha,\ha)-a(4;0,0)=-\frac{31}{360}$ to the $a$-anomaly (see Appendix \ref{7002}).}}
\be 
a(\De; 1,0) 
 =  \frac{3}{ 1440}\, \, (\Delta-2)\,\big[  3\,(\Delta-2)^{4}  - 25(\Delta-2)^2   + 60\big]\ .\label{3.86}
\ee
The    fermion contribution is 
\be 
a(\De; \ha ,0)  = a(\De; 0 ,\ha )= 
 -  \frac{2}{ 1440}\, \, (\Delta-2) \,\big[  3\,(\Delta-2)^{4}  - \tfrac{25}{2} (\Delta-2)^2  +    \tfrac{135}{16}\big]\ .\label{3.89}
\ee
Let  us   first consider  the   case of  (2,0)  multiplet of 5 conformally  coupled scalars, self-dual tensor and 4 fermions 
propagating  on AdS$_5 \times S^1$   with   all 
  scalars and fermions  not coupled to a $U(1)$ potential  in $\hat 5$   direction, i.e.  
   without shift of  $S^1$ mode number $n$. 
   
%%% v4:
   {This is a formal sum in which both the bosons and the fermions have integer $n$ and the $n=0$ tensor contribution
   is defined by the continuation of \rf{3.86}.}
    Expanding in Fourier modes in  this case we get system of AdS$_5$ fields
    with  all the fields  having  $\D-2= |n|$. Thus  we get  for the
      free energy in  AdS$_5 \times S^1$    (cf. \rf{36},\rf{001}) 
   \ba
 & \qquad \qquad  F  = \frac{2}{ \pi^3}  \, a\,  \vol(\AdS_5 \times S^1) = 4 a \log (r\LR) 
 \ ,  \la{863} \\
  & a_{(2,0)} =  \sum^{\infty}_{n=-\infty}  \Big[5a(\De; 0,0)  + a(\De; 1,0) + 4 a(\De; \ha ,0) \big] = \tfrac{1}{1440} P \ , \qquad 
   \ \ P= \sum^{\infty}_{n=-\infty} P_n \ ,\la{3580} \\
   &P_n=  5   |n|   (3 n^4    - 5  n^2 )  + 3   |n|   (3 n^4    - 25  n^2 + 60 ) 
   -     8 |n| ( 3 n^4    - \tfrac{25}{ 2}   n^2   + \tfrac{135}{16} ) =  \tfrac{225}{ 2} |n| \ , \la{999}\\
&  \qquad \qquad  P= 225 \sum_{n=1}^\infty n = 225\,    \zeta_R(-1) =-  \tfrac{225}{12} \ , \qquad \qquad 
a_{(2,0)} = -\tfrac{5}{ 384} \ . 
 \la{3590}
 %   {5\ov 16 \times 12 } \ , \ \ \ \ \ \ \   a= -\ha \widehat{\rm a}_{(2,0)}  = - {5\ov 32 \times 12 } \ . 
\ea
In   contrast to the  case of the $\N=4$   multiplet in AdS$_3\times S^1$  in \rf{36},\rf{37}  where  the  1-loop  free energy was  UV 
finite  and  vanishing  here $F^{(1)}$  is  quadratically divergent. As 
  in other similar examples  (see, e.g., 
\ci{Giombi:2023vzu,Beccaria:2023ujc,Beccaria:2023sph})
here we used the 
Riemann $\zeta$-function regularization to compute the resulting sum. 

The presence of the quadratic UV divergence  was, in fact, expected.  
 In 6d  the free energy   $F^{(1)}= \ha \log \det (-\na^2 + X)$ 
has a  UV divergent part  given  by    (in heat kernel regularization)
 \be F^{(1)}_\infty= - 
 %v3
  {1\ov (4 \pi)^3} \int d^6 \xi \sqrt g \Big(\te {1\ov 6} b_0 \Lambda^6 + {1\ov 4}  b_2 \Lambda^4 + {1\ov 2} b_4 \Lambda^2 + b_6 \log \Lambda\Big)\ , \la{9900}\ee 
where $b_k$ are the Seeley's coefficients.
 $b_0 = \tr 1$  counts total number of degrees of freedom 
 and thus  vanishes  for a supersymmetric model. One can   check   that 
 $b_2=\tr ( {1\ov 6} R - X)$  also vanishes  in a combination of 
   5 conformally  coupled  scalars ($\hat \Delta_0=-\na^2+ {1\ov 5} R$),
    4  massless  fermions  ($\hat \Delta_{1/2}=-\na^2+ {1\ov 4} R$)  and the 
        self-dual antisymmetric tensor in \rf{550},\rf{551}.  This  is 
        consistent with  the cancellation of $n^3$ terms in \rf{999}.
    
    At the same time,  
    one finds that $b_4$ and $b_6$   coefficients do not vanish in general.
 The $b_6$ coefficient  for the (2,0)  multiplet    (that determines its conformal anomaly) 
 was explicitly  
 computed in \ci{Bastianelli:2000hi} and is given by a combination  of the 6d Euler density and 3 cubic   invariants  built out of the 6d  Weyl tensor. The Euler density   vanishes for a space like $M^5 \times S^1$  and the Weyl tensor vanishes  in the case of the conformally  flat AdS$_5 \times S^1$ space.  This  is consistent with the absence of the log UV divergence in \rf{3580}. 
Complementing the discussion in  \ci{Bastianelli:2000hi}    and  computing $b_4$  for  the (2,0)  multiplet  on a general curved 6d space 
we get\foot{In a general number of dimensions  for an operator 
    $\hat \Delta=-\na^2(A)+ X$  defined on a vector  bundle   with connection $A_\m$  one has (up to  a total derivative term) 
    $b_4= \tr \big[ {1\ov 12} F_{\m\n}^2  + {1\ov 180} (R^2_{\m\n\l \r} -  R^2_{\m\n}) + {1\ov 2} ( {1\ov 6} R - X)^2\big]$.}
\be \la{4101}
b_4 =   {1\ov 4}  R^2_{\m\n\l \r}  -
%v2   {1\ov 2}   
%v3
{1\ov 2}   
    R^2_{\m\n}  + {1\ov 10} R^2 \ . 
\ee  
Computing this  for  AdS$_5 \times S^1$ we get non-vanishing result  for $b_4$, in agreement with non-cancellation of order $n$ terms in \rf{999} leading to quadratic  UV divergence.\foot{The fact  that $b_4$ is non-vanishing for 6d (2,0)  multiplet is analogous to 
non-vanishing of $b_4$ for the 4d $\N=4$  multiplet. }
%%% v4:
{Explicitly, for the unit-radius AdS$_5 \times S^1$ one has $R^2_{\m\n\l \r}=40$, $R^2_{\m\n}=80$ and $R^2=400$, so that $b_4=10$.
This value is related to the coefficient of the linear term in \rf{999} as $P_n=\frac{45}{4}\, b_4\, |n|=\frac{225}{2}|n|$.}
%[??? $b_4=10$   in fact --  {\bf why no  match to 225  coeff in \rf{999} ?}  but already spectral zeta  was used in \rf{3532} 
%that possibly dropped some  would-be  power divergences in 
% 6d   so may be direct  matching to heat kernel is not obvious. ]
 %%%%%%%%%%%%%%%%%%

Let us now  turn to the case of our interest when  the (2,0) multiplet  on AdS$_5 \times S^1$  originates  from the supersymmetric M5  brane embedded  into AdS$_7 \times S^4$   and is  thus coupled also  to  an effective  constant  $U(1)$ gauge potential  in $\hat 5$ direction
(with 2 scalars   having  charge  $\pm  2$ and the fermions charge $\pm 1$
which results  in the  shifts of $n$ in \rf{76},\rf{299}) and when the fermions { are antiperiodic.

%v4
As the fermionic modes now have half-integer $n$, the contributions cannot be combined at fixed $n$ as in \rf{999}.
They should be organized into  the supermultiplets.
Let us denote by $p_0(\nu)$, $p_T(\nu)$ and $p_F(\nu)$ the expressions \rf{3.85}, \rf{3.86} and \rf{3.89} multiplied by 1440, as functions of $\nu\equiv \Delta-2$.
The chiral multiplet ${\cal E}_m$ contains the modes of the 2 scalars in \rf{76} with $\nu=m-2$ and $\nu=m$, the modes of the 3 scalars in \rf{1445} and of the tensor with $\nu=m-1$,
and the fermions with $\nu=m-\tfrac{3}{2}$ and $\nu=m-\ha$ (see Table \ref{7011}).
The contribution of ${\cal E}_m$ and of its conjugate to $P$ in \rf{3580} is then
\ba
P_m=&\ 2p_0(m-2)+2p_0(m)+6p_0(m-1)+2p_T(m-1) \no \\
&+4p_F\big(m-\tfrac{3}{2}\big)+4p_F\big(m-\ha\big)=360\,(m-1) \ , \qquad \qquad m=2,3, ... \ . \la{1000}
\ea
The remaining modes are the $n=1$ mode of the complex scalar, the $n=0$ modes of the 3 scalars, the fermions with $n=\pm\ha$ and $\nu=\ha$, and the $n=0$ mode of the tensor.
They form the flavour current multiplet, with the contribution
\be \la{011}
P_1=2p_0(1)+3p_0(0)+4p_F(\ha)-124=-4+0-22-124=-150 \ ,
\ee
where $-124$ is the contribution of the massless 5d vector discussed in the footnote above \rf{3.86}.
The contribution \rf{1000} is again linear in the mode label, as in \rf{999},
consistent with expected presence of a quadratic UV divergence.}\foot{Note that  coupling to a constant {$U(1)$}   gauge field
 should  not   a priori  change the  values of the   coefficients $b_p$  of the UV divergent terms \rf{9900}.
 This is true, however, if one  uses
 a  covariant 6d regularization  which is not the case  here. Here  we first  expand in $S^1$ modes,
 then define the resulting 5d determinants using spectral $\zeta$-function
 and at the end  sum over {the modes}. A cut off on {the mode number} is obviously not covariant in 6d
 and thus the structure of (subleading) power divergences here is a priori sensitive to  shifts of $n$. }
%v4
{The total sum is
\be\la{7012}
P=\sum_{m=2}^\infty 360\,(m-1)-150 \ .
\ee
Its finite part depends on the prescription. 
The Riemann $\zeta$-function one  $\sum_{m=2}^\infty (m-1)\to \zeta_R(-1)=- {1\ov 12}$ used in \rf{3590} would give $P=-180$.

The sharp cutoff $m\le M$, with the terms growing with $M$ dropped, gives $P=-150$.}\foot{This corresponds to the finite assignment $\sum_{p=1}^\infty p\to 0$, which is the one that reproduces the 1-loop correction to the conformal anomaly in the case of the 10d supergravity
on $S^5$ \ci{Mansfield:2002pa,Beccaria:2014xda}. 
%v4
As shown in Appendix \ref{7003}, with this prescription  applied to free energy  of 10d type IIB  theory 
 on AdS$_5 \times S^5/\mathbb Z_k$
one also reproduces the  known conformal anomalies
of the dual  ${\mathbb Z}_k$ orbifold gauge theories, whose twisted sector contains the same multiplets as found here.}
 Using it we get
\be \la{201}
P= -150 \ ,  \qquad \qquad  a^{(1)}=\tfrac{1}{ 1440}P= -\tfrac{5}{ 48} \ .
\ee
This value assumes the standard boundary condition for the current multiplet.
It also depends on the prescription. 
The orbifold anomalies  mentioned in the previous footnote are reproduced equally well if the sums are defined
by the finite part of the superconformal index \ci{ArabiArdehali:2014cey}.
In that case the tower of ${\cal E}_m$ contributes $P=30$ instead of zero, so that
\be \la{7052}
P=-120 \ , \qquad \qquad a^{(1)}=-\tfrac{1}{12} \ .
\ee
Note that in  the orbifold case the difference between the two prescriptions is compensated by the untwisted sector, which is absent here.
%Thus the constant \rf{201} is defined only together with the order $N^0$ part of the classical response of the background.

%v4
{Let us now combine this 1-loop value of $a$ with the classical one in \rf{001}. The M5 brane tension scales as $N^2$, so that the loop corrections on the world volume
are of relative order $1/N^2$. However, as in the D3 brane case, the classical response of the background to the probe should also be taken into account.

Since the bulk action scales as $N^3$ while the M5 brane action scales as $N^2$, this response is of relative order $1/N$
and may contribute a term of order $N$ to the defect anomaly (see Appendix \ref{7004}).
Thus the  $S^4$ defect   anomaly coefficient is expected to have the form}
 \be \la{202}
 a= N^2 + {\rm e_1} N + {\rm e_2}  \ ,\ee
%v4
{where the M5 brane fluctuations contribute \rf{201} or \rf{7052}  to the term $\rm e_2$ of order $N^0$.
A natural conjecture for the term of order $N$, discussed in Appendix \ref{7004}, is  that $\rm e_1=-1$.}
% $a=N^2-N+\OO(N^0)$.}
 %%%%%%%%%%%%%%%%%%
 
 \iffa
 Let us  comment on  the validity of this   computation.
One should    account for the boundary conditions and   supermultiplet structure.
We assume Dirichlet b.c.  that seems  to be consistent   with the interpretation of  defect anomaly computation.
For example,  at level $n=0$   we have 2 scalars  with $(\Delta-2)^2 = (\pm 2)^2= 4$
so that  we have $\Delta_+ =4$ (the physical  mode  for   the Dirichlet b.c.)
  and $\Delta_- =0$   (the unphysical mode   for the  Neumann  b.c.).
  As $|\Delta_+ -2|= |\Delta_- -2|$  the two give the same contribution to \rf{3.85}  reflecting the fact that $J(z)$  in \rf{3531}   depends 
  only on $(\Delta-2)^2$. 
  Similarly, for $n=1$   and $(\Delta-2)^2 = (n-2)^2= 1$  we have $|\Delta-2|=\Delta_+ -2 =1$.
  In general, if we were  to drop some particular scalar mode we would need 
   also to drop some fermionic modes as well to preserve supersymmetry. 
   But   it looks like  there is no problem with this  in the present case. 
  \fi
  %%%%%%%%%%%%%%%%%%%%%%%%%
 % The number $\tfrac{1}{ 384} $   looks somewhat strange   but this is the current status. 
%  One may try to explore other options like   changing to Neumann b.c.  but this seems  unnatural in the context of defect anomaly interpretation. 
%paradox:   cancellation of $n^5 $ and $n^3$ is check on shifts  being consistent with susy? 
 % But   constant gauge field should not change UV ? 
  %A10 %v2
 {This expression 
  can not be directly compared 
 with the defect anomaly coefficient computed in   \ci{Capuozzo:2023fll} 
 using  the bubbling  solution in  supergravity where 
   it  has   $N^3$ scaling at leading order.  
%v3:    The precise  limit  of the parameters  of the solution in   \ci{Capuozzo:2023fll} 
%v3:    that corresponds to the probe  limit  in which   the  anomaly coefficient should scale as $N^2$ 
%v3:     remains to be understood.\footnote{We thank 
%v3:    J. Estes, B. Suzzoni and P. Capuozzo for a correspondence on this issue.} 
%%% v4:
   {A specialization of the bubbling result of \ci{Capuozzo:2023fll} with one of the  gauge group ranks
    kept fixed has the leading $N^2$ scaling, as shown in Appendix \ref{7004}.
   Its precise identification with our probe defect, and the matching of the normalizations and of the subleading terms, remain to be established.}\footnote{We thank
   J. Estes, B. Suzzoni and P. Capuozzo for a correspondence on this issue.}
  % \nw{The question of which defect of the (2,0) theory is described by the probe brane is also discussed in Appendix \ref{7004}.}
 %  }
   
   %The $N^2$ scaling can be reached for some particular fine-tuned parameters of the defect, %\footnote{Nevertheless, we are not able to match the coefficient in front of $N^2$ in the fine-tuned case and coefficient  in front of $N^2$ in  \eqref{202}.  }
%    but the sub-leading term  is of order $O(N)$, rather than $O(1)$ here that one would expect from  the quantum correction as the M5 brane tension $T_5\sim N^2$. Note that a prior it is not clear whether  the bubbling geometry solution can be extrapolated to some probe limit or be trusted beyond leading order.  }

%v4
\section{Concluding remarks}

 {Let us   summarize   some    subtle    points in the above discussion.
   %the differences from the original version of this paper.
They concern the boundary conditions for the modes that do not depend on the $S^1$ coordinate,
the periodicity of the M5 brane fermions, and the definition of the sum over the $S^1$ mode number $n$.

(i) The modes with $n\not=0$ are massive fields in AdS$_p$ with the standard (Dirichlet) boundary conditions.
The sector with $n=0$ is special. It contains the fields dual to the exactly marginal parameters of the defect.
For the world-volume gauge field it is a massless gauge field in AdS$_p$, for which there are two conformal boundary conditions
related by a duality transformation.
In the D3 brane case the $n=0$ mode of the vector is equivalent to two massless scalars in AdS$_3$:
the component of the vector along $S^1$ and the scalar dual to the 3d vector. Their boundary values are the parameters $\alpha$ and $\eta$ of the defect.\foot{In the original version of this paper  the contribution of the vector was continued from $n\not=0$ to $n=0$.
This corresponds to the Dirichlet condition for the 3d vector, i.e. to the Neumann condition for the dual scalar, and gave $b^{(1)}=-2$.}
Here we impose the Dirichlet condition on the dual scalar, as required for a fixed value of $\eta$ and by supersymmetry. This gives $b^{(1)}=-3$.
Similarly, in the M5 brane case the $n=0$ mode of the self-dual tensor is a massless 5d vector dual to a
 conserved current of the defect theory, and 
its contribution  should be included.\foot{A dependence of the 1-loop result on the boundary conditions for massless modes is familiar from other examples.
For the string in AdS$_5\times S^5$ ending on a Wilson loop the Dirichlet or Neumann conditions for the $S^5$ scalars
correspond to the supersymmetric or the ordinary Wilson loop \ci{Alday:2007he,Polchinski:2011im,Beccaria:2017rbe}.
For a gauge field in AdS$_4$ the two boundary conditions are related by a duality transformation \ci{Witten:2003ya}.
%Such a choice does not appear in the 1-loop M2 brane computations, as there are no world-volume gauge fields in that case.
}}

(ii) In the M5 brane case the supersymmetry requires  the fermions 
%were previously taken to be periodic on $S^1$.
%Supersymmetry requires them
 to be antiperiodic on  $S^1$  (see Appendices \ref{kisp} and \ref{7002}).
The fluctuations then form multiplets of the 4d $\N=2$ superconformal algebra.

(iii) In the D3 brane case the sum over $n$ is finite and no regularization is needed.
In the M5 brane case the sum is quadratically divergent.
If the modes are summed multiplet by multiplet, the finite part still depends on the prescription.
We constrain the prescription by comparing with the computation of 1-loop correction to 
  free energy of type IIB theory  on the  orbifold AdS$_5\times S^5/{\mathbb Z}_k$,
  whose twisted sector contains the same multiplets
 and the result can be compared with the conformal anomaly coefficients of the dual gauge theory. Assuming the same prescription should apply here, 
this rules out the Riemann $\zeta$-function  regularization prescription and leaves two other  possible choices of summation prescription
which give $a^{(1)}=-\frac{5}{48}$ and $a^{(1)}=-\frac{1}{12}$.

(iv) We also point out that the classical response of the background to the probe contributes at order $N^0$ in the D3 brane case
and at order $N$ in the M5 brane case, i.e. at the same or lower order as the 1-loop term
and thus   should   be taken into account when comparing with the dual theory result.}

%%%%%%%%%%%%%%%%%%%%%%%%%%%
\section*{Acknowledgements}
We are grateful  to M. Beccaria, N. Drukker,  S. Giombi, C. Herzog   and R. Metsaev  for  useful 
 discussions  and comments,   and also thank 
O. Aharony,   A. Chalabi   and D. Sorokin for related  correspondence.
This work  was supported in part by the STFC Consolidated Grants ST/T000791/1 and ST/X000575/1.

%%%%%%%%%%%%%%%%%%%%%%%%%%%%%%%%%%%
\appendix

%\section{\la{scalarflu} Scalar  fluctuations and spin connection of a $p$-brane in AdS$_{p+2} \times S^q$} 

 \section{Scalar fluctuations of a $p$-brane in AdS$_{p+2} \times S^1$}\la{scalarflu}
 
 The scalar fluctuations   of a $p$-brane  embedded in a  supersymmetric  way   in  AdS$_{p+2} \times S^1$
  can be computed in a universal way for any $p$, thus covering the cases of $p=3$ and $5$ discussed   in the main text. 
 The relevant part of the background is 
\ba
&ds_{ }^2=L_{A} ^2\big( du^2+\cosh^2 u \; ds_{\AdS_p}^2 +\sinh^2 u \; d\psi^2 \big)
+ L^2 d\phi^2~,\la{a1}\\
&C_{p+1}= L_{A}^{p+1}(\cosh^ {p+1} u-1) \vol_{\AdS_p}\wedge d\psi\equiv  {\sf C}_{p+1} \vol_{\AdS_p}\wedge d\psi ~.\la{a2}
\ea
% where we  have suppressed possible extra terms in the flux which are not relevant for us. 
The   configuration of the probe  $p$-brane 
related to   co-dimension  2    half-supersymmetric 
defect    in the  boundary theory is  such that  it wraps AdS$_p$  and   also 
(see \ci{Gutperle:2020gez}
and refs. there)  %, which gives the co-dim-2 defect on the boundary,    takes the following configuration: 
\be\la{aa1}
u=u_0={\rm const}~, \qquad  \qquad  L_A \psi=  L \phi~ \ .
\ee
The induced metric on the probe $p$-brane is  then  that of the equal-radii   AdS$_p\times S^1$   space 
\be\label{defectdualgeometry}
ds^2_{p+1}=L_A^2\cosh^2 u_0\; ds^2_{\AdS_{p }}+L^2_{A}\sinh^2 u_0 d\psi^2+L_A^2 d\psi^2
 =L_A^2\cosh^2 u_0\;  \big(ds^2_{\AdS_{p }} +d\psi^2\big)~.
\ee
Starting with the standard action   $ S\sim  - \int d^{p+1} \xi \sqrt { - G} +\int C_{p+1} $ of 
 a
 probe $p$-brane in the  background \rf{a1},\rf{a2}    let us find the  resulting quadratic fluctuation action 
in the  static gauge %($\alpha$  labels AdS$_p$  directions) %=\hat 0, \cdots, \widehat  {p-1}$)
\be\la{aa3}
 AdS_p=\{\xi^\alpha\}~,  \qquad u =u_0+  \delta u~, 
 \qquad \psi=\xi^{\hat p}~ , \qquad \phi=\frac{L_A}{L}\xi^{\hat p} +  \delta \phi~.
\ee
Then the induced metric $G_{\a\b}$   has the following quadratic  fluctuation part   
\beqn
ds_G^2=
L_A^2 \cosh^2 (u_0+\delta u)   (ds_{\AdS_p}^2+d\psi^2)
+2  L_AL d\delta\phi d\psi
+ (L_A^2 d\delta u^2+L^2 d\delta\phi^2)+...~.
\eeqn
This leads to\footnote{We use  that for  $\tilde g_{\mu\nu}=g _{\mu\nu}+h_{\mu\nu}$   one has 
$
\sqrt{-\tilde g}=\sqrt{-g} \big(1+ \frac12 h^\mu{}_\mu+\frac18   (h^\mu{}_\mu)^2-\frac14 h_{\mu\nu}h^ {\mu\nu} +\cdots\big)$.
Let us note also  that given  a metric $ds_g^2=g_{\m\n} dx^\m dx^\n=g_{ab}dx^a dx^b+g_{\theta\theta} d\theta^2$ ($a\neq \theta$) and  %the fluctuations
$
ds_ {\tilde g}^2=ds_g^2+    dF d\theta
$
where $F=F(x)$ is a function of $x^\m$ then 
%If $g_{a\theta}=0$, then
$
\sqrt{-\tilde g}=\sqrt{-g}\big(1+\frac12  g^{\theta\theta}\p_\theta F-\frac18  g^{\theta\theta}g^{\m\n}\p_\m F\p_\n F \big)~.\nonumber 
$
}
\ba
&\sqrt{-G}  =
\sqrt{-\sf  g}\, l_0^{p+1}\Big[
1+   \Big(  \frac{ L }{ L_A \cosh^2 u_0 } \p_\psi\delta\phi +(p+1)  \tanh u_0\, \delta u\Big)\no 
\\&
%\qquad\qquad\quad
+\frac{1}{ 2\cosh^2 u_0}
\Big({\sf g}^{\alpha \beta}\p_\alpha \delta u\,  \p_\beta \delta u +\frac{ L^2 \tanh^2 u_0 }{ L_A^2}  {\sf g}^{\alpha \beta}\p_\alpha  \delta \phi\,  \p_\beta \delta\phi
%\\& &
%\qquad\qquad\qquad\qquad\qquad\quad
+(p+1) (\cosh^2 u_0+p\sinh^2 u_0)\,  \delta u^2 \no 
\\& 
\qquad\qquad\qquad\qquad\qquad\quad
+\frac{ 2L (p-1)  \tanh u_0  }{ L_A} \delta u\,  \p_\psi\delta\phi
 \Big)+\cdots
\Big] ~,\la{a3}
\ea
where $l_0=L_A\cosh u_0$ and ${\sf g}_{\alpha\beta}$ is the metric  of the 
 unit-radius  AdS$_{p}\times S^1$. 
For the  variation  of the potential ${\sf C}_{p+1}$ in \rf{a2} we get 
%On the other hand, the flux becomes  (here we write the flux as $C_{p+1}={\sf C}_{p+1}\vol_{\AdS_p}\wedge d\psi$)
\be
 {\sf C}_{p+1}=-L_A^{p+1}+l_0^{p+1} \Big[1+(p+1) \tanh u_0\,  \delta u
+\frac12 (p+1)   (1+p \tanh^2 u_0 )\,  \delta u^2+\cdots \Big] ~.
\ee
Combined with \rf{a3} this gives 
\beqn
\sqrt{ -G}-\sf C_{p+1}&=&L_A^{p+1}+l_0^{p+1}
 \Big[\frac{L}{L_{ A}\cosh^2 u_0}    \p_\psi\delta\phi
 +
 \frac{1} {2   \cosh^2 u_0}
\Big(
{\sf g}^{\alpha \beta}\p_\alpha \delta u\,  \p_\beta \delta u 
  \\&&\quad\qquad\qquad
  +\frac{ L^2\tanh^2 u_0}{ L_A^2}   {\sf g}^{\alpha \beta}\p_\alpha\delta\phi\,  \p_\beta \delta\phi
 +2 (p-1) \frac{  L\tanh u_0}{ L_A } \delta u\,  \p_\psi\, \delta\phi
 \Big)+\cdots
 \Big] ~.
 \nonumber
\eeqn
As expected, the linear  fluctuation part is a total derivative   and  
 the  quadratic fluctuation  part of the action  is given by 
\ba
& S-S_0 = %&\int d^{p+1}\xi\;\sqrt{ -G}-C_{p+1}
%\\&=&
-    \frac {  l_0^{p+1}  }{2\cosh^2 u_0}
 \int d^{p+1}\xi\;\sqrt{-{\sf g}}
\Big[{\sf g}^{\alpha \beta}\p_\alpha \chi\p_\beta \bar\chi  +
\frac i 2 (p-1)      ( \bar\chi\p_\psi \chi -\chi \p_\psi \bar\chi)
\Big]~,\\ 
&\qquad \qquad  \chi\equiv \delta u -i \frac{ L  }{ L_A} \tanh u_0\,  \delta\phi~.
\ea
Rescaling  $\chi$  we  find that  a canonically normalized scalar fluctuation  action  is
\be
\frac12 \int d^{p+1}\xi\sqrt{-{\sf g}}
\Big({\sf g}^{ij}\p_i\chi \p_j \bar \chi  + 
\big[  \p_\psi \chi  -\frac {i}{   2  }(p-1) \chi \big]
\big[\p_\psi\bar \chi  + \frac {i}{   2  }(p-1) \bar \chi \big]
-\frac{1}{4}(p-1)^2 \chi\bar\chi
\Big)\, , \la{a6}
\ee
where $i,j$ are   indices of the  AdS$_{p}$  part of the brane metric.

 This   action   describes a conformally  coupled complex scalar on AdS$_{p} \times S^1$ 
 coupled also to a constant $U(1)$ potential $A= d \xi^{\hat p}$  with charge $q= \ha(p-1)$. 
 Fourier expanding  in $\psi=\xi^{\hat p}$, i.e.  $\chi=\sum_n  e^{in\xi^{\hat p}}\, \chi_n $
 we get a tower   of  scalars  on AdS$_p$  with  masses 
\be\la{a4}
m^2 = \big(n - \tfrac{p-1}{2}\big)^2 - \big( \tfrac{p-1}{2}\big)^2\ , \qquad \qquad n=0, \pm 1, \pm 2, ...\ . 
%\Delta (\Delta - (p-1)) = n^2-\frac i 2 (p-1) 2 i n  =n(n- (p-1))~,
\ee
The corresponding  boundary dimensions  are defined by  $\Delta[\Delta - (p-1) ] = m^2$   or 
for the  Dirichlet  boundary condition  choice  ($\Delta=\Delta_+$)  we get 
\be\la{a5}
\Delta-\tfrac{p-1}{2}= | n- \tfrac{p-1}{2}|~.
\ee
Since $\chi$ is complex, we actually have two towers of states with opposite shift, namely $\Delta-\tfrac{p-1}{2}= | n\pm  \tfrac{p-1}{2}|$.

%%%%%%%%%%%%%%%%%%%%%%%%%%%%%%
\section{Spin connection  and  projected spinor covariant derivative}\label{spincon}

 Here   we  will compute  the  spin  connection 
  contribution  to the  induced fermionic  covariant derivative. % which is important in studying the fermionic fluctuations.
Let us  consider the following  metric (cf. \rf{a1}) 
\be\label{ambAdS2}
ds_{ }^2=L_A ^2\big( du^2+\cosh^2 u \; ds_{\AdS_p}^2 +\sinh^2 u \; d\psi^2 \big)
+  L^2 \big( d\theta^2+ \sin^2\theta d\phi^2 \big)~.
\ee
We will label the target space coordinates   as follows:
 \be
 X^{0,1,\cdots ,\, p-1}=AdS_p~,\quad X^{p }=u~,  \quad X^{p+1}=\psi~,  \quad 
X^{p+2}=\theta~, \quad X^{p+3}=\phi ~.
\ee
We shall  assume that a $p$-brane wraps $AdS_p\times S^1$  as in \rf{aa1}.
Since $\theta = {\pi \ov 2}$  the 
 resulting  induced metric on the brane is  the same as in \eqref{defectdualgeometry}. 
For this  classical   brane configuration we  have (we will ignore bosonic fluctuations here, cf.   \rf{aa3}) %  we have  %as before:
 \beqn
 &&
 X^{0,1,...,\, p-1}=AdS_p= X^{\hat 0,\hat 1,...,\, \widehat{ p-1}}~,\quad X^{p }=u=u_0~,
   \quad X^{\widehat{ p+1}}=\psi=\xi^{\widehat{ p }}~,  \quad 
   \\&& 
X^{p+2}=\theta=\frac{\pi }{ 2}~, \qquad\qquad X^{p+3}=\phi = \frac{ L_{A }}{ L}\xi^{\widehat{ p }}~.
\eeqn
The vielbein components  for \eqref{ambAdS2} can be chosen as 
\beqn
&&
E_{m}^{\underline m }=L_A\cosh u\,  \hat E_{m}^{\underline m }~,
%=L_A  \cosh u/X^2
\quad\quad\quad m =0, ..., p-1 \quad
\\&&
E_{p}^{\underline p }=L_A~,  \quad
E_{p+1}^{\underline {  {p+1} }}=L_A\sinh u~, \quad
  E_{p+2}^{\underline  {  { p+2}} } =L~,\quad
 E_{p+3}^{\underline{   {p+3}} }=L\sin\theta ~,
\eeqn
where  the underlined indices   correspond to  tangent space directions
 and  $\hat E_{m}^{\underline m }$ is the vielbein  for the unit-radius  AdS$_{p}$. 
 %\footnote{For example,  $\hat E_{m}^{\underline m }=1/X^ {  { p-1}}$ for unit Poincare AdS $ds^2=(-(dX^0)^2+\cdots+(dX^{  { p-1}})^2)/(X^{  { p-1}})^2$.  } 
 The corresponding  spin connection  components  along the normal directions to the brane world volume 
 are found to be %can then be calculated 
\beqn
&&
\Omega_{m}^{\underline m\, \underline p}=\frac{ \tanh u}{L_{A  }} \, E_{m}^{\underline m }   
=\sinh u \, \hat E_{m}^{\underline m } ~, 
\quad\quad\quad  m =0, ..., p-1~, \quad
\\&&
\Omega_{p+1}^{\underline {p+1}\, \underline p}
= \frac{ 1}{L_{A  }\tanh u} \,  E_{p+1}^{\underline {  {p+1} }} = \cosh u~, \qquad  
  \qquad  
\Omega_{p+3}^{\underline {p+3}\;\underline {p+2}}=\cos\theta~.
\eeqn
%where we only show  a subset of non-vanishing spin-connections which are relevant for us, namely those with non-zero components along normal  directions.
 %Now we compute various geometric quantities on the $p$-brane  \eqref{adsbrane} which is induced from the ambient geometry \eqref{ambAdS2}.
 The projected covariant derivative  and  $\Gamma$-matrix are   given by 
\be\la{b1}
%\widehat\nabla_\alpha=
\p_\alpha X^m \nabla_m =\p_\alpha X^m \Big( \p_m +\frac14 \Omega_{m}^{\underline a\underline b}\Gamma_{\underline a\underline b}\Big)\ , \qquad \qquad \Gamma_\alpha=\p_\alpha  X^m \Gamma_m ~.
\ee
When $\alpha =\hat 0, \cdots, \widehat  {p-1}$, we have $\p_\alpha X^m=\delta_\alpha^m$, so
\be
\p_\alpha X^m \nabla_m =  \p_\alpha +\frac14 \Omega_{\alpha}^{\underline a\underline b}\Gamma_{\underline a\underline b}
=  \p_\alpha +\frac14 \Omega_{\alpha}^{\underline \beta\underline \gamma}\Gamma_{\underline \beta\underline \gamma}
+\frac{1}{2} \sinh u_0 \hat E_{\alpha}^{\underline \alpha }\Gamma_{\underline \alpha \underline p}~.
\ee
Since 
 $
\Gamma^\alpha=E^{\alpha}_{\underline \alpha }\Gamma^{\underline\alpha}=\frac{1}{L_A\cosh  u_0}\hat E^{\alpha}_{\underline \alpha }\Gamma^{\underline\alpha}
$, 
we have
\be\label{gammapl}
 \Gamma^\alpha \p_\alpha X^m \nabla_m  = \slashed \nabla_{_{\AdS_p}}
+\frac{1}{2L_A\cosh  u_0} \sinh  u_0  \hat E_{\alpha}^{\underline \alpha }\hat E^{\alpha}_{\underline \alpha' } \Gamma^{\underline \alpha' }\Gamma_{\underline \alpha  \underline p }
= \slashed \nabla   _{_{\AdS_p}}
+\frac{p}{2L_A } \tanh  u_0 \Gamma_{  \underline p}~,
\ee
where $ \slashed \nabla_{_{\AdS_p}}=\Gamma^\alpha\p_\alpha +\frac14 \Omega_{\alpha}^{\underline \beta\underline \gamma}\Gamma_{\underline \beta\underline \gamma}   $ is the   Dirac operator on AdS$_p$.

When $\alpha =\hat p$, we have $\p_\alpha X^m=\delta_ {p+1}^m+{L_{A }\ov L}  \delta_{p+3}^m$  so that %\p_ {p+3} $, so  
\beqn
\p_{\hat p} X^m \nabla_m 
&=&   \nabla_ {p+1}+\frac{ L_{A} }{ L }  \nabla_ {p+3} 
%\\&=&  \p_ {p+1} +\frac14 \Omega_{p+1}^{\underline a\underline b}\Gamma_{\underline a\underline b}
%+ \frac{ L_{\AdS } }{ L }  \Big(\p_ {p+3} +\frac14 \Omega_{p+3}^{\underline a\underline b}\Gamma_{\underline a\underline b}\Big)
% \nonumber\\ &=& 
%   \p_{p+1} +\frac14 \Omega_{{p+1}}^{\underline \beta\underline \gamma}\Gamma_{\underline \beta\underline \gamma}
%+  \frac{ L_{\AdS } }{ L }\Big( \p_{p+3}+\frac14 \Omega_{{p+3}}^{\underline \beta\underline \gamma}\Gamma_{\underline \beta\underline \gamma}\Big)
% +\frac{ \cosh u}{2}\Gamma_{\underline  {p+1} \underline p} 
% \nonumber
%\\&=& 
 %\p_{ \hat p}
  +\frac{ 1}{2}\cosh  u_0\,   \Gamma_{\underline  {p+1} \underline p}~.
\eeqn
Since 
\be
\Gamma^{\hat p}=\frac{1}{L_A\cosh  u_0}\hat E^{{\hat p}}_{\underline {\hat p} }\Gamma^{\underline{\hat p}}
%=\frac{1}{ L_A^2  \cosh^2 u_0 }  \Big( L_A \sinh u_0\Gamma_{\underline  {p+1}}+  L_A\Gamma_{\underline  {p+3}}
%\Big)
=\frac{ \sinh u_0\Gamma_{\underline  {p+1}}+ \Gamma_{\underline  {p+3}} 
}{ L_A   \cosh^2 u_0 }  ~,
\ee
 we find 
\be
\Gamma^{\hat p}\p_{\hat p} X^m \nabla_m  =  \Gamma^{\hat p}\p_{\hat p} +\frac{ \cosh  u_0}{2} \Gamma^{\hat p}\Gamma_{\underline  {p+1} \underline p}
%= \Gamma^{\hat p}\p_{\hat p}+\frac{  1}{2L_A} \Gamma^{\underline{\hat p}}\Gamma_{\underline  {p+1} \underline {   p}}
 = %\\&=&
 \Gamma^{\hat p}\p_{\hat p}
+\frac{  \tanh  u_0}{2L_ {A}} \Gamma_{ \underline {   p}} 
+\frac{1}{2L_A\cosh  u_0 }  \Gamma_{\underline  {p+3}\;\underline  {p+1} \;\underline p}   ~.
\label{gammap}
\ee
 Combining \eqref{gammapl} and \eqref{gammap}, we finally get
\be\label{b15}
\Gamma^\alpha\p_\alpha X^m \nabla_m  
=\slashed \nabla _{_{\AdS_p\times S^1}}
+
\frac{p +1}{2 L_A} \tanh  u_0\Gamma_{  \underline p}
 +  \frac{1}{2L_A\cosh   u_0 }  \Gamma_{\underline  {p+3}\;\underline  {p+1} \;\underline p}  ~.
 \ee

 \section{Fermion mass matrix for general $u_0$} \la{D3massu}
 %%%%%%%%%%%%%%%%%%
 
 When studying  fermionic fluctuations in  sections 2.2 and 3.2  
  we considered the limiting case of $u_0\to 0$.  
 Here we show that the equivalent  fermion mass matrix  is 
 obtained for the general  value of $u_0$,  in  both D3 and M5 brane  cases: the $u_0$ dependence 
  can be eliminated  by a spinor rotation.  
 
\subsection*{D3 brane case}

The fermion mass matrix in the  D3 brane  case is given by \eqref{25}:
\be 
\MM= 2 \tanh u_0\,   \Gamma_{  \underline 3}  
 +  \frac{1}{2\cosh  u_0 } \Gamma_{\underline 6\underline 4 \underline 3}
+\frac{1 }{4  }  \Gamma_{D3} \Gamma^\alpha   \big(\Gamma^{\underline 0 \cdots \underline 4}+\Gamma^{\underline 5 \cdots \underline 9}\big) \Gamma_\alpha \ .  
\ee 
%Let us show   that the dependence on $u_0$ can be eliminated  by a  spinor rotation 
%We can related different $u_0$ through the following rotation:  
Let us  consider the rotation matrix 
\be\la{c4}
\cR=\exp(\g \Gamma_{\underline 4\underline 6})=\cos \g+\Gamma_{\underline 4\underline 6} \sin \g~, \qquad
\cR^{-1}=\exp(-\g\Gamma_{\underline 4\underline 6})=\cos \g-\Gamma_{\underline 4\underline 6} \sin \g~,
\ee
where  $(\Gamma_{\underline 4\underline 6})^2=-1$  and $\g$ is related to $u_0$ by 
\be
 \frac{1+ \tan   \g }{1-  \tan  \g} = e^{u_0} ~, \ \ \ \ \ \ \ \ \ \ \  \tan \g=   \tanh { u_0\ov 2} \ . 
\ee 
 Then  $\cR\Gamma_{m}\cR^{-1}=\Gamma_m$ for  $m\neq   4,  6$  and 
\be\la{c5}
\cR\Gamma_{\underline 4}\cR^{-1}=\frac{1}{\cosh u_0}\Big( \Gamma_{\underline 4}-\sinh u_0  \Gamma_{\underline 6}\Big)~, \qquad 
\cR\Gamma_{\underline 6}\cR^{-1}=\frac{1}{\cosh u_0}\Big( \Gamma_{\underline 6}+\sinh u_0\Gamma_{\underline 4}\Big)~, \ee
This enables us to write the world-volume  components of $\Gamma$-matrices as  
\ba
&\Gamma_{ \hat 3}=\Gamma_4+\Gamma_6= \sinh u_0\,\Gamma_{\underline 4}+ \Gamma_{\underline 6}
=\cosh u_0\, \cR\Gamma_{\underline 6}\cR^{-1}~, \qquad
\Gamma^{\underline{  \hat 3}}=\Gamma_{\underline{  \hat 3}}=  \cR\Gamma_{\underline 6}\cR^{-1}~, \qquad
\\ 
&\qquad \ \ \  \Gamma_{D3}\equiv  \Gamma_{\underline {  \hat 0}\underline {  \hat 1}\underline {  \hat 2}\underline{  \hat 3}}
 =  \Gamma_{\underline {    0}\underline {    1}\underline {    2}}\cR\Gamma_{\underline 6}\cR^{-1}
   =    \cR \Gamma_{\underline {    0}\underline {    1}\underline {    2}\underline 6}\cR^{-1}~.
    \ea
 Then after a detailed computation, one finds 
    \beqn
  \MM \cosh u_0&=& 2 \sinh u_0\,   \Gamma_{  \underline 3}  
 +  \frac{1}{2 } \Gamma_{\underline 6\underline 4 \underline 3}
+\frac{1 }{4  } \cosh u_0\, \Gamma_{D3} \Gamma^\alpha   \big(\Gamma^{\underline 0 \cdots \underline 4}+\Gamma^{\underline 5 \cdots \underline 9}\big) \Gamma_\alpha\no 
\\
%&=& \cR \Big[ \frac12  \Gamma_{\underline 0\underline 1\underline 2\underline 5\underline 7\underline 8\underline 9}  
%+\sinh u_0 \Big( \Gamma_{\underline 3}- \Gamma_{\underline 0\underline 1\underline 2\underline 4\underline 5\underline 6\underline 7\underline 8\underline 9} \Big) 
 %\Big] \cR^{-1}
 &=&
\cR \Big[
\frac12  \Gamma_{\underline 3\underline 4\underline 6}  \Gamma^{\underline 0 \underline 1\cdots\underline 9}
+\sinh u_0\,   \Gamma_{\underline 3} \Big( 1-\Gamma^{\underline 0 \underline 1\cdots\underline 9} \Big) 
 \Big] \cR^{-1}\ . 
  \eeqn
Rotating the  fermions as
 $\vtheta\to \vartheta'= \cR^{-1}\vtheta$ we get   for the   
  gauge-fixed fermionic action (cf. \rf{244}) % And the action  becomes
\be\la{c8}
S_f=   \int d^4 \xi\;\sqrt{-   {  g} } \,\bar\vtheta\Big( { \slashed   \nabla } +  {  \MM}\Big) \vtheta
  \to     \int d^4 \xi\;   \sqrt{-   {  \sf g} } \,\bar\vartheta'\Big( { \slashed   \nabla } +  {  \MM'}\Big) \vartheta' \ , 
  \ee
where we ignore the overall   constant   factor  
% we have scaled out the overall world-volume  scale  factor  and $\sf g$ is the unit-radius $AdS_3\times S^1$
%metric. 
and  the rotated mass matrix  is   %above is then
\be
 {\MM}' =\cR^{-1} (  \MM \cosh u_0) \cR=
\frac12  \Gamma_{\underline 3\underline 4\underline 6}  \Gamma^{\underline 0 \underline 1\cdots\underline 9}
+\sinh u_0\,   \Gamma_{\underline 3} \big( 1-\Gamma^{\underline 0 \underline 1\cdots\underline 9} \big) 
=\frac12  \Gamma_{\underline 3\underline 4\underline 6} ~. 
\ee 
In the final equality  we have used  the chirality constraint   $\Gamma^{\underline 0 \underline 1\cdots\underline 9}\vartheta'   =\vartheta' $ 
as   $\MM'$ is acting on a MW spinor. %$\vartheta$. 
 Thus  the fermionic action is independent of $u_0$, up to an overall  constant  factor that can be absorbed into a rescaling of the fermionic field  and  in the present   context  does not change the value of the fermionic determinant.

\subsection*{M5 brane case}\la{M5massu}

Here the fermion  action is given by (cf. \rf{3361},\rf{88})
\ba\la{c10}
& S_f=   \int d^6 \xi  
%v3: \sqrt{-g}\,  g^{\alpha\beta} \, \bar\vtheta (1-\Gamma_*)
\sqrt{-g}\,  \bar\vtheta (1-\Gamma_*)   %%% v4: the factor g^{\alpha\beta} removed
\Big(
\slashed\nabla+\MM 
\Big) \vtheta
\ea
where $\slashed\nabla$ is the Dirac operator on AdS$_5\times S^1$ and   
\be\la{c11}
\MM=\Big(\frac{3}{2 L } \tanh u_0\Gamma_{  \underline 5}
 +  \frac{1}{4L\cosh  u_0 }  \Gamma_{\underline  {8}\underline  {6} \underline 5} \Big)
 +\Big(- \frac{5}{4L}\Gamma^{\underline 7\underline 8\underline 9\, \underline{\oo}}{} 
 -{1\ov 4L}\; \Gamma^{\hat 5}\Gamma^{\underline 7\underline 8\underline 9\,\underline{\oo}}{}_ 6
-\frac{ 1}{L}\Gamma^{\hat 5} \Gamma^{\underline 7\underline 8\underline  9\,\underline{\oo}} {}_8\Big)~,
\ee
where the first bracket   is the   contribution of the 
normal part of the spin connection 
 \eqref{spinconM5}  while the second 
  is the  contribution of $F_4$  terms in $D_m$ in \rf{3.43}  (see \eqref{fluxAdS5},\rf{91}). 
Introducing the  $u_0$-dependent rotation matrix as in \rf{c4}   so that  $\cR\Gamma_{m}\cR^{-1}=\Gamma_m$ for  $m\neq \underline 6,\underline 8$  and 
\be
\cR\Gamma_{\underline 6}\cR^{-1}=\frac{1}{\cosh u_0}\Big( \Gamma_{\underline  6}-\sinh u_0\,  \Gamma_{\underline 8}\Big), \qquad 
\cR\Gamma_{\underline 8}\cR^{-1}=\frac{1}{\cosh u_0}\Big( \Gamma_{\underline 8}+\sinh u_0\, \Gamma_{\underline 6}\Big), \qquad 
\ee 
we find
after a detailed computation that   the mass matrix \rf{c11}  may be  written as
\beqn
2L \cosh u_0\,\MM&=&\cR\Big[3  \sinh  u_0\, \Gamma_{  \underline 5}
 +    \frac12 \Gamma_{\underline  {8}\underline  {6} \underline 5}  
- \frac{3}{ 2} \; \Gamma_{\underline 7\underline 8\underline 9\;\underline { 10}}
+3\sinh u_0  \; \Gamma_{\underline 7\underline 6\underline 9\;\underline { 10}} 
\Big]\cR^{-1}\no 
\\&=&
\cR\Big[3  \sinh  u_0\, \Gamma_{  \underline 5} \Big( 1-\Gamma_{\underline  5\underline 6\underline 7\underline 9\;\underline { 10}} \Big) 
 +    \frac12 \Gamma_{\underline  {8}\underline  {6} \underline 5}  
 \Big( 1+3\Gamma_{\underline  5\underline 6\underline 7\underline 9\;\underline { 10}} \Big) 
\Big]\cR^{-1}\ . 
\eeqn
Let us  also rotate the fermions  so that   $\cR^{-1}\vtheta=\vartheta'$    and fix the  $\kappa$-symmetry  gauge  as in  \eqref{77}, i.e. 
 \be\label{c14}
 (1-\Gamma_*)\vtheta=0 =(1- \cR  \Gamma_{ \underline {  0}\cdots  \underline {  4}  \underline {  8}} \cR^{-1})\cR\vartheta'
 =\cR (1-\Gamma_{ \underline {  0}\cdots  \underline {  4}  \underline {  8}} )\vartheta'=0\ \  \to \ \ 
 (1-\Gamma_{ \underline {  0}\cdots  \underline {  4}  \underline {  8}} )\vartheta'=0~.
 \ee
 We also  have  $\bar\vartheta' (1+\Gamma_{ \underline {  0}\cdots  \underline {  4}  \underline {  8}} )=0 $.
    The fermion action  \eqref{c10} becomes 
\be\la{cc15}
%v3:   \int d^4 \xi\;   \sqrt{-   {  \sf g} } \,\bar\vartheta'\Big( { \slashed   \nabla } +  {  \MM'}\Big) \vartheta'\ , 
  \int d^{{6}} \xi\;   \sqrt{-   {  \sf g} } \,\bar\vartheta'\Big( { \slashed   \nabla } +  {  \MM'}\Big) \vartheta'\ ,    %%% v4: d^4 -> d^6
\ee
where we have scaled out the overall constant factor (so that  $\sf g$ is the metric of the unit-radius AdS$_5\times S^1$)
% world-volume size factor  
   and  
%the mass matrix  is  
\be\la{c16}
 {\MM}' =\cR^{-1} (2L\cosh u_0\, \MM)     \cR=
3  \sinh  u_0\, \Gamma_{  \underline 5} \Big( 1-\Gamma_{\underline  5\underline 6\underline 7\underline 9\;\underline { 10}} \Big) 
 +    \frac12 \Gamma_{\underline  {8}\underline  {6} \underline 5}  
 \Big( 1+3\Gamma_{\underline  5\underline 6\underline 7\underline 9\;\underline { 10}} \Big) 
  ~.
\ee 
Note that  here the first  term   multiplied  by 
% by $\Gamma_{  \underline 5} $  does not contribute to the action  \rf{cc15}   as 
$\Gamma_{  \underline 5}  -\Gamma_{\underline 6\underline 7\underline 9\;\underline { 10}}  $
does not contribute to the  gauge-fixed action as it commutes with $1-\G_*$.

From   the  analysis of   supersymmetry preserved  by the M5 brane embedding  in Appendix~\ref{kisp}
it follows that  assuming  that the  fermionic action  contains   the projector $1- \G_*$  as in \rf{765}   then $s_1=1$  in \rf{d15} 
 and  thus preservation of  supersymmetry is consistent with  the choice of $\G$-matrix representation  were 
   $s_2=1$ in \rf{ddd}. Then    the gauge-fixed  fermion should be subject  to \rf{555}, i.e.  
\be \la{5000}
\Gamma_{\underline 5\underline 6\un{7} \un{9} \, \underline {10}}\, \vtheta'= - \vtheta' \ , 
\ee
and thus finally we can replace \rf{c16} by 
\be 
 {\MM}' =     \frac12 \Gamma_{\underline  {8}\underline  {6} \underline 5}  
 \Big( 1+3\Gamma_{\underline  5\underline 6\underline 7\underline 9\;\underline { 10}} \Big) 
 = - \G_{\un 8 \un 6 \un 5} \la{666} \ , 
\ee 
which is equivalent to \rf{3.39}.

\section{\la{kisp}%Killing spinor  for      and 
Supersymmetry of M5 embedding  into AdS$_7\times S^4$ } 
%%%%%%%%%%%%%%%%%%%%%%%

Given the definition  of  covariant derivative $D_m$ in  \eqref{3.43} and the $F_4$   background 
 in  \eqref{79}, we find  that (cf. \rf{3322},\rf{3333})
\ba\la{d1}
D_m = \nabla_m-\frac{1}{ 4L}\Gamma _m \GS~, &\qquad m<7; \qquad 
D_m = \nabla_m+\frac{1}{ 2L}\Gamma _m \GS~, \qquad m\ge 7 \ ; \\
& \GS\equiv \Gamma_{\underline 7\underline 8\underline 9\, \underline {10}}. \la{dd2}
\ea
  Note that $[\Gamma _m, \GS]=0$ when $m<7$, and $\{\Gamma _m, \GS\}=0$ when $m\ge 7$.
The 11d Killing spinor equation  follows from the condition of the vanishing of the  local supersymmetry variation 
of the 11d gravitino 
(see, e.g., \ci{Lu:1998nu}) %transformation 
\be\la{d2}
\delta\psi_m=D_m \epsilon=0~.
\ee
Since $[\Gamma _m\GS, \Gamma_n\GS]=0$ when $m<7$ and $n\ge 7$, the AdS$_7$  and   $S^4$ parts of $\epsilon$ 
  factorize  % In particular, this means we   have
\be\la{d3}
\epsilon=\epsilon_{_{\AdS_7}}  \, \epsilon_{_{S^4}} = \cM_{\AdS_7}\cM_{S^4}\,  \epsilon_0~,\qquad \qquad [\cM_{\AdS_7},\cM_{S^4} ]=0 \ , \ \ \ \ \ \   
\ee
where $\epsilon_0$ is a constant spinor.

Let us first consider the AdS$_7$ part  and set 
%We can solve different components of Killing spinor equation in a specific order. For simplicity we set 
$L_A=2L= 1$  for simplicity.  Let us  first consider the $m=u=5$  (cf. \rf{78},\rf{60}) component of \rf{d2}, i.e.
\be\label{usol}
D_u\epsilon=\Big(\p_u -\frac12\Gamma_{\underline u}\GS\Big) \epsilon=0 \qquad \xrightarrow{}\qquad
\epsilon=e^{ \frac12 u\Gamma_{\underline u}\GS }\epsilon'~,
\ee
where $\epsilon'$ is independent of $u$.  Next, for $m=\psi=6$ we get\foot{We use the following relations 
 which are  valid for $i,j<7$: \ 
$
e^{\frac12\alpha \Gamma_i\GS}\Gamma_{ij}e^{-\frac12\alpha \Gamma_i\GS}
=\cosh \alpha \Gamma_{ij}+\sinh\alpha \Gamma_j\GS$ and $
e^{\frac12\alpha \Gamma_i\GS}\Gamma_{ j}\GS e^{-\frac12\alpha \Gamma_i\GS}
=\sinh \alpha \Gamma_{ij}+\cosh\alpha \Gamma_j\GS.$
}
   %Then we consider  $\psi$ component, 
\be\label{psieq}
D_\psi\epsilon=\Big[\p_\psi -\frac12 \Big( \cosh u\,  \Gamma_{\underline u\underline \psi}+\sinh u\, \Gamma_{\underline \psi}\GS\Big)\Big]\epsilon
=\Big[\p_\psi -\frac12 e^{  \frac12 u\,\Gamma_{\underline u}\GS }\Gamma_{ \underline u\underline \psi} e^{ -\frac12 u\Gamma_{\underline u}\GS } \Big] \epsilon=0~,
%, \qquad\epsilon=e^{ \frac12 u\Gamma_{\underline u}\GS }\epsilon'
\ee
Substituting \eqref{usol} into \eqref{psieq}, we get that 
\be
 \Big(\p_\psi -\frac12  \Gamma_{ \underline u\underline\psi}  \Big) \epsilon'=0
 \qquad \xrightarrow{}\qquad
  \epsilon'=e^{\frac12\psi \Gamma_{ \underline u\underline \psi} }\epsilon''~.
\ee
For $m=0, \cdots, 4$, we have
\be
D_m =\tilde \nabla_m-\frac12 \Big( \cosh u\,  \Gamma_m\GS- \sinh u\, \Gamma_{m \underline u}\Big)
=\tilde \nabla_m-\frac12 e^{  \frac12 u\Gamma_{\underline u}\GS }   \Gamma_m\GS e^{-  \frac12 u\Gamma_{\underline u}\GS }~,
\ee
where $ \tilde \nabla_m$   has spin connection 
 components   along AdS$_5$ only. This means we can write the Killing spinor in AdS$_7$  parametrized as in \rf{60} in terms of the Killing spinor on AdS$_5$ (independent of $u$ and $\psi$) as 
\be\label{777}
\epsilon_{_{\AdS_{7}}}= e^{  \frac12 u\Gamma_{\underline u}\GS } e^{\frac12\psi \Gamma_{ \underline u\underline \psi} }
\, \epsilon_{_{\AdS _5}}~, \qquad \qquad \Big( \hat \nabla_m- \frac12   \Gamma_m\GS  \Big)\epsilon_{_{\AdS _5}}=0~.
\ee
Similarly, for $S^4$ components of \rf{d2} we get ($\theta=7, \ \phi=8$)\foot{We use that  for $i,j\ge 7$: \ 
$
e^{\frac12\alpha \Gamma_i\GS}\Gamma_{ij}e^{-\frac12\alpha \Gamma_i\GS}
=\cos  \alpha \Gamma_{ij}+\sin \alpha \Gamma_j\GS $ and $
e^{ \frac12 \alpha\Gamma_{i}\GS}  \Gamma_{ j}\GS e^{- \frac12 \alpha\Gamma_{  i}\GS}
=   \cos\alpha\Gamma_{j}\GS-\sin\alpha \Gamma_{ij} 
$.}
% We can use the same strategy to study the Killing spinor on $S^4$. We start with the $\theta$ component 
\ba\label{KStheta}
& D_\theta\epsilon=\Big(\p_\theta+\frac12 \Gamma_{\underline \theta}\GS \Big) \epsilon=0
 \qquad \xrightarrow{}\qquad
\epsilon=e^{-\frac12 \theta\Gamma_{\underline \theta}\GS}\epsilon'~,
\\ 
\label{piheq}
&D_\phi\epsilon=\Big[\p_\phi+\frac12(-\cos\theta \Gamma_{\underline \theta\underline \phi}
+\sin\theta \Gamma_{\underline \phi}\GS) \Big] \epsilon=\Big(\p_\phi-\frac12e^{-\frac12\theta \Gamma_{\underline \theta}\GS}  \Gamma_{\underline \theta\underline \phi}e^{ \frac12\theta \Gamma_{\underline \theta}\GS} \Big) \epsilon=0~.
\ea
Substituting  \eqref{KStheta} into \eqref{piheq}, we find
\be
 \Big(\p_\phi-\frac12   \Gamma_{\underline \theta\underline \phi} \Big)\epsilon'=0  \qquad \xrightarrow{}\qquad
\epsilon'=e^{ -\frac12\phi \Gamma_{\underline \phi \underline   \theta}  }\epsilon''~.
\ee
For  the remaining   $m=9,10$   components of $D_m$    (corresponding to the $S^2\subset S^4$   angles 
$\vp_1=9,\,  \vp_2=10$) we get 
\be 
D_m=\hat  \nabla_m+ \frac12e^{-\frac12 \Gamma_{\underline \theta}\GS} 
 \Gamma_ m\GS e^{ \frac12 \Gamma_{\underline \theta}\GS} ~,
\ee
where $ \hat \nabla_m$ contains only the  $S^2$  spin connection. 
As a result,\footnote{The explicit solution for  $\epsilon_{_{S^2}}$   is 
$e^{-\frac12\vp_1 \Gamma_{\underline  {\vp_1}}\GS} e^{ -\frac12\vp_2 \Gamma_{\underline  {\vp_2} \underline    {\vp_1}}  }$.}
  %we find $S^4$ spinor can be written as
\be\label{S4spinor}
\epsilon_{_{S^4}}=e^{-\frac12\theta \Gamma_{\underline \theta}\GS} e^{ -\frac12\phi \Gamma_{\underline \phi \underline   \theta}  }\, \epsilon_{_{S^2}}~,\qquad \qquad \Big( \hat  \nabla_m+ \frac12  \Gamma_ m\GS   \Big)\, \epsilon_{_{S^2}}=0~.
\ee
Combining \eqref{777} and  \eqref{S4spinor}, we find     the Killing spinor on AdS$_7\times S^4$ can be written  as
% $$ where $\epsilon_0$ is constant and  
\ba
\epsilon=\cM\epsilon_0 \ , \qquad 
\cM=\cM_{\AdS_7}\cM_{S^4}&= e^{  \frac12 u\Gamma_{\underline u}\GS } e^{\frac12\psi \Gamma_{ \underline u\underline \psi} }\cM_{\AdS_5}
e^{-\frac12\theta \Gamma_{\underline \theta}\GS} e^{ -\frac12\phi \Gamma_{\underline \phi\underline \theta}    }\cM_{S^2}
\no \\ &=
 e^{  \frac12 u\Gamma_{\underline u}\GS } e^{\frac12\psi \Gamma_{ \underline u\underline \psi} }
e^{-\frac12\theta \Gamma_{\underline \theta}\GS} e^{ -\frac12\phi \Gamma_{\underline \phi\underline \theta}    }\cM_{\AdS_5\times S^2}~,
\label{d14}
\ea
where $\cM_{\AdS_5\times S^2}$ depends only on  the  AdS$_5$ and $S^2$  coordinates.

Let us now consider the M5 brane   configuration in \rf{78}  and  find the  amount of 
global  supersymmetry it preserves
(for a general discussion see, e.g.,  \ci{Bergshoeff:1997kr,Lunin:2007ab,Simon:2011rw}). 
  The  supersymmetry  condition for the brane   embedding is 
   determined by the  projector $\ha (1+\G_*)$  orthogonal to the one 
 that enters the  $\kappa$-symmetry transformation  of $\vtheta$ 
  % $\delta \theta= (1 - \G_*) \kappa$ 
   in the  M5 brane action \rf{3.41},\rf{765},\rf{766}
  and thus  the  $\kappa$-symmetry  gauge on $\theta$ in  \rf{77}.\foot{In general   \ci{Bergshoeff:1997kr}, 
  the variation  of $\vtheta$ in \rf{765} 
   under  the $\kappa$-symmetry and  target space  supersymmetry is 
  $\delta \vtheta= (1 - \G_*) \kappa + \epsilon$.  Upon gauge fixing $(1 - \G_*) \vtheta=0$, i.e.
  $\vtheta= (1+ \G_*)\tilde \vtheta $. 
  The preservation of the gauge condition 
    implies  $(1-\G_*)\delta  \vtheta= (1+ \G_*) \delta\vtheta=0$
   and thus  the condition for unbroken 
  global  supersymmetry of the  
  brane embedding  is  $(1 + \G_*) \epsilon=0$.} 
  
  To account for possible   orientation choice  ambiguity let  us introduce the parameter  $s_1=\pm 1$  and 
  assuming  the  fermionic action  (and thus also the gauge fixing condition \rf{77})
   contains the projector $1- s_1 \G_*$ ($s_1=1$ in \rf{765}) 
  consider in general the condition   
  $(1+ s_1 \G_*) \epsilon=0$, i.e.  (cf. \rf{3.35})
\be\la{d15}
  \Gamma_ {*}\epsilon=- s_1\,  \epsilon~,   \qquad\ \ \ 
  \Gamma_ {*} =  \Gamma_{ \underline { \hat 0}\un{ \hat 1}\un{\hat  2}\un{\hat 3} \un{\hat 4}  \underline {\hat  5}}= 
  \frac{1}{\cosh u_0} \Gamma_{ \underline {  0}\cdots  \underline {  4}}\Big(\sinh u_0\Gamma_{\underline 6} 
  +  \Gamma_{\underline 8}\Big)~, \ \ \ \ \ \  s_1=\pm 1 \ . 
\ee
%where we used the expression for  $\Gamma_{\un 5}$ (cf. \rf{3.35}). 
Let us also set 
\be\la{ddd}  \G_{11}\equiv  \Gamma_{ \underline {  0}\cdots  \underline {  9} \, \underline { 10}}=  s_2 \ , \qquad \qquad 
\G^2_{11}=1 \ , \ \ \ \ s_2 =\pm 1 \ , 
\ee
where $s_2$ is introduced to account for  a freedom in choice of $\G$-matrix representation.
Note that under $\G_m \to -\G_m$  we have $\G_* \to \G_*$,  $\G_{11} \to - \G_{11}$
and  M5 brane   action \rf{3.41}  stays  the same (up to overall sign) 
 provided  one also  changes $F_{4} \to - F_{4}$  in the covariant derivative $D_m$ in \rf{3.43}. 
Equivalently, the  Killing  spinor  in this case is still given by \rf{d14} with $\G_m \to -\G_m$  (with $\hat \G$ in \rf{dd2}  staying invariant). 

In \rf{d15}
$\epsilon$ is the Killing spinor of the  AdS$_7\times S^4$ background  \eqref{d14}, 
specialised  to  the brane  solution \rf{51}, i.e.   $ u=u_0, \  \phi=2\psi,\ \theta={\pi\ov 2}$. Since the  brane extends along AdS$_5$ and is  localized at  a point in $S^2$, we 
do not need  track the dependence of the Killing spinor in  \rf{d14} on those coordinates
and   may effectively set  %  We may assume a particular choice of coordinates where
 $\cM_{\AdS_5\times S^2}=1$.  Then 
\be\la{d16}
\epsilon=\cbM\, \epsilon_0~, \qquad\qquad \ \ \ 
\cbM=
 e^{  \frac12 u_0\Gamma_{\underline u}\GS } e^{\frac12\psi \Gamma_{ \underline u\underline \psi} }
e^{-\frac\pi 4 \Gamma_{\underline \theta}\GS} e^{ -\psi \Gamma_{\underline \phi\underline \theta}    } ~.
\ee
Then   condition  \rf{d15}   may be written as
\be\la{d17}
 K\epsilon_0=0 \ , \qquad \qquad 
K\equiv  \cbM ^{-1}(1+s_1 \Gamma_* )\cbM     \ . 
\ee
%r $t=\pm 1$ and constant spinor $\epsilon_0$.
% Let us first consider the case of \aat{[This was already defined in \rf{d1} ?  why choice or case? ]}
Using \rf{ddd} we get  for $\hat \G$ in \rf{dd2} 
  \be \la{d18} %\Gamma_{ \underline {  0}\cdots  \underline {  9} \underline { 10}}=  1 \ , \qquad \ \  
  \GS \equiv\Gamma_{\underline 7\underline 8\underline 9\; \underline {  10}} = s_2 
  \Gamma_{\underline 0\cdots\underline 6} \ . \ee
  One can show that   in general 
\be\la{d2111}
K =1-s_1 s_2 
+s_1  s_2\Big(  1
-\tanh u_0\cos\psi\;\Gamma_{\underline 6\underline 9\,\underline{ 10}}
+  \tanh u_0\sin\psi\;  \Gamma_{\underline 6\underline 7\underline 8 \un{9}\,  \un{10} } \Big) \Big(1- \Gamma_{\underline 5\underline 6\underline 7\underline 8} \Big)  ~.
\ee
To have a non-trivial  constant $\epsilon_0$  solution   of \rf{d17}  we are thus to require 
\be 
\la{d200}
s_1 s_2 =1 \ , \ \ \ \ \ \ \ \qquad
\big(1- \Gamma_{\underline 5\underline 6\underline 7\underline 8} \big) \epsilon_0=0 \ . \ee
Since $\Gamma_{\underline 5\underline 6\underline 7\underline 8}^2=1$   we thus  get a projector 
implying preservation of  half of the original supersymmetry. 
  
  Note that performing the  rotation $\varepsilon =\cR^{-1}\epsilon$ discussed in  Appendix~\ref{M5massu}, we 
get $\Gamma_ {*}$ in \rf{d15} 
 transformed to its $u_0=0$ value $\Gamma_{\underline 0\cdots\underline 4\underline 8}$ 
and thus the condition \rf{d15} becomes (cf. \rf{ddd})
\be\label{BPSM5}  
\G_* \varepsilon=\Gamma_{\underline 0\cdots\underline 4\underline 8}\, \varepsilon=-s_1  \varepsilon ~, \ \ \ \ \ \ \ \ \ 
\Gamma_{\underline 5\underline 6\un{7} \un{9} \, \underline {10}}\, \varepsilon=-\Gamma_{\underline 0\cdots\underline 4\underline 8} \Gamma_{ \underline {  0}\cdots  \underline {  9} \, \underline { 10}} \varepsilon=s_1s_2   \varepsilon= \varepsilon \ . 
\ee
At the same  time,  the $\kappa$-symmetry gauge  condition \rf{77} 
 on the fermionic field  $\vtheta$  that   involves  the projector 
complementary to the one  in  \rf{d15}  reads
\be \la{555}
\G_* \vtheta' = \Gamma_{\underline 0\cdots\underline 4\underline 8}\, \vtheta'= s_1  \vtheta' ~, \ \ \ \ \ \ \ \ \ 
\Gamma_{\underline 5\underline 6\un{7} \un{9} \, \underline{ 10}}\, \vtheta'=- s_1s_2   \vtheta'= - \vtheta' \ . 
\ee
This is the condition we used in the main text to arrive at  the expression for the  mass operator in  \rf{3.39}. 

%%% v4: new paragraph
{Let us note that the Killing spinor \rf{d16} is antiperiodic in $\psi$. Using that $\G_{\un u \un \psi}=\G_{\un 5\un 6}$,
$\G_{\un\phi\un\theta}=\G_{\un 8\un 7}$ and $\G_{\un\theta}\GS=\G_{\un 8\un 9\,\un{10}}$, and moving the constant factor in \rf{d16} to the right, one finds
\be\la{7010}
\cbM=e^{  \frac12 u_0\Gamma_{\underline u}\GS }\  e^{\frac12\psi\, \Gamma_{ \un 5\un 6} }\
e^{\psi\,  \Gamma_{\un 7\un 9\, \un{10}}}\  e^{-\frac\pi 4 \Gamma_{\underline \theta}\GS}   \ .
\ee
The matrices $\Gamma_{ \un 5\un 6}$ and $\Gamma_{\un 7\un 9\, \un{10}}$ commute and square to $-1$.
Thus all components of $\epsilon$ on the brane depend on $\psi$ as $e^{i\psi(\pm\frac{1}{2}\pm 1)}$, and for the components selected by \rf{d200}
the dependence is $e^{\mp\frac{i}{2}\psi}$.

This reflects the fact that in the polar frames used here a single-valued spinor is antiperiodic in each of the angles $\psi$ and $\phi$, and that $\phi=2\psi$ on the brane.
In the D3 brane case, where $\phi=\psi$, the corresponding phases are $e^{i\psi(\pm\frac{1}{2}\pm\frac{1}{2})}$ and the preserved supersymmetry parameters
do not depend on $\psi$. As the fermionic field $\vtheta$ is shifted by $\epsilon$ under the supersymmetry transformation, it should be antiperiodic
in $\xi^{\hat 5}$ in the M5 brane case.}

% \newpage

 %v4
\section{{Comments on boundary conditions, supermultiplets and summation prescription}}\la{7000}

%\nw{This Appendix is new in v4.}

{This  Appendix   discusses  the boundary condition for the zero mode of the D3 brane vector,
the periodicity of the M5 brane fermions, 
 the supermultiplet structure of the M5 brane fluctuations, the prescription for the sum over the multiplets,
and possible  terms of order $N$ in the M5 brane case.}

\subsection{D3 brane: vector zero mode  and  prescription for the mode sum}\la{7001}

Let us denote $\nu\equiv\Delta-1$. According to \rf{333},\rf{34},\rf{35} the contributions to $P$ in \rf{38} of a real scalar, of a transverse vector and of a
2-component fermion are
\be\la{7013}
p_0(\nu)=\nu^3\ , \qquad\ \ p_1(\nu)=2\nu(\nu^2-3)\ , \qquad \ \ p_{1/2}(\nu)=-2\nu\big(\nu^2-\tfrac{3}{4}\big)\ .
\ee
%\paragraph{Relation to the Maxwell multiplet in flat space.}
The shifts of $n$ in \rf{111} and \rf{29} are removable.
The shift $n\to n\pm 1$ of the scalar $\chi$ is undone by the single-valued redefinition $\chi\to e^{i\xi^{\hat 3}}\chi$.
The shift $n\to n\pm \ha$ of the fermions is equivalent to making them antiperiodic in $\xi^{\hat 3}$.

{Let us stress that this does not change the boundary condition used in section \ref{s2D3}.
There the fermions are periodic in $\xi^{\hat 3}$ and are coupled to the constant potential in the operator
$i\slashed\nabla_{\AdS_3}+i\G^{\hat 3}\big(\del_{\hat 3}-\ha\Gamma_{\underline 4\underline 3}\big)$ below \rf{28},
which is the origin of the shift $n\to n\pm\ha$ in \rf{29}.
This potential can be removed by the redefinition
\be\la{7046}
\vtheta'=e^{-\frac{1}{2}\xi^{\hat 3}\, \Gamma_{\underline 4\underline 3}}\, \vtheta \ ,
\qquad\qquad \vtheta'(\xi^{\hat 3}+2\pi)=-\vtheta'(\xi^{\hat 3}) \ .
\ee
Then $\vtheta'$ has no shift of $n$ but is antiperiodic, so that its mode numbers are half-integer.
The two descriptions are equivalent and give the same AdS$_3$ masses $|n\pm\ha|$ with integer $n$.

Thus the set of AdS$_3$ modes in \rf{111},\rf{15},\rf{29},\rf{32} is the same as for the free $\N=4$ Maxwell multiplet on AdS$_3\times S^1$
with the circle of length $2\pi$, with no background potential, and with periodic bosons and antiperiodic fermions.
%This is the same as for a thermal ensemble on AdS$_3$ at the temperature $T=\tfrac{1}{2\pi}$, which is the only sense in which
%the word ``thermal'' is used below.
The supersymmetry is not broken, as in this description the supersymmetry parameters are antiperiodic as well.

This is different from the M5 brane case in section \ref{s3M5}.
There the shifts of $n$ of the fermions in \rf{299} are integer and cannot be traded for a change of the periodicity,
so that the periodicity of the fermions is a separate input.}

%The shift $n\to n\pm \ha$ of the fermions is equivalent to making them antiperiodic in $\xi^{\hat 3}$.
%Thus the set of AdS$_3$ modes in \rf{111},\rf{15},\rf{29},\rf{32} is the same as for the free $\N=4$ Maxwell multiplet on AdS$_3\times S^1$
%with the circle of length $2\pi$ and the ``thermal'' spin structure. 

The AdS$_3\times S^1$ space is conformal to flat space,
$
dx_1^2+dx_2^2+d\varrho^2+\varrho^2 (d\xi^{\hat 3})^2
=\varrho^2\big[\frac{dx_1^2+dx_2^2+d\varrho^2}{\varrho^2}+(d\xi^{\hat 3})^2\big].
$
The antiperiodic fermions correspond to single-valued spinors in flat space, and the Dirichlet conditions
$\Delta=\Delta_+$ correspond to fields that are regular at $\varrho=0$.
Thus the fluctuations are equivalent to the free $\N=4$ Maxwell multiplet in flat space, or on $S^4$ of radius $r$
with the defect $S^2$ removed.

For a 4d conformal theory on $S^4$ the logarithmic term in the free energy is $4\aa\log(r\LR)$,
where $\aa$ is the conformal anomaly coefficient.
For the $\N=4$ Maxwell multiplet of 6 scalars, 4 Weyl fermions and the vector one has
\be \la{7019}
\aa_{_{U(1)}}=6\times\tfrac{1}{360}+4\times\tfrac{11}{720}+\tfrac{62}{360}=\tfrac{1}{4} \ .
\ee
   With the sum over the modes taken at fixed $n$ as in \rf{38},\rf{39}, the 1-loop result \rf{40} agrees with this term,
   \be\la{7020}
   F^{(1)}=\log(r\LR)=4\, \aa_{_{U(1)}}\, \log(r\LR)\ .
   \ee
   This agreement supports the fixed-$n$ prescription but is not a derivation of it from the conformal map.
   
   This is also the form found in a similar 3d problem. For the vortex loops in the
ABJM theory the 1-loop M2 brane term of \ci{Drukker:2023bip} is the free energy
on $S^3$ of the free $\N=8$ multiplet, corrected by the defect contribution of
this multiplet \ci{Cuomo:2026kuo}.
   
%   If the sum over $n$ is instead taken field by field with the $\zeta$-function, or with a 4d proper-time cutoff,
 %  one finds $P=\tfrac{15}{4}$ instead of 6.   %The two values differ by a finite local term on the world volume.
If the $n=0$ mode of the vector is instead included with the formal zero value of the vector term in \rf{39},
one gets $P=4$, i.e. $F^{(1)}=\tfrac{2}{3}\log(r\LR)$.
This corresponds to counting the vector in \rf{7019} with $\tfrac{32}{360}$ instead of $\tfrac{62}{360}$,
i.e. with $\tfrac{16}{45}$ instead of $\tfrac{31}{45}$ in $4\aa$.
The same discrepancy is known in the entanglement entropy of the Maxwell field \ci{Dowker:2010bu},
where it is resolved by the edge modes \ci{Donnelly:2014fua}.
In the present context it is resolved by the Dirichlet condition for the dual scalar $\sigma$ in \rf{7007},
which contributes $p_0(1)=1$ instead of $p_0(-1)=-1$ to $P_0$.

  % If the sum over $n$ is instead taken field by field with the $\zeta$-function, or with a 4d proper-time cutoff,
   %one finds $P=\tfrac{15}{4}$ instead of 6. The two values differ by a finite local term on the world volume.

%Let us note also that the $n=0$ multiplet \rf{7006} contributes $P_0=4\, p_0(1)+4\, p_{1/2}(\ha)=6$.
%With the alternative quantization of this multiplet, i.e. with the four scalars having $\Delta=0$
%and the four fermions having $\Delta=\ha$, one would get $P_0=-6$.
%The corresponding values of $b^{(1)}=-\ha P$ differ by 6, which is the central charge of a free 2d hypermultiplet.
%Thus \rf{40} is minus one half of this central charge.

%\paragraph{The classical response.}

Let  us now comment  on relation to  ref.   \ci{Jensen:2018rxu}. 
For $N_1$ coincident probes the supergravity expression $b=3(N^2-\sum_i N_i^2)$ of \ci{Jensen:2018rxu}, obtained from the bubbling solutions of \ci{Gomis:2007fi}, gives
\be\la{7021}
b=3\big[N^2-(N-N_1)^2-N_1^2\big]=6NN_1-3N_1^2-3N_1^2 \ .
\ee
The first term is the probe action. One may interpret the second as the classical backreaction of the flux of the probes, i.e. as \rf{7009} for $N_1=1$,
and the third as $-12\,\aa(U(N_1))$, which for $N_1=1$ is the fluctuation term \rf{40}.
Both terms are, however, parts of the same classical supergravity expression,
 and the known total does not determine how it is divided between
the response to the probe and the fluctuations. The solutions of \ci{Gomis:2007fi} fix the individual fluxes $N_i$ and their sum $N$,
but do not by themselves imply the shift \rf{7009}, which   remains a conjecture.

\subsection{M5 brane:  anti-periodicity of the fermions and supermultiplets}\la{7002}

%\paragraph{Periodicity.}
There are three arguments  implying  that the fermions are antiperiodic in $\xi^{\hat 5}$ in the frame used in section \ref{s3M5}.
First, the frames in Appendix \ref{spincon} are polar frames in the planes $(u,\psi)$ and $(\theta,\phi)$.
A single-valued 11d spinor is antiperiodic under $\psi\to\psi+2\pi$ and under $\phi\to\phi+2\pi$ in such frames.
On the solution \rf{51} the shift $\xi^{\hat 5}\to\xi^{\hat 5}+2\pi$ gives $\psi\to\psi+2\pi$ and $\phi\to\phi+4\pi$, i.e. the total sign is $-1$.
In the D3 brane case both angles are shifted by $2\pi$ and the sign is $+1$.

Second, the Killing spinors restricted to the brane are antiperiodic, see \rf{7010}.
Third, let $J_\psi$ and $J_\phi$ be the generators of the shifts of $\psi$ and $\phi$. The generator of $\del_{\hat 5}$ is $J_\psi+2J_\phi$.
It is the $U(1)_r$ generator of the superconformal algebra $su(2,2|2)$ preserved by the defect.
Here $SU(2)_R\times U(1)_r$ is the R-symmetry of $su(2,2|2)$, and $r$ denotes the $U(1)_r$ charge,
normalized so that the supercharges have $r=\mp\ha$ and a chiral primary has $\Delta=r$
(the complex scalar of the (2,0) theory with $\Delta=2$ has $J_\phi=1$ and thus the $U(1)_r$ charge  $r=\Delta$, as required for a chiral primary). Thus the bosons and the fermions in a multiplet have values of $n$
 that differ by half-integers.

As in the D3 brane case, the constant $U(1)$ potential can be removed by a redefinition of the fields.
Here both the shifts $\pm 2$ of the scalar $\chi$ in \rf{76} and the shifts $\pm 1$ of the fermions in \rf{299} are integer.
Thus the redefinition is single-valued and does not change the periodicity of the fields,
in contrast to the D3 brane case where it made the fermions antiperiodic.
One is left with the free (2,0) tensor multiplet on AdS$_5\times S^1$ with no background potential,
with periodic bosons and antiperiodic fermions.

The AdS$_5\times S^1$ space is conformal to flat 6d space, with $\xi^{\hat 5}$ being the polar angle in the plane transverse to the defect.
A single-valued spinor in flat space is antiperiodic in this angle.
Thus the fluctuations are equivalent to the free tensor multiplet in flat space with the fields regular at the position of the defect,
as in the D3 brane case in Appendix \ref{7001}.

%\paragraph{Chiral multiplets.}
For each integer $m\ge 2$ the modes form a chiral multiplet ${\cal E}_m$ of $su(2,2|2)$, whose primary has $\Delta=r=m$, and its conjugate.
The content of ${\cal E}_m$ is given in Table \ref{7011}, where $\nu\equiv \Delta-2$ and $\chi$ is the complex scalar with the dimensions \rf{76}.
\begin{table}[h]
\begin{center}
\begin{tabular}{|c|c|c|c|l|}
\hline
field & $\Delta$ & $r=n$ & $SU(2)_R$ & mode \\
\hline
$\varphi$ & $m$ & $m$ & ${\bf 1}$ & $\chi$ with $\nu=|n-2|$ \\
$\lambda$ & $m+\ha$ & $m-\ha$ & ${\bf 2}$ & fermion with $\nu=|n-1|$ \\
$F^+$, $D$ & $m+1$ & $m-1$ & ${\bf 1}$, ${\bf 3}$ & tensor $(1,0)$ and 3 scalars with $\nu=|n|$ \\
$\psi$ & $m+\tfrac{3}{2}$ & $m-\tfrac{3}{2}$ & ${\bf 2}$ & fermion with $\nu=|n+1|$ \\
$\varphi''$ & $m+2$ & $m-2$ & ${\bf 1}$ & $\bar\chi$ with $\nu=|n+2|$ \\
\hline
\end{tabular}
\end{center}
\caption{\small Content of the chiral multiplet ${\cal E}_m$ with $m\ge 2$.\la{7011}}
\end{table}

With $p_0$, $p_T$, $p_F$ defined above \rf{1000}, i.e.
\ba
& p_0(\nu)=3\nu^5-5\nu^3\ , \quad \quad p_T(\nu)=9\nu^5-75\nu^3+180\nu\ , \quad \quad  p_F(\nu)=-6\nu^5+25\nu^3-\tfrac{135}{8}\nu \ , \la{7023}
\ea
the contribution of ${\cal E}_m$ and of its conjugate is $P_m=360(m-1)$ in \rf{1000}, i.e. $\aa=\tfrac{1}{4}(m-1)$. The terms of order $m^5$ and $m^3$ cancel.

Note that for a field in the representation $(\Delta;j_1,j_2)$ of the AdS$_5$ isometry group the difference of the two conformal anomaly coefficients $\cc$ and $\aa$ is given by
(see \ci{ArabiArdehali:2013jiu,Beccaria:2014xda})
\ba
\cc(\Delta;j_1,j_2)-\aa(\Delta;j_1,j_2)=&-\tfrac{1}{360}(-1)^{2(j_1+j_2)}(2j_1+1)(2j_2+1)\, \nu\, \big[1+h(j_1)+h(j_2)\big] \ , \la{7024}\\
&h(j)\equiv j(j+1)\big[6j(j+1)-7\big] \ , \la{7025}
\ea
which is to be used on complete supermultiplets only. It gives $\cc-\aa=-\tfrac{1}{4}(m-1)$ for the same pair of multiplets, so that they have $\cc=0$.

The remaining modes are the  $n=1$ one  of $\chi$ with $\nu=1$, the  $n=0$ of the 3 scalars in \rf{1445} with $\nu=0$,
the fermions with $n=\pm\ha$ and $\nu=\ha$, and the $n=0$ mode of the tensor.
With the Dirichlet conditions the dimensions are $\Delta=3$, $2$ and $\tfrac{5}{2}$.
This is the content of the flavour current multiplet of the 4d $\N=2$ supersymmetry, which contains the triplet of moment maps with $\Delta=2$,
the fermions with $\Delta=\tfrac{5}{2}$, a complex scalar with $\Delta=3$ and a conserved current with $\Delta=3$.
The conserved current with $\Delta=3$  corresponds 
in  AdS$_5$  to a massless vector with the boundary value of its potential kept fixed.
The only candidate is the $n=0$ mode of the tensor.
Its components $A_{i5}$ and $A_{ij}$ are related by the self-duality condition, so that it describes one massless 5d vector.
With this boundary condition it contributes
\be\la{7026}
P_V=1440\,\big[a(3;\ha,\ha)-a(4;0,0)\big]=-68-56=-124 \ ,
\ee
where the second term is the contribution of the ghost. 
This gives \rf{011}, and using also \rf{7024} one finds for the current multiplet
\be\la{7047}
\aa_{\rm cur}=\tfrac{1}{1440}P_1=-\tfrac{5}{48}\ , \qquad \qquad \cc_{\rm cur}=-\tfrac{1}{12} \ .
\ee
These are $-\ha$ of the values $\aa=\tfrac{5}{24}$, $\cc=\tfrac{1}{6}$ for the free 4d $\N=2$ vector multiplet,
as should be the case for a 5d vector multiplet with the standard boundary conditions.

The presence of this multiplet means that the 4d defect theory has a global $U(1)$ symmetry.
Its conserved current is dual to the $n=0$ mode of the tensor.
The 3 scalar operators with $\Delta=2$ in the same multiplet (a triplet of $SU(2)_R$)
are dual to the $n=0$ modes of the 3 scalars in \rf{1445}.

The defect also has a complex modulus. Its absolute value is determined by the position $u_0$ of the brane,
and its phase is the constant that can be added to the relation $\phi=2\psi$ in \rf{51}.
The corresponding two exactly marginal operators have $\Delta=4$.
They are dual to the $n=0$ mode of the complex scalar $\chi$, i.e. to the fluctuations of $u$ and of $\phi$,
for which \rf{76} gives $\Delta-2=2$.
In terms of the multiplets in Table \ref{7011} this mode is the top component $\varphi''$ of ${\cal E}_2$, with $\Delta=4$ and $r=0$,
together with its conjugate.

\subsection{Summation prescription:  comparison with 1-loop supergravity on AdS$_5\times S^5/{\mathbb Z}_k$ orbifold}\la{7003}

The sum \rf{7012} is quadratically divergent and its finite part depends on the prescription.
The Riemann $\zeta$-function in $m-1$ gives $P=-180$, the sharp cutoff gives $P=-150$,
and the factor $e^{-\epsilon m}$ with the poles in $\epsilon$ dropped gives $P=0$.
The spectral $\zeta$-function applied field by field, with the $n=0$ mode \rf{7026} included, gives $P=-\tfrac{313}{8}-124$.
The first three prescriptions assign a common weight to each complete supermultiplet but give different finite parts.
Thus the completeness of the supermultiplets alone does not select the answer.
%The field by field continuation is included as an additional comparison. 
%Its compatibility with the full supersymmetric boundary problem 
%is not established here.

There is an  analogous  problem with the same spectrum.
The twisted sector of type IIB string  theory on AdS$_5\times S^5/{\mathbb Z}_k$ consists of $k-1$ tensor multiplets of (2,0) supersymmetry
on AdS$_5\times S^1$ with equal radii \ci{Gukov:1998kk}.
The dual states are the twisted chiral operators with $\Delta=m\ge 2$ and the $k-1$ baryonic currents
(for $k=2$ the twisted multiplets are listed in \ci{ArabiArdehali:2013jiu}).
Thus each twisted sector has the same content as found above, i.e. the multiplets ${\cal E}_m$ with $m\ge 2$ and one current multiplet.

The dual theory is the $\N=2$ quiver with the gauge group $[SU(N)]^k$ and $k$ bi-fundamental hypermultiplets
\ci{Kachru:1998ys,Lawrence:1998ja,Gukov:1998kk}.
It has $n_v=k(N^2-1)$ vector multiplets and $n_h=kN^2$ hypermultiplets, so that from
$\aa=\tfrac{1}{24}(5n_v+n_h)$ and $\cc=\tfrac{1}{12}(2n_v+n_h)$ \ci{Anselmi:1997ys} one finds
\be\la{7027}
\aa=\tfrac{1}{4}kN^2-\tfrac{5}{24}k\ , \qquad \qquad \cc=\tfrac{1}{4}kN^2-\tfrac{1}{6}k \ .
\ee
The order $N^0$ terms should be reproduced by the 1-loop contributions of the untwisted and of the twisted sectors.

%\paragraph{Untwisted sector.}
The Kaluza-Klein multiplet of level $p\ge 2$ of type IIB supergravity on AdS$_5\times S^5$ is generated from the scalar with $\Delta=p$ in the
representation $[0,p,0]$ of $SU(4)$. Its fields are labelled by a pair of entries of Table \ref{7028}, one for the chiral and one for the antichiral supercharges.
A field has the dimension $\Delta=p+\delta+\bar\delta$, the spins $(j,\bar j)$ and the $SU(4)$ Dynkin labels equal to $[0,p,0]$ plus the two shifts.
For the antichiral part the shifts are the mirror images of those in the table.

\begin{table}[h]
\begin{center}
\begin{tabular}{|c|c|c|c|c|c|c|}
\hline
$\delta$ & 0 & $\ha$ & 1 & 1 & $\tfrac{3}{2}$ & 2 \\
\hline
$j$ & 0 & $\ha$ & 1 & 0 & $\ha$ & 0 \\
\hline
shift & $[0,0,0]$ & $[0,-1,1]$ & $[0,-1,0]$ & $[0,-2,2]$ & $[0,-2,1]$ & $[0,-2,0]$ \\
\hline
\end{tabular}
\end{center}
\caption{\small The chiral half of the level $p$ multiplet.\la{7028}}
\end{table}

The generator of ${\mathbb Z}_k$ acts on the fundamental representation of $SU(4)$ as ${\rm diag}(\omega,\omega^{-1},1,1)$ with $\omega=e^{2\pi i/k}$.
The dimensions of the representations and their characters on this generator can be computed from the Jacobi-Trudi determinant, with the complete symmetric functions
$h_n(\omega) $ defined by
\be\la{7029}
\sum_{n=0}^\infty h_n(\omega) \, t^n=\frac{1}{(1-\omega t)(1-\omega^{-1}t)(1-t)^2} \ , \qquad \qquad \ \  \omega=e^{2\pi i/k} \ . 
\ee
This also accounts for the states that are absent at low $p$. 
Using \rf{3.3} and \rf{7024} the full multiplet gives
\be\la{7030}
\aa_p=\cc_p=\tfrac{1}{4}p \ .
\ee
For $k=2$ the ${\mathbb Z}_2$ invariant part of the multiplet  at level $p$ gives
\be\la{7031}
\aa_p^{\rm inv}=\tfrac{1}{4}p-\tfrac{1}{24}(-1)^p\, p\,(6p^2-7)\ , \qquad \qquad
\cc_p^{\rm inv}=\tfrac{1}{8}p-\tfrac{1}{24}(-1)^p\, p\,(6p^2-5) \ .
\ee
The oscillating sums are convergent in the Abel sense, i.e. they are defined as the limit $x\to 1$
of the sums with the extra factor $x^p$, which is equivalent to the $\zeta$-function value. This gives 
\be\la{7032}
\sum_{p=2}^\infty(-1)^p\, p\,(6p^2-7)=\tfrac{3}{2}\ , \qquad\qquad \sum_{p=2}^\infty(-1)^p\, p\,(6p^2-5)=3 \ .
\ee
Let $\Sigma$ denote the regularized value of the level sum  
$\sum_{p\ge 2}p$. Then the untwisted sector contributes
\be\la{7033}
\aa_{\rm untw}=\tfrac{1}{4}\Sigma-\tfrac{1}{16}\ , \qquad \qquad \cc_{\rm untw}=\tfrac{1}{8}\Sigma-\tfrac{1}{8} \ , \ \ \ \ \ \ \ \ \ \ 
\Sigma \equiv \sum^\infty _{p= 2}p \ . 
\ee
The sharp cutoff prescription, in which $\sum_{p=1}^{\infty}p\to 0$, gives $\Sigma=-1$. This is the value required in the case of
$k=1$, i.e.  $\AdS_5\times S^5$,
where $\sum_{p\ge 2}\tfrac{1}{4}p$ should be equal to $-\tfrac{1}{4}$ to match the ${1\ov 4} (N^2-1)$ conformal anomaly of the $\N=4$ SYM theory  \ci{Mansfield:2002pa,Beccaria:2014xda}. 

Then for $k=2$   we get 
\be\la{7034}
k=2: \qquad \qquad  \aa_{\rm untw}=-\tfrac{5}{16}\ , \qquad\qquad \cc_{\rm untw}=-\tfrac{1}{4} \ .
\ee
For $k=3,4,...$ the invariant parts are sums of terms of the form $\omega^{jp}$ times a polynomial in $p$.
Summing the terms with $j\not=0$ by the Abel summation and the terms with $j=0$ with the sharp cutoff one finds the values in Table \ref{7035}, i.e.\foot{We have  checked this numerically  for $k=2,\dots,10$.}
\be\la{7036}
k\geq 2: \qquad \qquad  \aa_{\rm untw}=-\tfrac{5}{48}(k+1)\ , \qquad\qquad \cc_{\rm untw}=-\tfrac{1}{12}(k+1) \ .
\ee
\begin{table}[h]
\begin{center}
\begin{tabular}{|c|c|c|c|c|}
\hline
$k$ & $\aa_{\rm untw}$ & $\cc_{\rm untw}$ & order $N^0$ term of $\aa$ & order $N^0$ term of $\cc$ \\
\hline
1 & $-\tfrac{1}{4}$ & $-\tfrac{1}{4}$ & $-\tfrac{1}{4}$ & $-\tfrac{1}{4}$ \\
2 & $-\tfrac{5}{16}$ & $-\tfrac{1}{4}$ & $-\tfrac{5}{12}$ & $-\tfrac{1}{3}$ \\
3 & $-\tfrac{5}{12}$ & $-\tfrac{1}{3}$ & $-\tfrac{5}{8}$ & $-\tfrac{1}{2}$ \\
4 & $-\tfrac{25}{48}$ & $-\tfrac{5}{12}$ & $-\tfrac{5}{6}$ & $-\tfrac{2}{3}$ \\
6 & $-\tfrac{35}{48}$ & $-\tfrac{7}{12}$ & $-\tfrac{5}{4}$ & $-1$ \\
\hline
\end{tabular}
\end{center}
\caption{\small  The untwisted sector contributions and the exact values of the order $N^0$ terms.
For $k\ge 2$ the latter are given by \rf{7027}. For $k=1$ they are those of the $\N=4$ $SU(N)$ theory,
where the hypermultiplet is in the adjoint representation and there is no twisted sector.\la{7035}}
\end{table}

As for the twisted sectors, 
comparing \rf{7036} with \rf{7027}, each of the $k-1$ of them  must contribute
\be\la{7037}
\aa_{\rm tw}=-\tfrac{5}{48}\ , \qquad \qquad \cc_{\rm tw}=-\tfrac{1}{12} \ .
\ee
This is exactly the contribution \rf{7047} of the current multiplet alone.

The exact order $N^0$ terms in \rf{7027} are thus reproduced by the sum of the 1-loop contributions of the two sectors
if the sums are defined as follows:

(i) In the untwisted sector the sum over the Kaluza-Klein level $p$ is regularized by the factor $z^p$,
and the poles at $z=1$ are dropped, as in \ci{Mansfield:2002pa,ArabiArdehali:2013jiu}.
For the non-oscillating terms, which are linear in $p$, this is equivalent to $\sum_{p=1}^\infty p\to 0$.

(ii) In each twisted sector the sum over the multiplets ${\cal E}_m$ in \rf{1000} is regularized in the same way,
with $p=m-1$ being the Kaluza-Klein level on $S^1$. Then the tower of ${\cal E}_m$ does not contribute.
One is left with the current multiplet, in which the $n=0$ mode of the tensor is included as in \rf{7026}.

With this prescription the twisted sectors contribute \rf{7047} each, and together with \rf{7036} one gets for $k\geq 2 $
\be\la{7048}
\aa=-\tfrac{5}{48}(k+1)-\tfrac{5}{48}(k-1)=-\tfrac{5}{24}k\ , \qquad\qquad
\cc=-\tfrac{1}{12}(k+1)-\tfrac{1}{12}(k-1)=-\tfrac{1}{6}k \ ,
\ee
in agreement with  the  expected dual orbifold  gauge theory  values in   \rf{7027}.

The agreement for $\cc$ is a check of the treatment of the $n=0$ mode of the tensor.
Each pair of ${\cal E}_m$ and its conjugate has $\cc=0$, so that the twisted value of $\cc$ does not depend on how the tower is summed.
Without the $n=0$ mode of the tensor one would get $\cc_{\rm tw}=-\tfrac{1}{30}$ instead of $-\tfrac{1}{12}$.

The agreement for $\aa$-anomaly  fixes the prescription for the twisted tower only relative to the one used in the untwisted sector.
For example, the Riemann $\zeta$-function value of the sum over $m$ would give $\aa_{\rm tw}=-\tfrac{1}{8}$,
which does not lead to \rf{7027} if (i) is used in the untwisted sector.
However, the split of the total between the two sectors depends on the regulator.
If the sum over the states is defined as in \ci{ArabiArdehali:2014cey},
i.e. by the superconformal index with the finite part at $t=1$ of the fugacity $t$ kept,
the tower of ${\cal E}_m$ contributes $(\aa,\cc)=(\tfrac{1}{48},0)$ instead of zero, so that
\be\la{7051}
\aa_{\rm tw}=-\tfrac{1}{12}\ , \qquad\qquad \cc_{\rm tw}=-\tfrac{1}{12} \ .
\ee
The untwisted sector then gives $\aa_{\rm untw}=-\tfrac{1}{24}(3k+2)$ instead of \rf{7036},
and the total is again \rf{7027}.

For $k=2$ the difference $\cc-\aa$ was computed with the prescription (i),(ii) in \ci{ArabiArdehali:2013jiu}.
The finite parts found there, $\tfrac{1}{16}$ for the untwisted sector, $\tfrac{1}{48}$ for the $p=0$ twisted multiplet
and $0$ for the twisted tower, agree with \rf{7036} and \rf{7037}.
Only $\cc-\aa$ was considered there, as the long multiplets of $SU(2,2|1)$ do not contribute to it,
while they were expected to contribute to $\aa$ and $\cc$ separately.
In fact, the long multiplets do not contribute to $\aa$ and $\cc$ separately \ci{Beccaria:2014xda}:
with the expressions \rf{3.3} and \rf{7024} the sum over the component fields of a long multiplet,
with a primary of any dimension $\Delta$ and any spins $(j_1,j_2)$, vanishes identically.

Using this, the matching of the order $N^0$ terms in $\aa$ and $\cc$ separately was reported in \ci{ArabiArdehali:2014cey},
for the $\mathbb Z_2$ orbifold with the index prescription,
and for the $\N=1$ orbifolds of \ci{ArabiArdehali:2013jiu} with the factor $z^p$.
The index approach was extended in \ci{ArabiArdehali:2015iow},
and the values \rf{7027} for general $k$ were discussed in \ci{Nakayama:2026bls}.
The computation above adds the explicit sum over the spectrum for the $\N=2$ orbifolds with $k>2$
and the split of the result between the untwisted and the twisted sectors.

Let us comment on the relation to the computation in terms of characters.
The contribution \rf{3.3} of a field to the $a$-anomaly can be obtained from the character of the corresponding
representation of $so(2,4)$ \ci{Bae:2016rgm,Basile:2018zoy}.
In AdS$_5$ the character integral reduces to a few coefficients in the expansion of the character at small $\tau$,
where $e^{-\tau}$ is the weight of the AdS energy. Applied to each multiplet separately this reproduces \rf{1000} and \rf{011}.
One may instead first sum the characters over the tower, which converges for $\tau>0$, and then extract these coefficients.
This is a definite prescription. It is equivalent to the spectral $\zeta$-function of the complete tower,
i.e. to the field by field assignment mentioned at the beginning of this section,
and for the M5 brane it gives $P=-\tfrac{1305}{8}$, i.e. $a^{(1)}=-\tfrac{29}{256}$.
Thus the use of characters does not remove the dependence on the prescription.
The question of how to sum over the levels is replaced by the question of whether the sum is done
before or after the extraction of the anomaly coefficient.

Let us now return to the present case of the M5 brane in AdS$_7 \times S^4$.
Its fluctuations on AdS$_5 \times S^1$ form the same multiplets as one twisted sector of the $S^5/\mathbb Z_k$ orbifold,
i.e. the tower of ${\cal E}_m$ with $m\ge 2$ and one current multiplet.
We adopt the prescription (ii), i.e. the finite part at $z=1$ of the sum in which each complete multiplet is weighted by $z^p$ with $p=m-1$.
This choice is consistent with the orbifold comparison when the prescription (i) is used in the untwisted sector.
The total orbifold anomaly does not select it uniquely, as the index prescription gives the same total
with a different split between the sectors.
Its use for the M5 brane fluctuations is therefore an additional assumption.

%We thus  define the sum \rf{7012} by the prescription (ii)  above  that is required in the orbifold case.
The contribution $360(m-1)$ of ${\cal E}_m$ and its conjugate is multiplied by $z^{p}$,
where $p=m-1$ is the mode number on $S^1$ of the tensor and of the 3 scalars in this multiplet, and the poles at $z=1$ are dropped,\foot{The finite part of the sum regularized in this way depends on the power of $z$ used.
For the factor $z^{p+v}$ with a constant $v$ one finds
$\sum_{p=1}^\infty p\, z^{p+v}=\frac{1}{(z-1)^2}+\frac{v+1}{z-1}+\ha v(v+1)+{\cal O}(z-1)$,
so that the finite part vanishes for $v=0$ but is equal to 1 for $v=1$, i.e. for the factor $z^{m}$.
The choice $v=0$, where $p=m-1$ is the Kaluza-Klein level on $S^1$, is the one made in \ci{ArabiArdehali:2013jiu}
and required for the agreement with \rf{7027} in the orbifold case.
It is also the one that agrees with the sharp cutoff, for which $\sum_{m=2}^{M}(m-1)=\ha M(M-1)$
has no constant term whether it is written in terms of $M$ or of $M-1$.}
\be\la{7049}
\sum_{p=1}^\infty 360\, p\, z^{p}=\frac{360}{(z-1)^2}+\frac{360}{z-1}\ \to\ 0 \ .
\ee
The same result is found with the sharp cutoff $m\le M$ if the terms growing with $M$ are dropped.
Thus the tower does not contribute, and the 1-loop result is that of the current multiplet,
\be\la{7050}
P=P_1=-150\ , \qquad\qquad a^{(1)}=\tfrac{1}{1440}P=-\tfrac{5}{48} \ ,
\ee
which is the value in \rf{201}.
The Riemann $\zeta$-function prescription, i.e. the factor $e^{-\epsilon p}$ with the poles at $\epsilon=0$ dropped, would give instead
$P=-180$ and $a^{(1)}=-\tfrac{1}{8}$. It is excluded in the orbifold case by the requirement of matching the value of $\aa$.

The index prescription of \ci{ArabiArdehali:2014cey} is not excluded in this way.
Applied to the M5 brane it would give \rf{7051}, i.e. $P=-120$ and $a^{(1)}=-\tfrac{1}{12}$.
In the orbifold case the difference of $\tfrac{1}{48}$ between the two prescriptions is compensated by the untwisted sector.
In the M5 brane case there is no such sector, so that the two values are different definitions of the 1-loop term.
%The choice between them should be correlated with the definition of the order $N^0$ part of the classical response of the background,
%which is not computed here.

\subsection{Possible term of order $N$ in defect anomaly in M5 brane case}\la{7004}

The bulk  11d supergravity action contribution 
 scales as $N^3$ and the M5 brane action contribution 
  as $N^2$, so that the response $\delta g$ of the background fields to the probe is of order $1/N$.
The quadratic term $(\delta g)^2$  in the bulk action and the linear term in the probe action evaluated on this response can then be of order $N$,
while the cubic and the quadratic terms respectively can be of order $N^0$,
\be\la{7045}
N^3(\delta g)^2\sim N^2\, \delta g\sim N\ , \qquad\qquad N^3(\delta g)^3\sim N^2(\delta g)^2\sim N^0 \ .
\ee
This is  the reason for allowing the terms of order $N$ in \rf{202}.
In the D3 brane case the bulk  10d  action scales as $N^2$ and the probe action as $N$, and the same estimate gives the terms of order $N^0$.

The coefficient of the term of order $N$ in \rf{202} can be guessed from the 6d analog of \rf{7008}.
For a 6d theory  let $\bar \aa$ be the conformal $a$-anomaly coefficient, 
 normalized so that  %the universal part of the entanglement entropy
%of a spherical region of radius $R$ is $4A\log(R/\epsilon)$, where $\epsilon$ is a short-distance cutoff.
 the free (2,0) tensor multiplet has $\bar \aa=\tfrac{7}{48}$. This is $-24$ times the coefficient $-\tfrac{7}{1152}$ in the normalization of \ci{Bastianelli:2000hi}.
For the (2,0) theory of type $A_{N-1}$ one has, in the same units (see, e.g., \ci{Cordova:2015fha})
\be\la{7038}
\bar \aa(SU(N))=\tfrac{1}{48}\big(16N^3-9N-7\big) \ .
\ee
The conjecture is that the anomaly of the defect which breaks  group $G$ to $L$ is
\be\la{7039}
a=\bar \aa (G)-\bar \aa (L) \ .
\ee
For $G=SU(N)$ and $L=S(U(1)\times U(N-1))$ this gives
\be\la{7040}
a=\bar \aa(SU(N))-\bar \aa(SU(N-1))-\tfrac{7}{48}=N^2-N \ .
\ee
The leading term agrees with the classical M5 brane result \rf{001}, including the coefficient. The term $-N$ corresponds to $N^2\to  N(N-1)$.
    The relation \rf{7039} is a proposed relation between a difference of 6d anomalies and the 4d defect anomaly.
   % The tensor branch results of \ci{Cordova:2015fha} do not by themselves establish it.
    %We use it only to motivate the value $-1$ of the coefficient of the term of order $N$.  %The term of order $N^0$ is not determined.
    It  is the 6d analog of the relation $b=12\,\Delta\aa$
derived in 4d in \ci{Cuomo:2026kuo}, where  an  extension to the (2,0) theory was
suggested. A similar relation was found there in 3d: the
defect free energy of the vortex loops in the ABJM theory is the difference of
the free energies on $S^3$ of the theories with $N$ and $N-M$ colours, up to the
contribution of the free multiplets, in agreement with the M2 brane results of
\ci{Drukker:2008jm,Drukker:2023bip}.

Let us note that if \rf{7040} is exact, the  term ${\rm e}_2$  in \rf{202} vanishes.
In \rf{7040} the contribution of the free tensor multiplet is $-\tfrac{7}{48}$,
which is the coefficient of the logarithmic term in its free energy on $S^6$,
while the remaining part is $\bar\aa(SU(N))-\bar\aa(SU(N-1))=N^2-N+\tfrac{7}{48}$.
This differs from \rf{201} and \rf{7052}, which should therefore not be viewed as the full constant term.

%The relation \rf{7039} is a new relation between a 6d and a 4d anomaly, and the tensor branch results of \ci{Cordova:2015fha} do not by themselves establish it.
For a general Levi group $L=S(U(N_1)\times \cdots \times U(N_k))$ eq. \rf{7039} would give
\be\la{7041}
a=\tfrac{1}{3}\big(N^3-\sum_{i=1}^k N_i^3\big) \ .
\ee
%The term of order $N^0$ remains open. The fluctuations contribute \rf{201}.
%The nonlinear response of the background and the higher-derivative terms may also contribute at this order. These contributions have not been calculated here.
%\paragraph{Comparison with the bubbling solutions.}
The expression \rf{7041} scales as $N^3$ when all $N_i$ are of order $N$ and as $N^2$ in the probe limit.
Let us compare it with the supergravity expression in  \ci{Capuozzo:2023fll}, which depends on the ranks $N_a$ and on the additional integers $k_a$.
For the solution with two ordered ranks $N_1=p<N_2=N-p$ and with $k_1=k_2=1$ it gives
\ba
a&=\tfrac{1}{32}N^3-\tfrac{1}{96}\big[3p^3+3(N-p)^3+p\,\big(p^2+3(N-p)^2\big)\big] =\tfrac{1}{16}pN^2-\tfrac{1}{32}p^2N-\tfrac{1}{24}p^3 \ , \la{7042}
\ea
while \rf{7041} gives $pN^2-p^2N$.
Thus at fixed $p$ the supergravity expression has the $N^2$ scaling, and it contains a term of order $N$, in agreement with \rf{202}.\foot{The two expressions do not agree by an overall normalization only: multiplying \rf{7042} by 16 gives the term $-\tfrac{1}{2}p^2N$ instead of $-p^2N$.
Thus the equality of the scaling is established, while the equality of the defects and of the anomaly coefficients is not.
The identification of this member of the bubbling family with the defect described by the probe is not known.
    Without an identification of this bubbling solution with the probe defect and a matching of the normalizations,
    \rf{7042} does not yet provide a test of the conjecture \rf{7040} for the probe, while it does not support the general expression \rf{7041}.}

%\paragraph{Which defect is described by the probe.}
\iffa 
Let $z$ be the complex coordinate in the plane transverse to the defect in ${\mathbb R}^6$ and $w$ the complex coordinate in the plane of the two scalars
that form the complex scalar $\Phi$ of the (2,0) theory. The solution \rf{51} with $\phi=2\psi$ is the holomorphic curve
\be\la{7043}
z^2\, w=c \ , \qquad\qquad {\rm i.e.}\qquad \Phi\sim \frac{c}{z^2}
\ee
for one eigenvalue of $\Phi$.
To see this consider a planar defect, and let $v$ be the Poincar\'e radial coordinate of AdS$_5$ in \rf{622}.
The Poincar\'e radial coordinate $y$ of AdS$_7$ and the distance $\rho$ from the defect in the boundary are $y=v/\cosh u$ and $\rho=v\tanh u$.
The value of the scalar that corresponds to a brane at the position $y$ is proportional to $y^{-2}$, so that with a suitable choice of the orientations
$z=\rho\, e^{i\psi}$ and $w=C\, y^{-2}e^{-i\phi}$, where $C$ is a normalization constant.
Then \rf{51} gives \rf{7043} with $c=C\sinh^2u_0$.

The power of $z$ is fixed by scale invariance as $\Phi$ has dimension 2. This is the analog of the profile $\Phi\sim c/z$ of the surface defect of \ci{Gukov:2006jk} in 4d.
At large $|z|$ the brane is one of the $N$ branes of the stack, and near $z=0$ it wraps the plane of $w$ twice.
This is a description of the abelian probe, and it does not by itself give the complete classification of the corresponding defect of the non-abelian theory.

In particular, the relation of this defect to the regular punctures of class S is not established.
The simple regular puncture is described by a single transverse brane at $z=0$ and has no modulus $c$.
Its effective numbers of vector multiplets and hypermultiplets are $N^2-1$ and $N^2$ \ci{Chacaltana:2010ks}, which gives the additive contribution
$\delta \aa=\tfrac{1}{4}N^2-\tfrac{5}{24}$ to the conformal anomaly of the compactified theory. This differs from \rf{202} already at the leading order.
Note, however, that the relation between such a contribution of a puncture and the anomaly of the defect in flat space would require a further argument.
\fi
 
 \newpage 
 \small
 \ed